\documentclass[showpacs,showkeys,preprintnumbers,amsmath,amssymb,latexsym]{revtex4}

\usepackage{graphicx}
\usepackage{dcolumn}
\usepackage{bm}

\makeatletter    % make @ like a letter

\def\meqalign#1{\null\,\vcenter{\openup\jot\m@th \let\\=\crcr % more latexlike
  \ialign{\strut\hfil$\displaystyle{##}$&&$\displaystyle{{}##}$\hfil
      \crcr#1\crcr}}\,}

\makeatother   % back to nonletter category

\begin{document}

\title{Charged massive particle at rest in the field of a Reissner-Nordstr\"om  black hole}

\author{D. Bini}
  \email{binid@icra.it}
 \affiliation{Istituto per le Applicazioni del Calcolo ``M. Picone'', CNR I-00161 Rome, Italy and\\ 
International Center for Relativistic Astrophysics - I.C.R.A.\\
University of Rome ``La Sapienza'', I-00185 Rome, Italy}
\author{A. Geralico}
  \email{geralico@icra.it}
 \affiliation{International Center for Relativistic Astrophysics - I.C.R.A.\\
University of Rome ``La Sapienza'', I-00185 Rome, Italy}
\author{R. Ruffini}
 \email{ruffini@icra.it}
\affiliation{Physics Department and \\
International Center for Relativistic Astrophysics - I.C.R.A.\\
University of Rome ``La Sapienza'', I-00185 Rome, Italy}

\date{\today}

\begin{abstract}
The interaction of a Reissner-Nordstr\"om black hole and a charged massive particle is studied in the framework of perturbation theory. 
The particle backreaction is taken into account, studying the effect of general static perturbations of the hole following the approach of Zerilli.
The solutions of the combined Einstein-Maxwell equations for both perturbed gravitational and electromagnetic fields at first order of the perturbation are exactly reconstructed by summing all multipoles, and are given explicit closed form expressions. 
The existence of a singularity-free solution of the Einstein-Maxwell system requires that the charge to mass ratios of the black hole and of the particle satisfy an equilibrium condition which is in general dependent on the separation between the two bodies. 
If the black hole is undercritically charged (i.e. its charge to mass ratio is less than one), the particle must be overcritically charged, in the sense that the particle must have a charge to mass ratio greater than one. If the charge to mass ratios of the black hole and of the particle are both equal to one (so that they are both critically charged, or \lq\lq extreme''), the equilibrium can exist for any separation distance, and the solution we find coincides with the linearization in the present context of the well known Majumdar-Papapetrou solution for two extreme Reissner-Nordstr\"om black holes. In addition to these singularity-free solutions, we also analyze the corresponding solution for the problem of a massive particle at rest near a Schwarzschild black hole, exhibiting a strut singularity on the axis between the two bodies. The relations between our perturbative solutions and the corresponding exact two-body solutions belonging to the Weyl class are also discussed. 
\end{abstract}



\maketitle

\section{Introduction}

The problem of the effect of gravity on the electromagnetic field of a charged particle leading to the consideration of the Einstein-Maxwell 
equations has been one of the most extensively treated in the literature, resulting in exact solutions 
(see \cite{exactsols} and references therein) as well as in a variety of approximation methods 
\cite{fermi,whittaker,copson,hannijth,CoW,HR,bicdev,linet,leaute}.

We address in this article the problem of a massive charged particle of mass $m$ and charge $q$ in the field of a Reissner-Nordstr\"om geometry describing a static charged black hole, with mass $\mathcal{M}$ and charge $Q$. We solve this problem by the first order perturbation approach formulated by Zerilli \cite{Zerilli} using the tensor harmonic expansion of the Einstein-Maxwell system of equations. 
Note that a similar approach was used by Sibgatullin and Alekseev \cite{sibgatullin} to study electrovacuum perturbations of a Reissner-Nordstr\"om black hole. A manifestly gauge invariant (hamiltonian) formalism was developed by Moncrief \cite{moncrief1,moncrief2,moncrief3}. The relations between these two different treatments of perturbations have been extensively investigated by Bi{\v c}\'ak \cite{bicak}.

Both the mass and charge of the particle are described by delta functions. 
The source terms of the Einstein equations contain the energy-momentum tensor associated with the particle's mass, the electromagnetic energy-momentum tensor associated with the background field as well as additional interaction terms, to first order in $m$ and $q$, proportional to the product of the square of the charge of the background geometry and the particle's mass ($\sim Q^2m$) and to the product of the charges of both the particle and the black hole ($\sim qQ$). These terms give origin to the so called \lq\lq electromagnetically induced gravitational perturbation'' \cite{jrz2}. 
On the other hand, the source terms of the Maxwell equations contain the electromagnetic current associated with the particle's charge as well as interaction terms proportional to the product of the black hole's charge and the particle's mass ($\sim Qm$), giving origin to the \lq\lq gravitationally induced electromagnetic perturbation'' \cite{jrz1}.  
We give in Section II the basic equations to be integrated in order to solve this problem. 

Before proceeding to the solution of the most general problem, we recall a variety of particular approaches followed  in the current literature within the simplified framework of test field approximations \cite{HR,linet,leaute,bonnor}. 
In Section III, the test field solutions for a charged particle at rest in the field of a Schwarzschild as well as a Reissner-Nordstr\"om black hole are summarized, under the conditions $q/m\gg1$, $m\approx0$ and $q\ll{\mathcal M}$, $q\ll Q$. The case of a test charge in the Schwarzschild black hole is reviewed first in Section III A; the corresponding solution is obtained by using the vector harmonic expansion of the electromagnetic field in curved space \cite{HR}. The conditions above imply the solution of the Maxwell equations only in a fixed Schwarzschild metric, since the perturbation to the background geometry given by the electromagnetic stress-energy tensor is second order in the particle's charge. As a result, no constraint on the position of the test particle follows from the Einstein equations and the Bianchi identities: any position of the particle is allowed, and the solution for the corresponding electromagnetic field has been widely analyzed \cite{HR,linet}. 

In Section III B we recall this same test field approximation applied to the case of a Reissner-Nordstr\"om black hole.
This treatment is due to Leaute and Linet \cite{leaute}; in analogy with the Schwarzschild case, they used the vector harmonic expansion of the electromagnetic field holding the background geometry fixed. However, this \lq\lq test field approximation'' is not valid in the present context. In fact, in addition to neglecting the effect of the particle mass on the background geometry, this treatment also neglects the electromagnetic induced gravitational perturbation terms linear in the charge of the particle which would contribute to modifying the metric as well. 
In Section III C we recall how, within this same approximation still neglecting these same feedback terms, Bonnor \cite{bonnor} studied the condition for the equilibrium involving the black hole and particle parameters $Q, {\mathcal M}, q, m$ as well as their separation distance $b$. He finds the constraint
\begin{equation}
\label{bonnoreqcond0}
m=\frac{qQb}{{\mathcal M}b-Q^2}\left(1-\frac{2{\mathcal M}}{b}+\frac{Q^2}{b^2}\right)^{1/2}\ ,
\end{equation}  
If the black hole is \lq\lq extreme'' or \lq\lq critically charged'' satisfying $Q/\mathcal{M}=1$, then the particle must also have the same ratio $q/m=1$, and equilibrium exists independent of the separation.
In the general non-extreme case $Q/\mathcal{M}<1$ there is instead only one position of the particle which corresponds to equilibrium, for given values of the charge-to-mass ratios of the bodies. In this case the particle charge-to-mass ratio satisfies $q/m>1$. This condition corresponds to the overcritically charged case for a Reissner-Nordstr\"om solution with the same parameters $m$ and $q$ corresponding to a naked singularity.

Already the Leaute and Linet generalization modifies in a significant way 
the results obtained in the case of a Schwarzschild black hole. Indeed a new phenomenon occurs in this case: as the hole becomes extreme an effect analogous to the Meissner effect for the electric field arises, with the electric field lines of the test charge being forced outside the outer horizon \cite{elbaproc}. 
This result by itself justifies the need of addressing this problem in the more general Zerilli approach duly taking into account all the first order perturbations.

The correct way to attack the problem is, thus, to solve the linearized Einstein-Maxwell equations following Zerilli's first order tensor harmonic analysis \cite{Zerilli}.   
In Section \ref{schwpert} we first consider the simplest special case of a massive neutral particle at rest near a Schwarzschild black hole. We first show explicitly that a perturbative solution for this problem free of singularities cannot exist. We then give the explicit form of the perturbation  corresponding to a stable configuration when there is the presence of a \lq\lq strut'' between the particle and the black hole, corresponding to a conical singularity. This result is a direct consequence of the classical work by Einstein and Rosen \cite{einstein}, 
who considered the Weyl class double Chazy-Curzon solution \cite{curzon} for two point masses placed on the symmetry axis at a fixed distance. 
We find it very helpful to use a new gauge condition particularly adapted to this problem which differs from the corresponding Regge-Wheeler gauge in the Schwarzschild case, by adding additional terms to the third Regge-Wheeler gauge condition to take into account the additional terms in the field equations.
The resulting perturbed metric we obtain is the linearized form of the exact solution representing two collinear uncharged black holes in a static configuration belonging to the Weyl class \cite{exactsols}.
We review some of their properties and peculiar features of these solutions in the Appendix. 

We then turn to the general case of a charged massive particle at rest in a Reissner-Nordstr\"om background. 
This problem needs the simultaneous analysis of both the vectorial perturbation to describe the electromagnetic perturbation and the tensorial perturbation necessary to analyze the perturbation by the particle mass as well as
the electromagnetically induced gravitational perturbation and the gravitationally induced electromagnetic perturbation.

The Einstein-Maxwell equations reduce to a system of $6$ coupled ordinary differential equations for $4$ unknown functions determining both the gravitational and electromagnetic perturbed fields. 
Then, by imposing the compatibility of the system (or equivalently, requiring that the Bianchi identities be fulfilled), we obtain an equilibrium condition for the system which coincides with the condition (\ref{bonnoreqcond0}) obtained by Bonnor in his simplified approach. 
This is surprising, since our result has been obtained within a more general framework, and both the gravitational and electromagnetic  fields are different from those used by Bonnor.

We then succeed in the exact reconstruction of both the perturbed gravitational and electromagnetic fields by summing all multipoles.
The perturbed metric we derive is spatially conformally flat, free of singularities and thus cannot belong to the Weyl class two-body solutions, which are characterized by the occurrence of a conical singularity on the axis between the bodies.
The asymptotic mass measured at large distances by the Schwarzschild-like behaviour of the metric of the whole system consisting of black hole and particle is given by
\begin{equation}
M_{\rm eff}={\mathcal M}+m+E_{\rm int}\ ,
\end{equation} 
where the interaction energy turns out to be
\begin{equation}
E_{\rm int}=-m\left[1-\left(1-\frac{{\mathcal M}}{b}\right)\left(1-\frac{2{\mathcal M}}{b}+\frac{Q^2}{b^2}\right)^{-1/2}\right]\ .
\end{equation} 

Recently Perry and Cooperstock \cite{perry} and Bret\'on, Manko and S\'anchez \cite{manko} have used multi-soliton techniques to study exact electrostatic solutions of the Einstein-Maxwell equations representing the exterior field of two arbitrary charged Reissner-Nordstr\"om bodies in equilibrium.
They have shown by numerical methods that gravitational-electrostatic balance without intervening tension or strut can occur for non-critically charged spherically symmetric bodies in the case of one black hole with $Q/\mathcal{M}<1$ and one naked singularity. However, they have not obtained an explicit dependence (in algebraic form) of the balance condition on the separation of the bodies, so we cannot compare our analytical formula with their numerical results.
Attention is devoted also to the limiting case in which $Q/\mathcal{M}=q/m=1$. In this case our solution coincides with the linearized form of the exact solution of Majumdar and Papapetrou \cite{maj,pap}, where equilibrium exists independent of the separation between the bodies. 

Within our approximation for $Q/\mathcal{M}<1$, we find that a horizon seems to form around the particle very close to it where the gravitational perturbation is not small compared to the background metric and hence the approximation is no longer valid. 
The study of the full nonlinear problem, namely the corresponding exact two-body solution of the Einstein-Maxwell equations is therefore required.
Attempts in this direction are being pursued by Belinski and Alekseev \cite{belinski} by using soliton techniques.

\section{Perturbations on a Reissner-Nordstr\"om spacetime}

The Reissner-Nordstr\"om solution is an electrovacuum solution of the Einstein-Maxwell field equations
\begin{eqnarray}
\label{evEinMaxeqs}
G_{\mu \nu }&=&8\pi T_{\mu \nu }^{\rm em} ,\nonumber\\
F^{\mu \nu }{}_{;\,\nu }&=& 0, \quad {}^*  F^{\alpha\beta}{}_{;\beta}=0\ ,
\end{eqnarray}
where 
\begin{eqnarray}
T_{\mu \nu }^{\rm em}=\frac1{4\pi}\left[g^{\rho \sigma }F_{\rho \mu }F_{\sigma \nu } - \frac14 g_{\mu \nu }F_{\rho \sigma } F^{\rho \sigma }\right]
\end{eqnarray}
is the electromagnetic energy-momentum tensor.
In standard Schwarzschild-like coordinates, the corresponding metric is given by
\begin{eqnarray}
\label{RNmetric}
ds^2&=&- f(r)dt^2 + f(r)^{-1}dr^2+r^2(d\theta^2 +\sin ^2\theta d\phi^2)\ ,
\nonumber\\
f(r)&=&1 - \frac {2\mathcal{M}}{r}+\frac{Q^2}{r^2}\ ,
\end{eqnarray}
and the electromagnetic field by
\begin{equation}
\label{RNemfield}
F_{\rm{RN}}=-\frac{Q}{r^2}dt\wedge dr\ ,
\end{equation}
where $\mathcal{M}$ and $Q$ are respectively the mass and charge of the black hole, in terms of which the horizon radii are $r_\pm={\mathcal M}\pm\sqrt{{\mathcal M}^2-Q^2}={\mathcal M}\pm\Gamma$. 
We consider the case $ |Q|\leq {\mathcal M} $ and the region $r>r_+$ outside the outer horizon, with an extremely charged hole corresponding to $|Q|={\mathcal M}$ (which implies $\Gamma=0$) where the two horizons coalesce into one.

Let us consider a perturbation of the Reissner-Nordstr\"om solution due to some external source described by a matter energy-momentum tensor $T_{\mu \nu }$ as well as an electromagnetic current $J^{\mu }$.
The combined Einstein-Maxwell equations are 
\begin{eqnarray}
\label{EinMaxeqs}
\tilde G_{\mu \nu }&=&8\pi \left(T_{\mu \nu } + \tilde T_{\mu \nu }^{\rm em}\right)\ ,\nonumber\\
\label{EM}
\tilde F^{\mu \nu }{}_{;\,\nu }&=& 4\pi  J^{\mu }\ , \quad {}^* \tilde F^{\alpha\beta}{}_{;\beta}=0\ ,
\end{eqnarray}
where the quantities denoted by the tilde refer to the total electromagnetic and gravitational fields, to first order of the perturbation:
\begin{eqnarray}
\label{pertrelations}
\tilde g_{\mu \nu }&=&g_{\mu \nu } + h_{\mu \nu }\ ,\nonumber\\
\tilde F_{\mu \nu }&=&F_{\mu \nu }+  f_{\mu \nu }\ ,\nonumber\\
\tilde T_{\mu \nu }^{\rm em}&=&\frac1{4\pi}\left[\tilde g^{\rho \sigma }\tilde F_{\rho \mu }\tilde F_{\sigma \nu } - \frac14\tilde g_{\mu \nu }\tilde F_{\rho \sigma }\tilde F^{\rho \sigma }\right]\ ,\nonumber\\
\tilde G_{\mu \nu }&=&\tilde R_{\mu \nu }-\frac12\tilde g_{\mu \nu }\tilde R\ ;
\end{eqnarray}
note that the covariant derivative operation makes use of the perturbed metric $\tilde g_{\mu \nu }$ as well.
The corresponding quantities without the tilde refer to the background Reissner-Nordstr\"om geometry (\ref{RNmetric}), and electromagnetic field (\ref{RNemfield}).

Following Zerilli \cite{Zerilli}, we then consider first-order perturbations; the Einstein tensor and the electromagnetic energy-momentum tensor (\ref{pertrelations}) up to the linear order can be expanded as follows:
\begin{eqnarray}
\tilde G_{\mu \nu }&\simeq&G_{\mu \nu }+\delta G_{\mu \nu }\ , \nonumber\\
\tilde T_{\mu \nu }^{\rm em}&\simeq&T_{\mu \nu }^{\rm em}+\delta T_{\mu \nu }^{\rm em}\ , \qquad 
\delta T_{\mu \nu }^{\rm em}=\delta T_{\mu \nu }^{(h)}+\delta T_{\mu \nu }^{(f)}\ ,
\end{eqnarray}
where
\begin{eqnarray}
\delta G_{\mu \nu }&=&-\frac12 h_{\mu\nu;\alpha}{}^{;\alpha}+k_{(\mu;\nu)}-R_{\alpha\mu\beta\nu}h^{\alpha\beta}-\frac12 h_{;\mu;\nu}+R^{\alpha}{}_{(\mu}h_{\nu)\alpha}-\frac12 g_{\mu\nu}\left[k_{\lambda}{}^{;\lambda}-h_{;\lambda}{}^{;\lambda}-h_{\alpha\beta}R^{\alpha\beta}\right]-\frac12 h_{\mu\nu}R\ , \nonumber\\
\delta T_{\mu \nu }^{(h)}&=&-\frac1{4\pi}\left[\left(F^{\alpha}{}_{\mu}F^{\beta}{}_{\nu}-\frac12 g_{\mu\nu}F^{\alpha\lambda}F^{\beta}{}_{\lambda}\right)h_{\alpha\beta}+\frac14 F^{\rho\sigma}F_{\rho\sigma}h_{\mu\nu}\right]\ , \nonumber\\
\delta T_{\mu \nu }^{(f)}&=&-\frac1{4\pi}\left[2F^{\rho}{}_{(\mu}f_{\nu)\rho}+\frac12 g_{\mu\nu}F^{\rho\sigma}f_{\rho\sigma}\right]\ ,
\end{eqnarray}
with $k_{\mu}=h_{\mu\alpha}{}^{;\alpha}$ and $h=h_{\alpha}{}^{\alpha}$.
The expansion of the quantity $\tilde F^{\mu \nu }{}_{;\,\nu }$ appearing in the Maxwell equations turn out to be explicitly
\begin{equation}
\tilde F^{\mu \nu }{}_{;\nu}\simeq f^{\mu \nu }{}_{;\nu}-\delta J^{\mu}_{(h)}\ ,
\end{equation}
where the last term can be interpreted as an effective gravitational current and is given by
\begin{equation}
\delta J^{\mu}_{(h)}=F^{\mu\rho}{}_{;\sigma}h_{\rho}{}^{\sigma}+F^{\rho\sigma}h^{\mu}{}_{\rho;\sigma}+F^{\mu\rho}\left(k_{\rho}-\frac12 h_{;\rho}\right)\ .
\end{equation}
As a result, the terms $\delta T_{\mu \nu }^{(h)}\sim FFh$ and $\delta T_{\mu \nu }^{(f)}\sim Ff$ symbolically represent the electromagnetically induced gravitational perturbation, while the term $\delta J^{\mu}_{(h)}\sim Fh$ represents the gravitationally induced electromagnetic perturbation \cite{jrz1,jrz2}.

We now specialize this general framework to the case of a perturbing source being represented by a massive charged particle at rest. 
A point charge of mass $m$ and charge $q$ moving along a worldline $z^{\alpha}(\tau)$ with 4-velocity $U^{\alpha}=dz^{\alpha}/d\tau$ is described by the matter energy-momentum tensor
\begin{equation}
\label{Tmunupartgen}
T^{\mu \nu}_{\rm{part}}=\frac{m}{\sqrt{-g}}\int\delta^{(4)}(x-z(\tau))U^{\mu}U^{\nu}\,d\tau\ ,
\end{equation}
and electromagnetic current 
\begin{equation}
\label{Jmupartgen}
J^{\mu}_{\rm{part}}=\frac{q}{\sqrt{-g}}\int\delta^{(4)}(x-z(\tau))U^{\mu}\,d\tau\ ,
\end{equation}
where the normalization of the delta functions is defined by
\begin{equation}
\int\delta^{(4)}(x)\,d^4x=1\ .
\end{equation}
The perturbation equations are obtained from the Einstein-Maxwell equations (\ref{EM}), keeping terms to first order.
There are two parameters of smallness of the perturbation: the mass $m$ of the particle and its charge $q$, which 
can be expressed as $q=\epsilon m$, where $\epsilon$ is the charge to mass ratio of the particle. 
Therefore, the perturbation will be small if $m$ and so $q$ are sufficiently small with respect to the black hole mass and charge, the charge to mass ratio which, instead, need not to be small. 

Let us consider the special case of the charged particle at rest at the point $r=b $ on the polar axis $\theta=0$. 
The only nonvanishing components of the stress-energy tensor (\ref{Tmunupartgen}) and of the current density (\ref{Jmupartgen}) associated with the charged particle are given by
\begin{eqnarray}
\label{sorgenti}
T_{{00}}^{\rm{part}}&=&{\frac {m}{2\pi {b}^{2}}}f(b)^{3/2}\delta  \left( r-b \right) \delta  \left( \cos \theta -1 \right)\ ,\nonumber\\
J^{{0}}_{\rm{part}}&=&{\frac {q}{2\pi {b}^{2}}}\delta  \left( r-b \right) \delta  \left( \cos  \theta -1 \right)\ ,
\end{eqnarray}
since $U=f(r)^{-1/2}\partial_t$ is the corresponding 4-velocity.
 
Following Zerilli's \cite{Zerilli} procedure, we must now expand the fields $h_{\mu \nu }$ and $f_{\mu \nu }$ as well as the source terms (\ref{sorgenti}) in tensor and scalar harmonics respectively, obtaining a set of first order perturbation equations by linearizing the Einstein-Maxwell system (\ref{EinMaxeqs}).
Zerilli showed that in the Einstein-Maxwell system of equations electric gravitational multipoles couple only to electric electromagnetic multipoles, and similarly for magnetic multipoles.
The axial symmetry of the problem about the $z$ axis ($\theta=0$) allows us to put the azimuthal parameter equal to zero in the expansion, leading to a great simplification in the solving procedure.
Furthermore, it is sufficient to consider only electric-parity perturbations, since there are no magnetic sources \cite{jrz1,jrz2}.

\section{Test field approximation}
\label{testfield}

Before looking for a solution of the above general problem within the framework of first order perturbation theory it is interesting to recall the results in the simplest case of test field approximation, i.e. by neglecting the particle backreaction, namely the changes in the background metric due to the mass and charge of the particle. 
This test field approximation requires only the vector harmonic description of the electric field of the test particle alone.

\subsection{Electric test field solution on a Schwarzschild background}

For any elementary particle the charge to mass ratio is $q/m\gg1$, so it is natural to address the simplest problem of a test charge, neglecting the contribution of the mass, at rest in the field of a Schwarzschild metric, see e.g. the works by Hanni \cite{hannijth}, Cohen and Wald \cite{CoW}, Hanni and Ruffini \cite{HR}, Bi{\v c}\'ak and Dvo{\v r}\'ak \cite{bicdev}, Linet \cite{linet}.
In standard coordinates the Schwarzschild metric is given by
\begin{eqnarray}
ds^2&=&- f_{\rm s}(r)dt^2 + f_{\rm s}(r)^{-1}dr^2+r^2(d\theta^2 +\sin ^2\theta d\phi^2)\ ,
\nonumber\\
f_{\rm s}(r)&=&1 - \frac {2\mathcal{M}}{r}\ .
\end{eqnarray}
The Einstein-Maxwell equations are 
\begin{eqnarray}
G_{\mu \nu }&=&0\ ,\nonumber\\
F^{\mu \nu }{}_{;\,\nu }&=& 4\pi  J^{\mu }\ ,
\end{eqnarray}
where the only nonvanishing component of the current density associated with the charged particle at rest at the point $r=b$ on the polar axis $\theta=0$ is given by
\begin{equation}
\label{Jzero}
J^{{0}}={\frac {q}{2\pi {b}^{2}}}\delta  \left( r-b \right) \delta  \left( \cos  \theta -1 \right)\ .
\end{equation}
The electromagnetic stress-energy tensor is second order in the electromagnetic field and can be neglected, so it is enough to solve the Maxwell equations in a fixed Schwarzschild background. 
This solution has been discussed in detail by Hanni and Ruffini \cite{HR}. 
There is no constraint on the position of the particle in this case.

The (coordinate) components of the electric field are defined by $E_i=F_{i0}$. 
By introducing the vector potential $A_\mu$ defined by
\begin{equation}
F_{\alpha\beta}=2A_{[\beta;\alpha]}\ ,
\end{equation}
which in our case is determined by the electrostatic potential $V$ alone
\begin{eqnarray}
A_0=-V\ ,\qquad 
A_i=0\ ,
\end{eqnarray}
the electric field becomes
\begin{eqnarray}
\label{elecomp}
E_r=-V_{,r}\ ,\qquad
E_\theta=-V_{,\theta}\ ,\qquad
E_\phi=-V_{,\phi}\ .
\end{eqnarray}
The vector harmonic expansion of the relevant components of the electric field is given by
\begin{equation}
\label{erethetaharmexp}
E_r=-\sum_l \tilde {\mathcal E}_{1}^l(r) Y_{l0}\ , \qquad E_\theta=-\sum_l \tilde {\mathcal E}_{2}^l(r) \frac{\partial Y_{l0}}{\partial \theta}\ ,
\end{equation} 
where  
\begin{eqnarray}
\label{ypsilonl0}
Y_{l0}=\frac12\sqrt{{\frac {2l+1}{\pi }}}P_l(\cos\theta)\ 
\end{eqnarray}
are normalized spherical harmonics with azimuthal index equal to zero, because of the axial symmetry of the problem about the $z$ axis ($\theta=0$). 
The expansion of the source term (\ref{Jzero}) in scalar harmonics is given by
\begin{equation}
J^0=\sum_l \tilde {\mathcal J}^0_l(r) Y_{l0}\ , \qquad \tilde {\mathcal J}^0_l(r)=\frac 1{2\sqrt{\pi }}\frac{q\sqrt{2l+1}}{b^2}\delta(r-b)\ .
\end{equation} 
Therefore, after separating the angular part from the radial one, the Maxwell equations imply 
\begin{eqnarray}
  \label{RNeq1em} 
0&=&\tilde {\mathcal E}_{1}^l{}'+\frac2{r}\tilde {\mathcal E}_{1}^l -{\frac {l\left( l+1 \right)}{r^2f_{\rm s}(r)}}\tilde {\mathcal E}_{2}^l+4\pi \tilde {\mathcal J}^0_l\ , \\
  \label{RNeq2em} 
0&=&\tilde {\mathcal E}_{1}^l - \tilde {\mathcal E}_{2}^l{}'\ ,
\end{eqnarray}
where primes denote differentiation with respect to $r$. By solving Eq.~(\ref{RNeq2em}) for $\tilde {\mathcal E}_{1}^l$, and substituting into Eq.~(\ref{RNeq1em}), we obtain the following second order differential equation for $\tilde {\mathcal E}_{2}^l$:
\begin{equation}
 \label{RNeq3em} 
0=\tilde {\mathcal E}_{2}^l{}''+\frac2{r}\tilde {\mathcal E}_{2}^l{}'-\frac{l(l+1)}{r^2f_{\rm s}(r)}\tilde {\mathcal E}_{2}^l+4\pi \tilde {\mathcal J}^0_l\ .
\end{equation} 
The knowledge of the function $\tilde {\mathcal E}_{2}^l$ determines the radial part $\tilde {\mathcal V}_l$ of the expansion of the electrostatic potential $V$, from Eqs.~(\ref{elecomp}) and (\ref{erethetaharmexp}):
\begin{equation}
 \label{Vschwexp} 
V=\sum_l \tilde {\mathcal V}_l(r) Y_{l0}\ , \qquad \tilde {\mathcal V}_l(r)=\tilde {\mathcal E}_{2}^l\ .
\end{equation}
Putting $\tilde {\mathcal E}_{2}^l=f_{\rm s}(r)^{1/2}w(r)$ and making the transformation $z=r/{\mathcal M}-1$ Eq.~(\ref{RNeq3em}) becomes
\begin{equation}
 \label{eqleg1l} 
0=(1-z^2)w{}''-2zw{}'+\left[l(l+1)-\frac1{1-z^2}\right]w-2\sqrt{\pi}\frac{q}{\mathcal M}\sqrt{2l+1}\left(\frac{\beta-1}{\beta+1}\right)^{1/2}\delta(z-\beta)\ ,
\end{equation} 
where primes now denote differentiation with respect to the new variable $z$ and $\beta=b/{\mathcal M}-1$.
The general solutions of the corresponding homogeneous equation are the associated Legendre functions of the first and second kind $P_l^1(z)$ and $Q_l^1(z)$. Using this result and taking into account that $P_l^1(z)=\sqrt{z^2-1}dP_l(z)/dz$ and $Q_l^1(z)=\sqrt{z^2-1}dQ_l(z)/dz$, Whittaker \cite{whittaker} and Copson \cite{copson} then gave as the two linearly independent solutions of the homogeneous equation of (\ref{RNeq3em}) 
\begin{eqnarray}
\label{multipoliS}
f_l(r)&=&-\frac{(2l+1)!}{2^l(l+1)!l!{\mathcal M}^{l+1}}(r-2{\mathcal M}) 
      \frac{dQ_l(z(r))}{dr}\qquad\qquad l=0,1,2, ...\nonumber\\
g_l(r)&=&\left\{
     \begin{array}{ll}
     1&\qquad\qquad\quad l=0\\
     \noalign{\medskip}\displaystyle\frac{2^ll!(l-1)!{\mathcal M}^l}{(2l)!}(r-2{\mathcal M}) 
     \frac{dP_l(z(r))}{dr}&\qquad\qquad\quad l=1,2, ... 
     \end{array}
     \right. \ 
\end{eqnarray}
where $P_l$ and $Q_l$ are the two types of Legendre functions.
The solution for the electrostatic potential (\ref{Vschwexp}) is then given by
\begin{eqnarray}
\label{multipoli}
V=q\sum_l [f_l(b)g_l(r)\vartheta(b-r)+g_l(b)f_l(r) \vartheta(r-b)]P_l(\cos\theta)\ .
\end{eqnarray}
Copson \cite{copson} showed that this solution can be cast in closed form using the following representation formula
\begin{eqnarray}
\label{sommacopson}
\frac{xt-\cos\theta}{[x^2+t^2 - 2xt\cos\theta - \sin^2\theta]^{1/2}}&=&x\vartheta(t-x)+t\vartheta(x-t)-(x^2-1)(t^2-1)\nonumber\\
&&\times\sum_{l=1}^\infty \frac{2l+1}{l(l+1)}\left[\frac{dQ_l(x)}{dx}\bigg\vert_{x=t}\frac{dP_l(x)}{dx}\vartheta(t-x)\right.\nonumber\\
&&\left.+\frac{dP_l(x)}{dx}\bigg\vert_{x=t}\frac{dQ_l(x)}{dx}\vartheta(x-t)\right]P_l(\cos\theta)\ .
\end{eqnarray}
However, the solution he found did not satisfy the boundary conditions; the corrected version was presented about fifty years later by Linet \cite{linet}:
\begin{equation}
\label{solschwpot}
V_{S}= \frac q{b r} \frac{(r-{\mathcal M})(b-{\mathcal M})
 -{\mathcal M}^2\cos\theta}{D_{S}} + \frac{q{\mathcal M}}{b r}\ ,
\end{equation}
with
\begin{equation}
\label{denS}
D_{S}
= [(r-{\mathcal M})^2+(b-{\mathcal M})^2 
- 2(r -{\mathcal M})(b-{\mathcal M})\cos\theta 
- {\mathcal M}^2\sin^2\theta]^{1/2}\ . 
\end{equation} 
The components of the electric field are then easily evaluated
\begin{eqnarray} 
\label{Schwemtensorpert} 
E_r&=&\frac q{b r^2}\bigg\{ {\mathcal M} -\frac{{\mathcal M}(b-{\mathcal M})+{\mathcal M} ^2\cos\theta}{ D_{S}}+\frac{r[(r-{\mathcal M})(b-{\mathcal M})-{\mathcal M}^2\cos\theta]
[(r-{\mathcal M}) -(b-{\mathcal M})\cos\theta]}{D_{S}^3}\bigg\}\ , \nonumber\\ 
E_{\theta}&=&qbf_{\rm s}(b) rf_{\rm s}(r)
\frac{\sin\theta}{D_{S}^3}\ . 
\end{eqnarray} 

The properties of the above solution has been analyzed in detail by Hanni and Ruffini \cite{HR}.
They derived the lines of force by defining the lines of constant flux, and also introduced the concept of the induced charge on the surface of the black hole horizon, which indeed appears to have some of the properties of a perfectly conducting sphere terminating the 
electric field lines.

\subsection{Electric test field solution on a Reissner-Nordstr\"om background}

The generalization of the above treatment to the case of a charged test particle at rest near a Reissner-Nordstr\"om black hole was discussed by Leaute and Linet \cite{leaute}.
As in the Schwarzschild case, the problem is solved once the solution of Eq.~(\ref{RNeq3em}) with $f_{\rm s}(r)\longleftrightarrow f(r)$ is obtained. 
The corresponding solution for the electrostatic potential is again of the form (\ref{multipoli}) with functions 
\begin{eqnarray}
\label{multipoliRN}
f_l(r)&=&-\frac{(2l+1)!}{2^l(l+1)!l!\Gamma^{l+1}}\frac{(r-r_+)(r-r_-)}r 
      \frac{dQ_l(z(r))}{dr}\qquad\qquad\quad l=0,1,2, ...\nonumber\\
g_l(r)&=&\left\{
     \begin{array}{ll}
     1&\qquad\qquad\quad l=0\\
     \noalign{\medskip}\displaystyle\frac{2^ll!(l-1)!\Gamma^l}{(2l)!}\frac{(r-r_+)(r-r_-)}r 
     \frac{dP_l(z(r))}{dr}&\qquad\qquad\quad l=1,2, ... 
     \end{array}
     \right. \ 
\end{eqnarray}
where now $z=(r-{\mathcal M})/\Gamma$.
Leaute and Linet showed that this solution can be cast in the closed form expression
\begin{equation}
\label{solRNpot}
V_{\rm{RN}} = \frac q{b r} \frac{(r-{\mathcal M})(b-{\mathcal M})
 -\Gamma^2\cos\theta}{D_{\rm{RN}}} + \frac{q{\mathcal M}}{b r}\ ,
\end{equation}
with
\begin{equation}
\label{denRN}
D_{\rm{RN}}
= [(r-{\mathcal M})^2+(b-{\mathcal M})^2 
- 2(r -{\mathcal M})(b-{\mathcal M})\cos\theta 
- \Gamma^2\sin^2\theta]^{1/2}\ . 
\end{equation}
The components of the electric field of the test particle alone are then easily evaluated
\begin{eqnarray}
\label{eerreetheta} 
E_r&=&\frac q{b r^2}\left\{{\mathcal M} -\frac{{\mathcal M}(b-{\mathcal M})
+\Gamma^2\cos\theta}{D_{\rm{RN}}}
+\frac{r[(r-{\mathcal M})(b-{\mathcal M})-\Gamma^2\cos\theta]
[(r-{\mathcal M}) -(b-{\mathcal M})\cos\theta]}{D_{\rm{RN}}^3}\right\}
\ , \nonumber\\
E_\theta
&=&qbf(b)rf(r)\frac{\sin\theta}{D_{\rm{RN}}^3}\ .
\end{eqnarray}
The black hole has its own electric field and electrostatic potential
\begin{equation}
\label{BHpot}
E_r^{\rm{BH}}=\frac Q{r^2}\ , \qquad
V^{\rm{BH}}=\frac Q{r}\ .
\end{equation}

This test field approach has been largely used in the current literature \cite{leaute,bonnor,selfSchw,selfRN}.
It is interesting to analyze what conceptual differences are introduced in the properties of the electric field of the test particle in the context of a charged Reissner-Nordstr\"om geometry compared and contrasted with the Schwarzschild case. The concept of the induced charge on the horizon and the behavior of the electric lines of force have been addressed in \cite{noi}.
We recall here the main result: as the hole becomes extreme an effect analogous to the Meissner effect for the electric field arises, with the electric field lines of the test charge being forced outside the outer horizon. We are witnessing a transition from the infinite conductivity of the horizon in the limiting uncharged Schwarzschild case to the zero conductivity of the outer horizon in the extremely charged Reissner-Nordstr\"om case.

\subsection{Bonnor's equilibrium condition}

In this same test field approximation Bonnor \cite{bonnor} has addressed the issue of the equilibrium of a test particle of mass $m$ and charge $q$ at rest  at $r=b$, $\theta=0$ outside the horizon of a Reissner-Nordstr\"om  black hole. 
By considering the classical expression for the equation of motion of the particle 
\begin{equation}
m U^\alpha \nabla_\alpha U^\beta =q F^\beta{}_\mu U^\mu\ ,
\end{equation} 
with 4-velocity $U^\alpha =f(r)^{-1/2}\delta^\alpha_0$, he found the following equilibrium condition for such a system
\begin{equation}
\label{bonnoreqcond}
m=qQ\frac{b f(b)^{1/2}}{{\mathcal M}b-Q^2}\ .
\end{equation} 
From this equation it follows that there exist equilibrium positions which are separation-dependent, and require either $q^2<m^2$ and $Q^2>{\mathcal M}^2$ or $q^2>m^2$ and $Q^2<{\mathcal M}^2$.
A condition which is sufficient, but not necessary for the equilibrium is $|q|=m$ and $|Q|={\mathcal M}$; it represents a special case of the Newtonian condition $qQ=m{\mathcal M}$, so the equilibrium can occur at arbitrary separations.
We will again address this issue on the existence of equilibrium conditions in Section \ref{RNpert}, using the correct treatment which takes into proper account the perturbation of the Reissner-Nordstr\"om field induced by the charge and mass of the particle.

\section{Perturbation analysis: the Schwarzschild case}
\label{schwpert}

Let us consider now the problem consisting of a neutral particle of mass $m$ at rest on the polar axis near a Schwarzschild black hole in the framework of first order perturbation theory. 
The Einstein equations are 
\begin{equation}
\label{Eineqs}
\tilde G_{\mu \nu }=8\pi T_{\mu \nu }^{\rm part}\ ,
\end{equation}
where the perturbed Einstein tensor is defined in (\ref{pertrelations}), and the source term is given by (\ref{sorgenti}).
We need to expand the Einstein tensor as well as the source term in tensor and scalar harmonics respectively, obtaining a set of first order perturbation equations, once a gauge is specified.

The static perturbations due to a point mass in a Schwarzschild black hole background has been studied by Zerilli \cite{ZerilliSchw} in the dynamical case following the Regge-Wheeler \cite{ReggeW} treatment. 
The geometrical perturbations $h_{\mu \nu }$ corresponding to the electric multipoles are given by
\begin{eqnarray}
\label{gengeompertSchw}
 ||h_{\mu \nu }||=\left[ 
\begin {array}{cccc} 
{e^{\nu_{\rm s}}}H_0Y_{l0}&H_1Y_{l0}&h_0\displaystyle\frac{\partial Y_{l0}}{\partial \theta }&0
\\\noalign{\medskip}\rm{sym}&{e^{-\nu_{\rm s}}}H_2Y_{ l0}&h_1\displaystyle\frac{\partial Y_{l0}}{\partial \theta }&0
\\\noalign{\medskip} \rm{sym} &\rm{sym} &{r}^{2}\left(KY_{l0}+G\displaystyle\frac{\partial^2 Y_{l0}}{\partial \theta^2 }\right)&0
\\\noalign{\medskip} \rm{sym} & \rm{sym} & \rm{sym} &{r}^{2} \sin ^{2}\theta\left(KY_{ l0}+G\cot\theta \displaystyle\frac{\partial Y_{l0}}{\partial \theta }\right)
\end {array} 
\right],
\end{eqnarray}
where the symbol ``sym'' indicates that the missing components of $h_{\mu \nu }$ are to be found from the symmetry $h_{\mu \nu }=h_{\nu \mu }$, and $e^{\nu_{\rm s}}=f_{\rm s}(r)$
is Zerilli's notation.

This general form of the perturbation can be simplified by performing a suitable gauge choice.
Consider the infinitesimal coordinate transformation 
\begin{equation}
x{}'^\mu=x^\mu+\xi^\mu\ ,
\end{equation}
where the infinitesimal displacement $\xi^\mu$ is a function of $x^\mu$ and transforms like a vector. 
This transformation induces the following transformation of the perturbation tensor $h_{\mu\nu}$:
\begin{equation}
h^{(\rm new)}_{\mu\nu}=h_{\mu\nu}-2\xi_{(\mu;\nu)}\ ,
\end{equation} 
where the multipole expansion of the second term can be easily obtained by expanding $\xi^\mu$ in the electric-type vector harmonics
\begin{equation}
\xi=\sum_{l}\left[A_0Y_{l0}\partial_t+A_1Y_{l0}\partial_r+A_2\frac{\partial Y_{l0}}{\partial \theta }\partial_\theta\right]\ .
\end{equation}   
The new metric perturbation functions are then given by
\begin{eqnarray}
\label{gengauge}
&&H_0^{(\rm new)}=H_0+\frac{2{\mathcal M}}{r(r-2{\mathcal M})}A_1\ , \quad  H_1^{(\rm new)}=H_1+\left(1-\frac{2{\mathcal M}}{r}\right)A_0{}'\ , \quad h_0^{(\rm new)}=h_0+\left(1-\frac{2{\mathcal M}}{r}\right)A_0\ , \nonumber\\
&&H_2^{(\rm new)}=H_2+\frac{2{\mathcal M}}{r(r-2{\mathcal M})}A_1-2A_1{}'\ , \quad  h_1^{(\rm new)}=h_1-\frac{r}{r-2{\mathcal M}}A_1-r^2A_2{}'\ , \nonumber\\
&&K^{(\rm new)}=K-\frac{2}{r}A_1\ , \quad  G^{(\rm new)}=G-2A_2\ ,
\end{eqnarray}
where a prime denotes differentiation with respect to $r$.

\subsection{The Regge-Wheeler approach}

We use the Regge-Wheeler \cite{ReggeW} gauge to set
\begin{eqnarray*}
h_0^{(\rm RW)}\equiv h_1^{(\rm RW)}\equiv G^{(\rm RW)}\equiv 0\ .
\end{eqnarray*}
This specialization is accomplished through the gauge functions 
\begin{equation}
A_0=-\frac{r}{r-2{\mathcal M}}h_0\ , \quad A_1=\left(1-\frac{2{\mathcal M}}{r}\right)\left(h_1-\frac{r^2}{2}G{}'\right)\ , \quad A_2=\frac{G}{2}\ ,
\end{equation} 
so that the metric perturbation functions (\ref{gengauge}) expressed in the Regge-Wheeler gauge as combinations of metric perturbations expressed in an arbitrary gauge are given by
\begin{eqnarray}
\label{rwfuncts}
&&H_0^{(\rm RW)}=H_0+\frac{2{\mathcal M}}{r^2}h_1-{\mathcal M}G{}'\ , \quad  H_1^{(\rm RW)}=H_1+\frac{2{\mathcal M}}{r(r-2{\mathcal M})}h_0-h_0{}'\ , \nonumber\\
&&H_2^{(\rm RW)}=H_2+r(r-2{\mathcal M})G{}''+(2r-3{\mathcal M})G{}'-2\left(1-\frac{2{\mathcal M}}{r}\right)h_1{}'-\frac{2{\mathcal M}}{r^2}h_1\ , \nonumber\\
&&K^{(\rm RW)}=K+(r-2{\mathcal M})\left(G{}'-\frac{2}{r^2}h_1\right)\ .
\end{eqnarray}
The general perturbation (\ref{gengeompertSchw}) thus becomes
\begin{eqnarray}
\label{RWgeompertSchw}
 ||h_{\mu \nu }||=\left[ 
\begin {array}{cccc} 
{e^{\nu}}H_0Y_{l0}&H_1Y_{l0}&0&0
\\\noalign{\medskip}H_1Y_{l0}&{e^{-\nu}}H_2Y_{ l0}&0&0
\\\noalign{\medskip}0&0&{r}^{2}KY_{l0}&0\\\noalign{\medskip}0&0
&0&{r}^{2} \sin ^{2}\theta KY_{l0}
\end{array} 
\right]\ ,
\end{eqnarray}
where superscripts indicating the chosen gauge have been dropped for simplicity.

Therefore, the independent first order perturbations of the quantities appearing in the Einstein field equations (\ref{Eineqs}) are given by
\begin{eqnarray}
  \label{RWlambda00}  
{\tilde G}_{00}&=&-\frac12\bigg\{{e^{2\nu_{\rm s}}} \left[ 2K{}''-\frac2rH_2{}'+\left(\nu_{\rm s}{}'+\frac6r\right) { K{}'} -    2\left(\frac1{r^2}+\frac{\nu_{\rm s}{}'}r\right)(H_0+H_2) \right] \nonumber\\
&& -\frac{2 e^{\nu_{\rm s}}}{r^2}[(\lambda+1)H_2-H_0+\lambda K]\bigg\}Y_{l0}\ , \\
  \label{RWlambda11}  
{\tilde G}_{11}&=&-\frac12\bigg\{\frac 2rH_0{}'-\left(\nu_{\rm s}{}'+\frac2r\right)K{}' +\frac{2 e^{-\nu_{\rm s}}}{r^2}[H_2-(\lambda+1)H_0+\lambda K]\bigg\}Y_{l0}\ , \\
  \label{RWlambda22}  
{\tilde G}_{22}&=&\frac{r^2}2e^{\nu_{\rm s}}\bigg\{K{}''+\left(\nu_{\rm s}{}'+\frac2r\right)K{}'- H_0{}'' -\left(\frac{\nu_{\rm s}{}'}2+\frac1r\right)H_2{}' -\left(\frac{3\nu_{\rm s}{}'}2+\frac1r\right)H_0{}'
  +2(\lambda+1)\frac{e^{-\nu_{\rm s}}}{r^2}(H_0-H_2)\bigg\}Y_{l0}\nonumber\\
&&+\frac12\bigg\{H_0-H_2\bigg\}\frac{\partial^2 Y_{l0}}{\partial \theta^2}\ , \\
  \label{RWlambda12}  
{\tilde G}_{12}&=&-\frac12\bigg\{-H_0{}' + K{}' -\left(\frac{\nu_{\rm s}{}'}2+\frac1r\right)H_2-\left(\frac{\nu_{\rm s}{}'}2-\frac1r\right)H_0\bigg\}\frac{\partial Y_{l0}}{\partial \theta}\ , \\
  \label{RWlambda01}  
{\tilde G}_{01}&=&\bigg\{\left[\frac{\lambda}{r^2}+\frac{e^{\nu_{\rm s}}}r\left(\nu_{\rm s}{}'+\frac1r\right)\right]H_1\bigg\}Y_{l0}\ , \\
\label{RWlambda02}  
{\tilde G}_{02}&=&\frac{e^{\nu_{\rm s}}}{2}\left\{H_1{}' + {\nu_{\rm s}}{}'H_1\right\}\frac{\partial Y_{l0}}{\partial \theta}\ , \\ 
\label{RWdeltaT00}  
T_{00}^{\rm part}&=&\frac1{16\pi}A_{{00}}^{\rm s}Y_{l0}\ ,
\end{eqnarray}
where $\lambda=\frac12 \left( l-1 \right) \left( l+2 \right)$ and the quantities with the subscript/superscript $s$ stand for
\begin{eqnarray}
\label{sorgS}
e^{\nu_{\rm s}}\equiv f_{\rm s}(r)=1-\frac{2{\mathcal M}}r\ , \qquad
A_{{00}}^{\rm s}=8\sqrt{\pi} \frac{m\sqrt {2l+1}}{b^2}f_{\rm s}(b)^{3/2}\delta \left( r-b \right)\ .
\end{eqnarray}
The angular factors containing derivatives vanish for $l=0$; moreover, the two angular factors in the expression (\ref{lambda22}) for ${\tilde G}_{22}$ are not independent when $l=1$ (in fact, ${\partial^2 Y_{10}}/{\partial \theta^2}=-Y_{10}$).
Therefore, the cases $l=0,1$ must be treated separately.

For all higher values of $l$, the Einstein field equations (\ref{Eineqs}) imply that the corresponding curly bracketed factors on the left and right hand sides are equal, so that the system of radial equations we have to solve is the following:
\begin{eqnarray}
  \label{RWeq1S}  
0&=&{e^{2\nu_{\rm s}}} \left[ 2K{}''-\frac2rW{}'+\left(\nu_{\rm s}{}'+\frac6r\right) { K{}'}-    4\left(\frac1{r^2}+\frac{\nu_{\rm s}{}'}r\right)W \right]-\frac{2\lambda e^{\nu_{\rm s}}}{r^2}(W+K)   
    +A_{{00}}^{\rm s}\ , \\
  \label{RWeq2S} 
0&=&\frac 2rW{}'-\left(\nu_{\rm s}{}'+\frac2r\right)K{}' -\frac{2\lambda e^{-\nu_{\rm s}}}{r^2}(W-K)\ ,\\
  \label{RWeq3S} 
0&=&K{}''+\left(\nu_{\rm s}{}'+\frac2r\right)K{}'- W{}'' -2\left(\nu_{\rm s}{}'+\frac1r\right)W{}'\ ,\\
  \label{RWeq4S} 
0&=&-W{}' + K{}' -\nu_{\rm s}{}' W\ ,
\end{eqnarray}
since
\begin{eqnarray}
H_0= H_2\equiv W\ , \qquad 
H_1\equiv 0\ .
\end{eqnarray}

We are dealing with a system of $4$ ordinary differential equations for $2$ unknown functions: $K$ and $W$.
Compatibility of the system requires that these equations not be independent. 
The second order equations (\ref{RWeq1S}) and  (\ref{RWeq3S}) show that the first derivatives of both the functions $K$ and $W$ are not continuous at the particle position with the same jump 
\begin{eqnarray}
\label{jump}
[K{}']=[W{}']=-4\sqrt{\pi}\frac{m}{b^2}\sqrt{2l+1}f_{\rm s}(b)^{-1/2}\ ,
\end{eqnarray}
where the quantity $[K{}']\equiv K{}'_+-K{}'_-$, $K{}'_\pm=K{}'(b\pm\epsilon)$ is obtained by integrating both sides of Eq. (\ref{RWeq1S}) with respect to $r$ between $b-\epsilon$ and $b+\epsilon$ and then taking the limit $\epsilon\to0$.
Only the pair of first order equations (\ref{RWeq2S}) and  (\ref{RWeq4S}) thus remain to be considered.

It is useful to introduce the the new combinations 
\begin{eqnarray}
\label{newfunctionsS}
X=K-W\ , \qquad Y=K+W\ ,
\end{eqnarray}
so that the system we have to solve for the unknown functions $X$ and $Y$ is the following
\begin{eqnarray}
  \label{eq2Sxy} 
0&=&-\frac2r X{}'-\frac{\nu_{\rm s}{}'}2(X{}'+Y{}') +\frac{2\lambda e^{-\nu_{\rm s}}}{r^2}X\ ,\\
  \label{eq4Sxy} 
0&=&X{}'  -\frac{\nu_{\rm s}{}'}2(X-Y)\ .
\end{eqnarray}
Solving Eq.~(\ref{eq4Sxy}) for $Y$ and then substituting it into Eq.~(\ref{eq2Sxy}) gives the following equation for $X$
\begin{equation}
\label{eq2Sxynew} 
0=r(r-2{\mathcal M})X{}''+4(r-{\mathcal M})X{}'-(l+2)(l-1)X\ . 
\end{equation}
Putting $X=f_{\rm s}(r)^{-1/2}w(r)/r$ and making the transformation $z=r/{\mathcal M}-1$ the previous equation becomes
\begin{equation}
0=(1-z^2)w{}''-2zw{}'+\left[l(l+1)-\frac1{1-z^2}\right]w\ ,
\end{equation} 
where primes now denote differentiation with respect to the new variable $z$.
The general solutions are the associated Legendre functions of the first and second kind $P_l^1(z)$ and $Q_l^1(z)$, implying that 
\begin{equation}
X=(r-2{\mathcal M})^{-1}[c_1f_l(r)+c_2g_l(r)]
\equiv c_1X_1+c_2X_2\ ,
\end{equation}
where the functions $f_l(r)$ and $g_l(r)$ have been defined in (\ref{multipoliS}).
Choosing the arbitrary constants $c_1$ and $c_2$ in order that the function $X$ be continuous at the particle position and to satisfy regularity conditions on the horizon and at infinity implies that the solution can be written as 
\begin{equation}
\label{XsolS}
X={\mathcal N}_X[X_1(r)X_2(b)\vartheta(b-r)+X_1(b)X_2(r)\vartheta(r-b)]\ , 
\end{equation}
where ${\mathcal N}_X$ is an arbitrary constant.
The corresponding solution for $Y$ can be easily obtained from Eq.~(\ref{eq4Sxy}): 
\begin{eqnarray}
\label{YsolS}
Y=X+\frac{r}{{\mathcal M}}(r-2{\mathcal M})X{}'\ . 
\end{eqnarray}
Inverting the relations (\ref{newfunctionsS}) yields immediately the solutions for the functions $K$ and $W$:
\begin{eqnarray}
\label{KWsolS}
K=X+\frac{r}{2{\mathcal M}}(r-2{\mathcal M})X{}'\ , \qquad 
W=-\frac{r}{2{\mathcal M}}(r-2{\mathcal M})X{}'\ .
\end{eqnarray}
The value of the arbitrary constant ${\mathcal N}_X$ is determined by imposing the condition (\ref{jump}):
\begin{eqnarray}
{\mathcal N}_X=8\sqrt{\pi} \frac{m}{b^4}{\mathcal M}\sqrt {2l+1}f_{\rm s}(b)^{-3/2}[X_2{}''(b)-X_1{}''(b)]^{-1}\ .
\end{eqnarray}

Next consider the case $l=0$.
The relevant equations come from quantities (\ref{RWlambda00})--(\ref{RWdeltaT00}) which do not contain angular derivatives:
\begin{eqnarray}
  \label{RWlambda00leq0}  
0&=&r(r-2{\mathcal M})K{}''+(3r-5{\mathcal M}) K{}'-(r-2{\mathcal M}) H_2{}'+K-H_2+\frac12\frac{r^3}{r-2{\mathcal M}}A_{{00}}^{\rm s}\ , \\
  \label{RWlambda11leq0} 
0&=&(r-2{\mathcal M})H_0{}'-(r-{\mathcal M})K{}'+H_2-K\ , \\
  \label{RWlambda22leq0} 
0&=&r(r-2{\mathcal M})[K{}''-H_0{}'']+ (r-{\mathcal M})[2K{}'-H_2{}']-(r+{\mathcal M})H_0{}'\ .
\end{eqnarray}
It is easy to show that looking for solutions of the form $H_0=H_2\equiv W$ leads to a system of equations which coincide with Eqs. (\ref{RWeq1S})--(\ref{RWeq3S}) for $l=0$ (or $\lambda=-1$).
An analogous situation occurs in the remaining case $l=1$: the relevant equations coming from quantities (\ref{RWlambda00})--(\ref{RWdeltaT00}) 
\begin{eqnarray}
  \label{RWlambda00leq1}  
0&=&r(r-2{\mathcal M})K{}''+(3r-5{\mathcal M}) K{}'-(r-2{\mathcal M}) H_2{}'-2H_2+\frac12\frac{r^3}{r-2{\mathcal M}}A_{{00}}^{\rm s}\ , \\
  \label{RWlambda11leq1} 
0&=&(r-2{\mathcal M})H_0{}'-(r-{\mathcal M})K{}'+H_2-H_0\ , \\
  \label{RWlambda22leq1} 
0&=&r(r-2{\mathcal M})[K{}''-H_0{}'']+ (r-{\mathcal M})[2K{}'-H_2{}']-(r+{\mathcal M})H_0{}'+H_0-H_2\ , \\
 \label{RWlambda12leq1} 
0&=&r(r-2{\mathcal M})[K{}'-H_0{}']-(r-{\mathcal M})H_2+(r-3{\mathcal M})H_0\ ,
\end{eqnarray}
reduce to the system (\ref{RWeq1S})--(\ref{RWeq4S}) where we set $l=1$ (or $\lambda=0$).
Therefore we can take the solutions (\ref{KWsolS}) for all values of $l$.

The reconstruction of the solution summing over all multipoles is nontrivial. Furthermore, the first derivatives of both the functions $K$ and $W$ are not continuous at the particle position implying that the perturbed Riemann tensor is singular there. 
Indeed, a singularity-free solution for this problem is obviously impossible, since there is no external force to oppose the infall of the particle towards the black hole, so that equilibrium cannot be reached in any way.  
We will see in Section IV B how adopting a gauge different from the Regge-Wheeler one gives rise to a more convenient form of the gravitational perturbation functions, yielding a closed form expression for the perturbed metric by summing over all multipoles. The singular character of the solution will be manifest in this new gauge.
 
Of course, one can get regular solutions even in the Regge-Wheeler gauge either by modifying the symmetry of the problem e.g. by the introduction of angular momentum or transverse stresses to balance the gravitational attraction making the resulting configuration as stable.
Consider, for instance, the solution describing a thin spherical shell of matter with suitable isotropic pressure at rest around a (concentric) Schwarzschild black hole. An exact solution for this problem has been found by Frauendiener, Hoenselaers and Konrad \cite{hoenselaers}. 
The spherical symmetry of the problem (with only two additional components $T_{\theta\theta}$ and $T_{\phi\phi}$ of the stress-energy tensor of the source, while $T_{rr}$ is assumed identically zero) allows to get a regular solution, to which only the monopole $l=0$ contributes. 
A problem with less symmetry would require the addition of anisotropic transverse stresses, and all higher values of $l$ may contribute in this case.   

\subsection{The perturbation analysis of the Weyl class}

As it is well known, there exist exact solutions to the vacuum Einstein's field equations representing the nonlinear superposition of individual static gravitating bodies in an axially symmetric configuration: the solutions belonging to the Weyl class (see e.g. \cite{exactsols}).
These solutions are not singularity-free, but exhibit singular structures as ``struts'' and ``membranes'' necessary to balance the bodies.
We refer to the Appendix for a summary on these many-body solutions belonging to the Weyl class. In particular, we are interested in the exact solutions corresponding to a point particle at rest above the horizon of a Schwarzschild black hole as well as to a pair of collinear Schwarzschild black holes. We thus expect to find a solution of the first order perturbation equations obtained following Zerilli's approach that represents just a linearization of these known exact solutions.

Let us start again with the general form of the perturbed metric (\ref{gengeompertSchw}); first of all, we use the available gauge freedom to eliminate two of the off-diagonal terms, corresponding to the first two Regge-Wheeler conditions
\begin{eqnarray}
\label{wgauge1}
h_0\equiv h_1\equiv 0\ .
\end{eqnarray}
We then have
\begin{eqnarray}
\label{Chandrageompert}
 ||h_{\mu \nu }||=\left[ 
\begin {array}{cccc} 
{e^{\nu_{\rm s}}}H_0Y_{l0}&H_1Y_{l0}&0&0
\\\noalign{\medskip}H_1Y_{l0}&{e^{-\nu_{\rm s}}}H_2Y_{ l0}&0&0
\\\noalign{\medskip} 0&0&{r}^{2}\left(KY_{l0}+G\displaystyle\frac{\partial^2 Y_{l0}}{\partial \theta^2 }\right)&0
\\\noalign{\medskip} 0 &0 &0 &{r}^{2} \sin ^{2}\theta\left(KY_{ l0}+G\cot\theta \displaystyle\frac{\partial Y_{l0}}{\partial \theta }\right)
\end {array} 
\right]\ .
\end{eqnarray}
The remaining gauge freedom will be exploited later.

The independent first order perturbations of the quantities appearing in the Einstein field equations (\ref{Eineqs}) are
\begin{eqnarray}
  \label{lambda00}  
{\tilde G}_{00}&=&-\frac12\bigg\{{e^{2\nu_{\rm s}}} \bigg[ 2K{}''-2(\lambda+1)G{}''-\frac2rH_2{}'+\left(\nu_{\rm s}{}'+\frac6r\right)[K{}'-(\lambda+1)G{}']\nonumber\\
&& -    2\left(\frac1{r^2}+\frac{\nu_{\rm s}{}'}r\right)(H_0+H_2) \bigg]-\frac{2 e^{\nu_{\rm s}}}{r^2}[(\lambda+1)H_2-H_0+\lambda K]\bigg\}Y_{l0}\ , \\
  \label{lambda11}  
{\tilde G}_{11}&=&-\frac12\bigg\{\frac 2rH_0{}'-\left(\nu_{\rm s}{}'+\frac2r\right)[K{}'-(\lambda+1)G{}'] +\frac{2 e^{-\nu_{\rm s}}}{r^2}[H_2-(\lambda+1)H_0+\lambda K]\bigg\}Y_{l0}\ , \\
  \label{lambda22}  
{\tilde G}_{22}&=&\frac{r^2}2e^{\nu_{\rm s}}\bigg\{K{}''+\left(\nu_{\rm s}{}'+\frac2r\right)K{}'- H_0{}'' -\left(\frac{\nu_{\rm s}{}'}2+\frac1r\right)H_2{}' -\left(\frac{3\nu_{\rm s}{}'}2+\frac1r\right)H_0{}'\bigg\}Y_{l0}\nonumber\\
&&-\frac{\cot\theta}2\bigg\{H_0-H_2-r^2 e^{\nu_{\rm s}}\bigg[G{}''+\left(\nu_{\rm s}{}'+\frac2r\right)G{}'\bigg]\bigg\}\frac{\partial Y_{l0}}{\partial \theta}\ , \\
  \label{lambda12}  
{\tilde G}_{12}&=&-\frac12\bigg\{-H_0{}' + K{}'- G{}' -\left(\frac{\nu_{\rm s}{}'}2+\frac1r\right)H_2-\left(\frac{\nu_{\rm s}{}'}2-\frac1r\right)H_0\bigg\}\frac{\partial Y_{l0}}{\partial \theta}\ , \\
  \label{lambda01}  
{\tilde G}_{01}&=&\bigg\{\left[\frac{\lambda}{r^2}+\frac{e^{\nu_{\rm s}}}r\left(\nu_{\rm s}{}'+\frac1r\right)\right]H_1\bigg\}Y_{l0}\ , \\
  \label{lambda02}  
{\tilde G}_{02}&=&\frac{e^{\nu_{\rm s}}}{2}\left\{H_1{}' + {\nu_{\rm s}}{}'H_1\right\}\frac{\partial Y_{l0}}{\partial \theta}\ , \\ 
\label{T00partS}  
T_{00}^{\rm part}&=&\frac1{16\pi}A_{{00}}^{\rm s}Y_{l0}\ ,
\end{eqnarray}
where the gravitational source term coming from the point particle is still given by Eq. (\ref{sorgS}).
The angular factors containing derivatives vanish for $l=0$; moreover, the two angular factors in the expression (\ref{lambda22}) for ${\tilde G}_{22}$ are not independent when $l=1$.
Therefore, the cases $l=0,1$ will be treated separately.

For all higher values of $l$, the Einstein field equations (\ref{Eineqs}) imply that the corresponding curly bracketed factors on the left and right hand sides are equal, so that the system of radial equations to be solved is the following:
\begin{eqnarray}
  \label{eq1S}  
0&=&{e^{2\nu_{\rm s}}} \left[ 2K{}''-2(\lambda+1)G{}''-\frac2rH_2{}'+\left(\nu_{\rm s}{}'+\frac6r\right)[K{}'-(\lambda+1)G{}']  - 2\left(\frac1{r^2}+\frac{\nu_{\rm s}{}'}r\right)(H_0+H_2) \right]\nonumber\\
&& -\frac{2 e^{\nu_{\rm s}}}{r^2}[(\lambda+1)H_2-H_0+\lambda K]+A_{{00}}^{\rm s}\ , \\
  \label{eq2S} 
0&=&\frac 2rH_0{}'-\left(\nu_{\rm s}{}'+\frac2r\right)[K{}'-(\lambda+1)G{}'] +\frac{2 e^{-\nu_{\rm s}}}{r^2}[H_2-(\lambda+1)H_0+\lambda K]\ ,\\
  \label{eq3Sa} 
0&=&K{}''+\left(\nu_{\rm s}{}'+\frac2r\right)K{}'- H_0{}'' -\left(\frac{\nu_{\rm s}{}'}2+\frac1r\right)H_2{}' -\left(\frac{3\nu_{\rm s}{}'}2+\frac1r\right)H_0{}'\ ,\\
  \label{eq3Sb} 
0&=&H_0-H_2-r^2e^{\nu_{\rm s}}\left[G{}''+\left(\nu_{\rm s}{}'+\frac2r\right)G{}'\right]\ ,\\
  \label{eq4S} 
0&=&-H_0{}' + K{}'- G{}' -\left(\frac{\nu_{\rm s}{}'}2+\frac1r\right)H_2-\left(\frac{\nu_{\rm s}{}'}2-\frac1r\right)H_0\ ,
\end{eqnarray}
since $H_1\equiv 0$. 

\subsubsection{The new gauge condition}

Consider first the case $l\geq2$. 
We are dealing with a system of $5$ ordinary differential equations for $4$ unknown functions: $H_0$, $H_2$, $K$ and $G$.
Compatibility of the system requires that these equations not be independent. It is easy to show that Eq.~(\ref{eq3Sa}) is identically satisfied by substituting into it the quantities $K{}'$, $K{}''$ and $H_2$ obtained from Eq.~(\ref{eq4S}), the same equation after differentiation with respect to $r$ and Eq.~(\ref{eq3Sb}), respectively.

By imposing the following further gauge condition on the perturbation functions
\begin{equation}
\label{furthercond}
H_0+H_2=2[K-(\lambda+1)G]\ ,
\end{equation}
instead of the third Regge-Wheeler gauge condition $G=0$, we show in the following that with this specific choice we can obtain in a manifest way the linearized form (\ref{weylsol}) of the metric of two collinear Schwarzschild black holes belonging to the Weyl class. In fact, direct inspection of the metric components (\ref{weylgeomfuncts}) below clearly shows that they certainly satisfy the above condition, once expanded in multipoles according to the general form (\ref{gengeompertSchw}).
It is clear that this remarkable result cannot be obtained by using the Regge-Wheeler gauge, since in that case the perturbation analysis remains within the framework of a single Schwarzschild solution.
We will refer to this new gauge as BGR gauge.

Summarizing, our gauge choice is completely specified by the three conditions
\begin{eqnarray}
\label{bgrgauge}
h_0^{(\rm BGR)}\equiv h_1^{(\rm BGR)}\equiv 0\ , \quad H_0^{(\rm BGR)}+H_2^{(\rm BGR)}=2[K^{(\rm BGR)}-(\lambda+1)G^{(\rm BGR)}]\ .
\end{eqnarray}
The metric perturbation functions (\ref{gengauge}) expressed in this gauge are thus given by
\begin{eqnarray}
&&H_0^{(\rm BGR)}=H_0+\frac{2{\mathcal M}}{r(r-2{\mathcal M})}A_1\ , \quad  H_1^{(\rm BGR)}=H_1+\frac{2{\mathcal M}}{r(r-2{\mathcal M})}h_0-h_0{}'\ , \nonumber\\
&&H_2^{(\rm BGR)}=2(\lambda+1)(2A_2-G)-H_0+2K-\frac{2}{r}\frac{2r-3{\mathcal M}}{r-2{\mathcal M}}A_1\ , \nonumber\\
&&K^{(\rm BGR)}=K-\frac{2}{r}A_1\ , \quad G^{(\rm BGR)}=G-2A_2\ ,
\end{eqnarray}
since
\begin{equation}
A_0=-\frac{r}{r-2{\mathcal M}}h_0\ ,
\end{equation} 
while the remaining gauge functions must satisfy the following equations:
\begin{equation}
A_1{}'=\frac{2}{r}\frac{r-{\mathcal M}}{r-2{\mathcal M}}A_1-(\lambda+1)(2A_2-G)+\frac12(H_0+H_2)-K\ , \quad 
A_2{}'=-\frac{A_1}{r(r-2{\mathcal M})}+\frac{h_1}{r^2}\ .
\end{equation} 
Differentiating both sides of the second equation above with respect to $r$ and using the first equation in it leads to
\begin{equation}
A_2{}''=\frac{2(\lambda+1)}{r(r-2{\mathcal M})}A_2+\Lambda\ , \quad 
\Lambda=\frac{h_1{}'}{r^2}-2\frac{h_1}{r^3}- \frac{1}{2r(r-2{\mathcal M})}[2(\lambda+1)G+H_0+H_2-2K]\ ,
\end{equation} 
which involves the function $A_2$ only.

\subsubsection{Explicit solution for the Weyl-type perturbations}

Let us look for a solution of the system consisting of equations (\ref{eq1S}), (\ref{eq2S}), (\ref{eq3Sb}) and (\ref{eq4S}), taking into account the condition (\ref{furthercond}). 
By solving the gauge condition (\ref{furthercond}) for $G$, and substituting it into all the other equations, we have:
\begin{eqnarray}
  \label{unoSw}
0&=&r(r-2{\mathcal M})^2[H_2{}''+H_0{}'']+(r^2-3{\mathcal M}r+2{\mathcal M}^2)H_2{}'+(3r^2-11{\mathcal M}r+10{\mathcal M}^2)H_0{}'\nonumber\\
&&-(r-2{\mathcal M})[(l^2+l+2)H_2-(l+2)(l-1)K]-r^3A_{{00}}^{\rm s}\ , \\
  \label{dueSw}
0&=&(r-{\mathcal M})H_2{}'-(r-3{\mathcal M})H_0{}'-2H_2+l(l+1)H_0-(l+2)(l-1)K\ , \\ 
  \label{treSw}
0&=&(r-2{\mathcal M})[2K{}''-H_2{}''-H_0{}'']+(r-{\mathcal M})[2K{}'-H_2{}'-H_0{}']+H_2-H_0\ , \\
  \label{quattroSw}
0&=&r(r-2{\mathcal M})[H_2{}'-(l^2+l-1)H_0{}'+(l+2)(l-1)K{}']-l(l+1)[(r-{\mathcal M})H_2-(r-3{\mathcal M})H_0]\ .
\end{eqnarray}
Next we eliminate the function $K$ by solving Eq.~(\ref{quattroSw}) for $K{}'$, substituting it and its derivative $K{}''$ into Eq.~(\ref{treSw}), and then by solving Eq.~(\ref{dueSw}) for $K$, substituting it into Eq.~(\ref{unoSw}), thus obtaining a pair of equations involving the functions $H_0$ and $H_2$ only
\begin{eqnarray}
\label{unoSwn}
0&=&r(r-2{\mathcal M})[H_0{}''+H_2{}'']+4(r-2{\mathcal M})H_0{}'-l(l+1)[H_0+H_2]+\frac{r^3}{r-2{\mathcal M}}A_{{00}}^{\rm s}\ ,\\
\label{treSwn}
0&=&r(r-2{\mathcal M})[H_2{}''-H_0{}'']-4{\mathcal M}H_0{}'-l(l+1)[H_2-H_0]\ .
\end{eqnarray}
By subtracting the previous equations, we obtain the following second order differential equation for the function $H_0$:
\begin{equation}
\label{unoSwnew}
0=r(r-2{\mathcal M})H_0{}''+2(r-{\mathcal M})H_0{}'-l(l+1)H_0+\frac{r^3}{2(r-2{\mathcal M})}A_{{00}}^{\rm s}\ .
\end{equation}
Making the transformation $z=r/{\mathcal M}-1$ leads to
\begin{equation}
\label{unoSwnewz}
0=(1-z^2)H_0{}''-2zH_0{}'+l(l+1)H_0-4\sqrt{\pi}\frac{m}{\mathcal M}\sqrt{2l+1}\left(\frac{\beta-1}{\beta+1}\right)^{1/2}\delta(z-\beta)\ ,
\end{equation} 
where primes now denote differentiation with respect to the new variable $z$ and $\beta=b/{\mathcal M}-1$.
The general solutions of the corresponding homogeneous equation are the two types of Legendre functions $P_l$ and $Q_l$.
After imposing regularity conditions on the horizon and at infinity, the solution of Eq. (\ref{unoSwnewz}) is then given by
\begin{eqnarray}
\label{H0ullgeq2}
H_0=4\sqrt\pi\sqrt{2l+1}\frac{m}{\mathcal M}f_{\rm s}(b)^{1/2}\left[P_l(z)Q_l(\beta)\vartheta(b-r)+P_l(\beta)Q_l(z)\vartheta(r-b)\right]\ . 
\end{eqnarray}
From Eq.~(\ref{treSwn}), the corresponding solution for $H_2$ is given by
\begin{eqnarray}
\label{H2vllgeq2}
H_2&=&\left(3-2\frac{r}{\mathcal M}\right)H_0-8\sqrt\pi\frac{\sqrt{2l+1}}{l(l+1)}\frac{m}{{\mathcal M}^2}(b-{\mathcal M})f_{\rm s}(b)^{1/2}r(r-2{\mathcal M})\nonumber\\
&&\times\left[\frac{dQ_l(z(r))}{dr}\bigg\vert_{r=b}\frac{dP_l(z(r))}{dr}\vartheta(b-r)+\frac{dP_l(z(r))}{dr}\bigg\vert_{r=b}\frac{dQ_l(z(r))}{dr}\vartheta(r-b)\right]\ .
\end{eqnarray}
Expressions for the remaining functions $G$ and $K$ can be easily obtained from Eq.~(\ref{eq3Sb}) and the relation (\ref{furthercond}) respectively. 

Finally we have to analyze the cases $l=0,1$ separately. 
Consider first the $l=0$ case. The relevant equations come from quantities (\ref{lambda00})--(\ref{lambda12}) which do not contain angular derivatives. They now have the form
\begin{eqnarray}
  \label{unoSwleq0}
0&=&r(r-2{\mathcal M})K{}''+(3r-5{\mathcal M})K{}'-(r-2{\mathcal M})H_2{}'+K-H_2+\frac{r^3}{2(r-2{\mathcal M})}A_{{00}}^{\rm s}\ , \\
  \label{dueSwleq0}
0&=&(r-{\mathcal M})K{}'-(r-2{\mathcal M})H_0{}'+K-H_2\ ,\\ 
  \label{treSwleq0}
0&=&r(r-2{\mathcal M})[K{}''-H_0{}'']+(r-{\mathcal M})[2K{}'-H_2{}']-(r+{\mathcal M})H_0{}'\ .
\end{eqnarray}
The algebraic gauge condition (\ref{furthercond}) then becomes simply
\begin{equation}
\label{furthercondleq0}
H_0+H_2=2K\ .
\end{equation}
Thus the function $G$ drops out of the equations and can be chosen to be equal to the corresponding one for $l\geq2$ evaluated at $l=0$ without loss of generality: in fact inspection of the perturbed metric tensor (\ref{gengeompert}) shows that the corresponding angular factors vanish identically for $l=0$.

Then by solving Eq.~(\ref{furthercondleq0}) for $K$, and substituting it into equations (\ref{unoSwleq0})--(\ref{treSwleq0}), we obtain 
the following three equations involving the functions $H_0$ and $H_2$ only:
\begin{eqnarray}
  \label{unoSwleq0n}
0&=&r(r-2{\mathcal M})[H_0{}''+H_2{}'']+(3r-5{\mathcal M})H_0{}'+(r-{\mathcal M})H_2{}'+H_0-H_2 +\frac{r^3}{r-2{\mathcal M}}A_{{00}}^{\rm s}\ ,\\
  \label{dueSwleq0n}
0&=&(r-{\mathcal M})H_2{}'-(r-3{\mathcal M})H_0{}'+H_0-H_2\ , \\ 
  \label{treSwleq0n}
0&=&r(r-2{\mathcal M})[H_2{}''-H_0{}'']-4{\mathcal M}H_0{}'\ .
\end{eqnarray}
By subtracting the last two of the previous equations from the first one, we obtain the following second order differential equation for the function $H_0$:
\begin{equation}
0=r(r-2{\mathcal M})H_0{}''+2(r-{\mathcal M})H_0{}'+\frac{r^3}{2(r-2{\mathcal M})}A_{{00}}^{\rm s}\ ,
\end{equation}
which coincides with the (\ref{unoSwnew}) for $l=0$. So, its solution is simply 
\begin{equation}
\label{H0ulleq0}
H_0=-2\sqrt\pi\frac{m}{\mathcal M}f_{\rm s}(b)^{1/2}\left[\ln{\left(1-\frac{2{\mathcal M}}b\right)}\vartheta(b-r)+\ln{\left(1-\frac{2{\mathcal M}}r\right)}\vartheta(r-b)\right]\ ; 
\end{equation}
from Eq.~(\ref{treSwleq0}), $H_2$ results to be given by
\begin{eqnarray}
\label{H2vlleq0}
H_2&=&\left(3-2\frac{r}{\mathcal M}\right)H_0+8\sqrt\pi\frac{m}{\mathcal M}f_{\rm s}(b)^{1/2}
\left[\frac{(r-{\mathcal M})(b-{\mathcal M})-{\mathcal M}^2}{b(b-2{\mathcal M})}\vartheta(b-r)+\vartheta(r-b)\right]\ ,
\end{eqnarray}
while the function $K$ can be easily obtained from relation (\ref{furthercondleq0}). 

Only the case $l=1$ remains to be considered. The two angular factors in the expression (\ref{lambda22}) for the $\tilde G_{22}$ component of the Einstein tensor
are not independent when $l=1$, since $Y_{10}=\cos\theta$, so the two curly bracketed factors must be considered together. 
The relevant equations coming from quantities (\ref{lambda00})--(\ref{lambda12}) are thus given by
\begin{eqnarray}
  \label{unoSwleq1}
0&=&r(r-2{\mathcal M})[G{}''-K{}'']+(3r-5{\mathcal M})[G{}'-K{}']+(r-2{\mathcal M})H_2{}'+2H_2-\frac{r^3}{2(r-2{\mathcal M})}A_{{00}}^{\rm s}\ ,  \\
  \label{dueSwleq1}
0&=&(r-{\mathcal M})[K{}'-G{}']-(r-2{\mathcal M})H_0{}'+H_0-H_2\ ,\\ 
  \label{treSwleq1}
0&=&r(r-2{\mathcal M})[K{}''-G{}''-H_0{}'']+(r-{\mathcal M})[2(K{}'-G{}')-H_2{}']-(r+{\mathcal M})H_0{}'+H_0-H_2\ ,\\ 
 \label{quattroSwleq1}
0&=&r(r-2{\mathcal M})[K{}'-G{}'-H_0{}']-(r-{\mathcal M})H_2-(r-3{\mathcal M})H_0\ .
\end{eqnarray}
The gauge condition (\ref{furthercond}) becomes
\begin{equation}
\label{furthercondleq1}
H_0+H_2=2(K-G)\ .
\end{equation}
By solving Eq.~(\ref{furthercondleq1}) for $G$, and substituting it into Eqs. (\ref{unoSwleq1})--(\ref{quattroSwleq1}), we obtain 
the following four equations involving only the functions $H_0$ and $H_2$:
\begin{eqnarray}
\label{neweqsSwleq1}
0&=&r(r-2{\mathcal M})[H_0{}''+H_2{}'']+(3r-5{\mathcal M})H_0{}'+(r-{\mathcal M})H_2{}'-4H_2+\frac{r^3}{r-2{\mathcal M}}A_{{00}}^{\rm s}\ ,  \nonumber\\
0&=&(r-{\mathcal M})H_2{}'-(r-3{\mathcal M})H_0{}'+2(H_0-H_2)\ , \nonumber\\ 
0&=&r(r-2{\mathcal M})[H_2{}''-H_0{}'']-4{\mathcal M}H_0{}'+2(H_0-H_2)\ , \nonumber\\ 
0&=&r(r-2{\mathcal M})[H_0{}'-H_2{}']+2(r-{\mathcal M})H_2-2(r-3{\mathcal M})H_0\ .
\end{eqnarray}
By subtracting the second and third of the previous equations from the first one, we obtain the following second order differential equation for the function $H_0$
\begin{equation}
0=r(r-2{\mathcal M})H_0{}''+2(r-{\mathcal M})H_0{}'-2H_0+\frac{r^3}{2(r-2{\mathcal M})}A_{{00}}^{\rm s}\ ,
\end{equation}
which coincides with Eq. (\ref{unoSwnew}) for $l=1$; therefore its solution is simply 
\begin{eqnarray}
\label{H0ulleq1}
H_0=4\sqrt{3\pi}\frac{m}{\mathcal M}f_{\rm s}(b)^{1/2}\left[P_1(z)Q_1(\beta)\vartheta(b-r)+P_1(\beta)Q_1(z)\vartheta(r-b)\right]\ , 
\end{eqnarray}
where
\begin{equation}
P_1(z)=z\ , \qquad Q_1(z)=\frac{z}{2}\ln{\left(\frac{z+1}{z-1}\right)}-1\ .
\end{equation}
Furthermore, the third of Eqs. (\ref{neweqsSwleq1}) coincides with Eq. (\ref{treSwn}) for $l=1$; the corresponding solution for $H_2$ is then given by Eq. (\ref{H2vllgeq2}) evaluated at $l=1$.
The function $K$ remains undetermined, and can be chosen to be equal to the corresponding one for $l\geq2$ evaluated at $l=1$ without loss of generality. Then the function $G$ can be easily obtained from Eq. (\ref{furthercondleq1}).

\subsubsection{The analytic solution summed over all values of $l$}

We want now to reconstruct the solution for the gravitational perturbation functions $H_0$, $H_2$, $K$ and $G$ for all values of $l$. We will denote by a bar the corresponding quantities summed over all multipoles.

Consider first the solution (\ref{H0ullgeq2}) for $H_0$. The sum over all multipoles turns out to be
\begin{eqnarray}
\label{barH0}
\bar H_0&=&\sum_{l=0}^\infty H_0Y_{l0}\nonumber\\
&=&2\frac{m}{\mathcal M}f_{\rm s}(b)^{1/2}\sum_{l=0}^\infty H_0Y_{l0}\left[P_l(z)Q_l(\beta)\vartheta(b-r)+P_l(\beta)Q_l(z)\vartheta(r-b)\right]P_l(\cos\theta)\nonumber\\
&=&2\frac{m}{D_{S}}f_{\rm s}(b)^{1/2}\ ,
\end{eqnarray}
where the quantity $D_{S}$ is defined in (\ref{denS}) and the following representation formula has been used:
\begin{eqnarray}
\label{sommaweyl}
\frac{1}{[x^2+t^2 - 2xt\cos\theta - \sin^2\theta]^{1/2}}&=&\sum_{l=0}^\infty (2l+1)\left[P_l(x)Q_l(t)\vartheta(t-x)+P_l(t)Q_l(x)\vartheta(x-t)\right]P_l(\cos\theta)\ .
\end{eqnarray}	

Consider then the solution (\ref{H2vllgeq2}) for $H_2$. The sum over all multipoles turns out to be
\begin{eqnarray}
\label{barH2}
\bar H_2&=&\sum_{l=0}^\infty H_2Y_{l0}=\sum_{l=1}^\infty H_2Y_{l0}+\frac{1}{2\sqrt{\pi}}H_2\vert_{l=0}\nonumber\\
&=&\left(3-2\frac{r}{\mathcal M}\right)\sum_{l=0}^\infty H_0Y_{l0}-4\frac{m}{{\mathcal M}^2}(b-{\mathcal M})f_{\rm s}(b)^{1/2}r(r-2{\mathcal M})\nonumber\\
&&\times\sum_{l=1}^\infty\frac{2l+1}{l(l+1)}\left[\frac{dQ_l(z(r))}{dr}\bigg\vert_{r=b}\frac{dP_l(z(r))}{dr}\vartheta(b-r)+\frac{dP_l(z(r))}{dr}\bigg\vert_{r=b}\frac{dQ_l(z(r))}{dr}\vartheta(r-b)\right]P_l(\cos\theta)\nonumber\\
&&+4\frac{m}{\mathcal M}f_{\rm s}(b)^{1/2}\left[\frac{(r-{\mathcal M})(b-{\mathcal M})-{\mathcal M}^2}{b(b-2{\mathcal M})}\vartheta(b-r)+\vartheta(r-b)\right]\nonumber\\
&=&2\frac{m}{D_{S}}f_{\rm s}(b)^{1/2}-\frac{4{\mathcal M}m}{b(b-2{\mathcal M})}f_{\rm s}(b)^{1/2}\left[1-\frac{r-{\mathcal M}-(b-{\mathcal M})\cos\theta}{D_{S}}\right]\ ,
\end{eqnarray}
where $z=r/{\mathcal M}-1$ and the representation formula (\ref{sommacopson}) has been used.

In order to find the sum over all multipoles of the remaining gravitational perturbation functions $K$ and $G$ it proves more convenient to proceed as follows, rather than deriving first the corresponding multipolar solutions and then trying to sum the series.
Consider Eqs. (\ref{eq3Sb}) and (\ref{eq4S}); since the parameter $l$ does not appear explicitly, they remain valid for the corresponding summed quantities as well and can be rewritten as 
\begin{eqnarray}
  \label{eq3Sbsum} 
0&=&\bar H_0-\bar H_2-\partial_r\left[r(r-2{\mathcal M})\partial_r\bar G\right]\ ,\\
  \label{eq4Ssum} 
0&=&r(r-2{\mathcal M})\partial_r[-\bar H_0 + \bar K- \bar G] -(r-{\mathcal M})\bar H_2+(r-3{\mathcal M})\bar H_0\ .
\end{eqnarray}
These equations must now be treated as partial differential equations rather than ordinary differential equations.
In fact summing over the spherical harmonics leads to the barred functions which are thus depending also on the angular variable $\theta$. After integration on the radial variable each function $\bar G$ and $\bar K$ will be determined up to an arbitrary function of $\theta$, which will be then chosen in order the perturbed metric to satisfy the Einstein field equations (\ref{Eineqs}).
Substituting the solutions (\ref{barH0}) and (\ref{barH2}) into Eq. (\ref{eq3Sbsum}) gives the following first integral \begin{equation}
\label{DrbarG}
\partial_r\bar G=\frac{4{\mathcal M}m}{b(b-2{\mathcal M})}f_{\rm s}(b)^{1/2}\frac{r-D_{S}+k_1(\theta)}{r(r-2{\mathcal M})}\ ,
\end{equation} 
where $k_1(\theta)$ is an arbitrary function of the polar angle. A further integration gives
\begin{eqnarray}
\label{barG}
\bar G&=&\frac{4{\mathcal M}m}{b(b-2{\mathcal M})}f_{\rm s}(b)^{1/2}\left\{
-\ln\left(\frac{z-\beta\cos\theta+J}{\sqrt{z^2-1}}\right)+\frac{1}{2}\beta\ln\left[1+\frac{2J(z\beta-\cos\theta+J)}{(\beta^2-1)(z^2-1)}\right]\right.\nonumber\\
&&\left.-\frac{1}{2}\cos\theta\ln\left[-4(\beta^2-1)\left(\sin^2\theta-\frac{2J(\beta-z\cos\theta+J)}{z^2-1}\right)\right]-k_1(\theta){\rm arctanh}(z)+k_2(\theta)
\right\}\ ,
\end{eqnarray}
where $z=r/{\mathcal M}-1$, $\beta=b/{\mathcal M}-1$, $J=D_{S}/{\mathcal M}$ and $k_2(\theta)$ is another arbitrary function of $\theta$.
The solution for $\bar K$ follows immediately by integrating Eq. (\ref{eq4Ssum}): 
\begin{eqnarray}
\label{barK}
\bar K=\bar H_0 + \bar G + \frac{4{\mathcal M}m}{b(b-2{\mathcal M})}f_{\rm s}(b)^{1/2}\left\{
-\ln\left(\frac{z-\beta\cos\theta+J}{\sqrt{z^2-1}}\right)+\frac{1}{2}\beta\ln\left[1+\frac{2J(z\beta-\cos\theta+J)}{(\beta^2-1)(z^2-1)}\right]+k_3(\theta)\right\}\ ,
\end{eqnarray}
where $k_3(\theta)$ is a third arbitrary function.

Therefore, starting from the expansion (\ref{gengeompertSchw}) for the gravitational field, the perturbed metric summed over all multipoles turns out to be 
\begin{equation}
\label{bgrsol}
d{\tilde s}^2=-e^{\nu_{\rm s}}[1-\bar H_0]dt^2+e^{-\nu_{\rm s}}[1+\bar H_2]dr^2+r^2\left[1+\left(\bar K+\frac{\partial^2 \bar G }{\partial \theta^2 }\right)\right]d\theta^2+r^2\sin\theta^2\left[1+\left(\bar K+\cot\theta \frac{\partial \bar G}{\partial \theta }\right)\right]d\phi^2\ .
\end{equation}
The Einstein field equations (\ref{Eineqs}) give the following constraints on the undetermined angular functions $k_1(\theta)$, $k_2(\theta)$ and $k_3(\theta)$:
\begin{eqnarray}
k_1(\theta)&=&c_1\cos\theta+c_2\left[1+\frac12\cos\theta\ln\left(\frac{1-\cos\theta}{1+\cos\theta}\right)\right]\ , \nonumber\\ 	
k_2(\theta)+k_3(\theta)&=&c_3\cos\theta+c_4\left[2+\frac12\cos\theta\ln\left(\frac{1-\cos\theta}{1+\cos\theta}\right)\right]+\cos\theta\ln(1-\cos\theta)\ ,
\end{eqnarray}
where $c_i$, $i=1,\ldots,4$ are arbitrary integration constants.
Making the choice $c_2=0=c_4$ and $k_3(\theta)=0$ brings the metric (\ref{bgrsol}) in the form 
\begin{equation}
\label{bgrsolW}
d{\tilde s}^2=-e^{\nu_{\rm s}}[1-\bar H_0]dt^2+e^{-\nu_{\rm s}}[1+\bar H_2]dr^2+r^2\left[1+\bar H_2\right]d\theta^2+r^2\sin\theta^2\left[1+\bar H_0\right]d\phi^2\ ,
\end{equation}
since in this case
\begin{equation}
\bar K+\frac{\partial^2 \bar G }{\partial \theta^2 }\equiv\bar H_2\ , \qquad
\bar K+\cot\theta \frac{\partial \bar G}{\partial \theta }\equiv\bar H_0\ .
\end{equation}

\subsubsection{Comparison with the Weyl class two-body solution}

The solution (\ref{bgrsolW}) we have found to first order in the perturbation is just the linearization with respect to $m$ of the exact solution given in Appendix A representing the superposition of a Chazy-Curzon particle and a Schwarzschild black hole or of two collinear Schwarzschild black holes (to linear order these two solutions agree) 
\begin{equation}
\label{weylsol}
d{\tilde s}^2=-e^{\nu_{\rm s}}[1-{\bar h}^{\rm w}_0]dt^2+e^{-\nu_{\rm s}}[1+{\bar h}^{\rm w}_1]dr^2+r^2[1+{\bar h}^{\rm w}_2]d\theta^2+r^2\sin\theta^2[1+{\bar h}^{\rm w}_3]d\phi^2\ ,
\end{equation}
where
\begin{eqnarray}
\label{weylgeomfuncts}
{\bar h}^{\rm w}_0&=&\frac{2m}{{\mathcal D}_{\rm w}}\ ,\nonumber\\
{\bar h}^{\rm w}_1&=&{\bar h}^{\rm w}_0-\frac{4{\mathcal M}m}{b^2-{\mathcal M}^2}\left[1-\frac{r-{\mathcal M}-b\cos\theta}{{\mathcal D}_{\rm w}}\right]\ ,\nonumber\\
{\bar h}^{\rm w}_2&=&{\bar h}^{\rm w}_1\ ,\nonumber\\
{\bar h}^{\rm w}_3&=&{\bar h}^{\rm w}_0\ ,
\end{eqnarray}
and
\begin{eqnarray*}
e^{\nu_{\rm s}}\equiv f_{\rm s}(r)=1-\frac{2{\mathcal M}}r\ , \qquad {\mathcal D}_{\rm w}= [(r-{\mathcal M})^2+b^2 
- 2(r -{\mathcal M})b\cos\theta - {\mathcal M}^2\sin^2\theta]^{1/2}\ . 
\end{eqnarray*}
In fact the metric (\ref{bgrsolW}) has the same form of the Weyl metric (\ref{weylsol}); direct comparison between Eqs. (\ref{barH0}) and (\ref{barH2}) and Eq. (\ref{weylgeomfuncts}) shows that the functions $\bar H_0$ and $\bar H_2$ coincide exactly with the corresponding ones ${\bar h}^{\rm w}_0$ and ${\bar h}^{\rm w}_1$ after shifting the location of the particle to $b\longrightarrow b-{\mathcal M}$ and recalling the definition of the \lq\lq active gravitational mass'' of the particle by \cite{membrane2} resulting in the overall factor $f_{\rm s}(b)^{1/2}$ (so that $m\longrightarrow mf_{\rm s}(b)^{-1/2}$).
 
It is worth noting again that this solution is characterized by the presence of a conical singularity on the polar axis between the bodies \cite{weyl,einstein}.

\subsubsection{Relation with the Regge-Wheeler approach}

Using Eq. (\ref{rwfuncts}) expressing the metric perturbation functions in the Regge-Wheeler gauge as combinations of metric perturbations expressed in an arbitrary gauge together with the conditions (\ref{bgrgauge}) identifying our gauge choice yields the explicit relation between the two gauges 
\begin{eqnarray}
&&H_0^{(\rm RW)}=H_0^{(\rm BGR)}-{\mathcal M}G^{(\rm BGR)}{}'\ , \quad  H_1^{(\rm RW)}=0=H_1^{(\rm BGR)}\ , \nonumber\\
&&H_2^{(\rm RW)}=H_2^{(\rm BGR)}+r(r-2{\mathcal M})G^{(\rm BGR)}{}''+(2r-3{\mathcal M})G^{(\rm BGR)}{}'\ , \nonumber\\
&&K^{(\rm RW)}=K^{(\rm BGR)}+(r-2{\mathcal M})G^{(\rm BGR)}{}'\ .
\end{eqnarray}
Since the parameter $l$ does not appear explicitly, the previous relations remain valid for the corresponding summed quantities as well, which will be denoted by a bar. The perturber metric written in the Regge-Wheeler gauge in then given by
\begin{equation}
\label{RWsolsum}
d{\tilde s}^2=-e^{\nu_{\rm s}}[1-\bar W^{(\rm RW)}]dt^2+e^{-\nu_{\rm s}}[1+\bar W^{(\rm RW)}]dr^2+r^2\left[1+\bar K^{(\rm RW)}\right](d\theta^2+r^2\sin\theta^2d\phi^2)\ ,
\end{equation}
where 
\begin{eqnarray}
\bar H_0^{(\rm RW)}&=&\bar H_2^{(\rm RW)}\equiv\bar W^{(\rm RW)}=\bar H_0-{\mathcal M}\partial_r\bar G=\bar H_0-\frac{4{\mathcal M}^2m}{b(b-2{\mathcal M})}f_{\rm s}(b)^{1/2}\frac{r-D_{S}+k_1(\theta)}{r(r-2{\mathcal M})}\ , \nonumber\\
\bar K^{(\rm RW)}&=&\bar K+(r-2{\mathcal M})\partial_r\bar G=\bar K+\frac{4{\mathcal M}m}{b(b-2{\mathcal M})}\frac{f_{\rm s}(b)^{1/2}}{r}[r-D_{S}+k_1(\theta)]\ ,
\end{eqnarray}
where $\bar H_0$ and $\bar K$ are given by Eqs. (\ref{barH0}) and (\ref{barK}) respectively and Eq. (\ref{DrbarG}) has been used.

\section{Perturbation analysis: the Reissner-Nordstr\"om case}
\label{RNpert}

\subsection{The general perturbation equations}

We now turn our attention to the charged Reissner-Nordstr\"om black hole case.
The geometrical perturbations $h_{\mu \nu }$ for the electric multipoles are given by
\begin{eqnarray}
\label{gengeompert}
 ||h_{\mu \nu }||=\left[ 
\begin {array}{cccc} 
{e^{\nu}}H_0Y_{l0}&H_1Y_{l0}&h_0\displaystyle\frac{\partial Y_{l0}}{\partial \theta }&0
\\\noalign{\medskip}\rm{sym}&{e^{-\nu}}H_2Y_{ l0}&h_1\displaystyle\frac{\partial Y_{l0}}{\partial \theta }&0
\\\noalign{\medskip} \rm{sym} &\rm{sym} &{r}^{2}\left(KY_{l0}+G\displaystyle\frac{\partial^2 Y_{l0}}{\partial \theta^2 }\right)&0
\\\noalign{\medskip} \rm{sym} & \rm{sym} & \rm{sym} &{r}^{2} \sin ^{2}\theta\left(KY_{ l0}+G\cot\theta \displaystyle\frac{\partial Y_{l0}}{\partial \theta }\right)
\end {array} 
\right],
\end{eqnarray}
where the symbol ``sym'' indicates that the missing components of $h_{\mu \nu }$ are to be found from the symmetry $h_{\mu \nu }=h_{\nu \mu }$, and $e^{\nu}=f(r)$
is Zerilli's notation.

In order to simplify the description of the perturbation, we use the Regge-Wheeler \cite{ReggeW} gauge to set
\begin{eqnarray*}
h_0\equiv h_1\equiv G\equiv 0\ ;
\end{eqnarray*}
thus we have
\begin{eqnarray}
\label{RWgeompert}
 ||h_{\mu \nu }||=\left[ 
\begin {array}{cccc} 
{e^{\nu}}H_0Y_{l0}&H_1Y_{l0}&0&0
\\\noalign{\medskip}H_1Y_{l0}&{e^{-\nu}}H_2Y_{ l0}&0&0
\\\noalign{\medskip}0&0&{r}^{2}KY_{l0}&0\\\noalign{\medskip}0&0
&0&{r}^{2} \sin ^{2}\theta KY_{l0}
\end{array} 
\right]\ .
\end{eqnarray}
The electromagnetic field harmonics $f_{\mu \nu }$ for the electric multipoles are given by
\begin{eqnarray}
\label{RWempert}
 ||f_{\mu \nu }||= \left[ 
\begin {array}{cccc} 
0&\tilde f_{{01}}Y_{l0}&\tilde f_{{02}}\displaystyle\frac{\partial Y_{l0}}{\partial \theta }&0
\\\noalign{\medskip} \rm{antisym} &0&\tilde f_{{12}}\displaystyle\frac{\partial Y_{l0}}{\partial \theta }&0
\\\noalign{\medskip}\rm{antisym}&\rm{antisym}&0&0
\\\noalign{\medskip}\rm{antisym}&\rm{antisym}&\rm{antisym}&0
\end {array} 
\right]\ , 
\end{eqnarray}
where $\tilde f_{{\mu \nu }}$ denotes the $\theta$-independent part of $ f_{{\mu \nu }}$, and the symbol ``antisym'' indicates components obtainable by antisymmetry.

The expansion of the source terms (\ref{sorgenti}) gives the relations
\begin{eqnarray}
\sum_l A_{{00}}Y_{l0}=16\pi T_{{00}}^{\rm{part}}\ , \qquad 
\sum_l vY_{l0}=J^{{0}}_{\rm{part}}\ ,
\end{eqnarray}
with 
\begin{eqnarray}
\label{sorgexp}
A_{{00}}=8\sqrt{\pi} \frac{m\sqrt {2l+1}}{b^2}f(b)^{3/2}\delta \left( r-b \right)\ , \qquad
v=\frac1{2\sqrt{\pi}} \frac{q\sqrt {2l+1}}{b^2}\delta \left( r-b \right)\ ,
\end{eqnarray}
where we have used the following expansion for $\delta(\cos\theta-1)$:
\begin{eqnarray}
\label{deltaexp}
\delta(\cos\theta-1)=\sqrt{\pi }\sum_l \sqrt{2l+1}Y_{l0}\ .
\end{eqnarray}

The independent first order perturbations of the quantities appearing in the Einstein-Maxwell field equations (\ref{EinMaxeqs}) are given by
\begin{eqnarray}
  \label{RNlambda00}  
{\tilde G}_{00}&=&-\frac12\bigg\{{e^{2\nu}} \left[ 2K{}''-\frac2rH_2{}'+\left(\nu{}'+\frac6r\right) { K{}'} -    2\left(\frac1{r^2}+\frac{\nu{}'}r\right)(H_0+H_2) \right] \nonumber\\
&& -\frac{2 e^{\nu}}{r^2}[(\lambda+1)H_2-H_0+\lambda K]\bigg\}Y_{l0}\ , \\
  \label{RNlambda11}  
{\tilde G}_{11}&=&-\frac12\bigg\{\frac 2rH_0{}'-\left(\nu{}'+\frac2r\right)K{}' +\frac{2 e^{-\nu}}{r^2}[H_2-(\lambda+1)H_0+\lambda K]\bigg\}Y_{l0}\ ,\\
  \label{RNlambda22}  
{\tilde G}_{22}&=&\frac{r^2}2e^{\nu}\bigg\{K{}''+\left(\nu{}'+\frac2r\right)K{}'- H_0{}'' -\left(\frac{\nu{}'}2+\frac1r\right)H_2{}' -\left(\frac{3\nu{}'}2+\frac1r\right)H_0{}'
  +2(\lambda+1)\frac{e^{-\nu}}{r^2}(H_0-H_2)\nonumber\\
&& +\left(\nu{}''+{\nu{}'}^2+\frac{2\nu{}'}r\right)(K-H_2)\bigg\}Y_{l0}+\frac12\bigg\{H_0-H_2\bigg\}\frac{\partial^2 Y_{l0}}{\partial \theta^2}\ ,\\
  \label{RNlambda12}  
{\tilde G}_{12}&=&-\frac12\bigg\{-H_0{}' + K{}' -\left(\frac{\nu{}'}2+\frac1r\right)H_2-\left(\frac{\nu{}'}2-\frac1r\right)H_0\bigg\}\frac{\partial Y_{l0}}{\partial \theta}\ ,\\
  \label{RNlambda01}  
{\tilde G}_{01}&=&\bigg\{\left[\frac{\lambda}{r^2}+\frac{e^{\nu}}r\left(\nu{}'+\frac1r\right)\right]H_1\bigg\}Y_{l0}\ , \\
  \label{RNlambda02}  
{\tilde G}_{02}&=&\frac{e^{\nu}}{2}\left\{H_1{}' + {\nu}{}'H_1\right\}\frac{\partial Y_{l0}}{\partial \theta}\ ,
\end{eqnarray}
\begin{eqnarray}
  \label{RNtildeT00}  
{\tilde T}_{00}&=&-\frac1{8\pi}\left\{{\frac {{Q}^{2}{e^{\nu}}H_2 }{{r}^{4}}}+2{\frac{Q{e^{\nu}}\tilde     f_{{01}}}{{r}^{2}}}\right\}Y_{l0}\ , \hspace{3cm} {\tilde T}_{11}=-\frac1{8\pi}\left\{ {\frac{{Q}^{2}{e^{-\nu}}H_0 }{{r}^{4}}}-2{\frac {Q{e^{-\nu}}\tilde f_{{01}} }{{r}^{2}}}\right\}Y_{l0}\ ,\\ 
  \label{RNtildeT22}  
{\tilde T}_{22}&=&\frac{r^2e^{\nu}}{8\pi}\left\{{\frac{{Q}^{2}{e^{-\nu}}K}{{r}^{4}}}
    -\frac{2Q{e^{-\nu}}}{{r}^{2}}  \tilde f_{{01}}\right\}Y_{l0}\ , \hspace{2.7cm}   
{\tilde T}_{12}=\frac1{8\pi}\left\{2{\frac {Q{e^{-\nu}}\tilde f_{{02}} }{{r}^{2}}}\right\}\frac{\partial Y_{l0}}{\partial \theta}\ , \\
  \label{RNtildeT01}  
{\tilde T}_{01}&=&-\frac1{8\pi}\left\{2{\frac {Q^2}{{r}^{4}}}H_1 \right\}Y_{l0}\ ,  \hspace{5cm}
{\tilde T}_{02}=\frac{e^{\nu}}{8\pi}\left\{2{\frac {Q}{{r}^{2}}}\tilde f_{{12}} \right\}\frac{\partial Y_{l0}}{\partial \theta}\ ,\\
  \label{RNdeltaT00}  
T_{00}^{\rm part}&=&\frac1{16\pi}A_{{00}}Y_{l0}\ ,  \hspace{6.2cm}
J^{0}_{\rm part}=vY_{l0}\ ,\\
 \label{RNdeltaF0nu}  
{\tilde F}^{0\nu}{}_{;\,\nu }&=&-\left\{\tilde f_{{01}}{}'+\frac2{r}\tilde f_{{01}} -{\frac {l\left( l+1 \right) {e^{-\nu}}\tilde     f_{{02}} }{{r}^{2}}}-{\frac {Q}{{r}^{2}}}K{}'\right\}Y_{l0}\ ,  \hspace{0.5cm} 
{\tilde F}^{1\nu}{}_{;\,\nu }=\left\{ -{\frac {l\left( l+1 \right) {e^{\nu}} }{{r}^{2}}}\tilde     f_{{12}}\right\}Y_{l0}\ , \\ 
  \label{RNdeltaF2nu}  
{\tilde F}^{2\nu}{}_{;\,\nu }&=&-\frac{e^{\nu}}{r^2}\left\{\tilde f_{{12}}{}' + {\nu}{}'\tilde     f_{{12}}\right\}\frac{\partial Y_{l0}}{\partial \theta}\ ,  \hspace{3.7cm}  
{}^*{\tilde F}^{3\nu}{}_{;\,\nu }=\frac1{r^2\sin\theta}\left\{\tilde f_{{01}} - \tilde f_{{02}}{}'\right\}\frac{\partial Y_{l0}}{\partial \theta}\ ,
\end{eqnarray}
where $\lambda=\frac12 \left( l-1 \right) \left( l+2 \right)$ and a prime denotes differentiation with respect to $r$.  

The angular factors containing derivatives vanish for $l=0$; moreover, the two angular factors in the expression (\ref{RNlambda22}) for ${\tilde G}_{22}$ are not independent when $l=1$ (in fact, ${\partial^2 Y_{10}}/{\partial \theta^2}=-Y_{10}$).
Therefore, the cases $l=0,1$ will be treated separately.

For all higher values of $l$, the Einstein-Maxwell field equations (\ref{EinMaxeqs}) imply that the corresponding curly bracketed factors on the left and right hand sides are equal, so that the system of radial equations we have to solve is the following:
\begin{eqnarray}
  \label{eq1RN}  
0&=&{e^{2\nu}} \left[ 2K{}''-\frac2rW{}'+\left(\nu{}'+\frac6r\right) { K{}'}-    4\left(\frac1{r^2}+\frac{\nu{}'}r\right)W \right]-\frac{2\lambda e^{\nu}}{r^2}(W+K)   
    -2{\frac {{Q}^{2}{e^{\nu}}W }{{r}^{4}}}-4{\frac{Q{e^{\nu}}\tilde     f_{{01}}}{{r}^{2}}}+A_{{00}}\ , \\
  \label{eq2RN} 
0&=&\frac 2rW{}'-\left(\nu{}'+\frac2r\right)K{}' -\frac{2\lambda e^{-\nu}}{r^2}(W-K)-        2{\frac{{Q}^{2}{e^{-\nu}}W }{{r}^{4}}}+4{\frac {Q{e^{-\nu}}\tilde f_{{01}} }{{r}^{2}}}\ ,\\
  \label{eq3RN} 
0&=&K{}''+\left(\nu{}'+\frac2r\right)K{}'- W{}'' -2\left(\nu{}'+\frac1r\right)W{}'
    +\left(\nu{}''+{\nu{}'}^2+\frac{2\nu{}'}r\right)(K-W) -2{\frac{{Q}^{2}{e^{-\nu}}K}{{r}^{4}}}
    +\frac{4Q{e^{-\nu}}}{{r}^{2}}  \tilde f_{{01}}\ ,\\
  \label{eq4RN} 
0&=&-W{}' + K{}' -\nu{}' W +4{\frac {Q{e^{-\nu}}\tilde f_{{02}} }{{r}^{2}}}\ ,\\
  \label{eq5RN} 
0&=&\tilde f_{{01}}{}'+\frac2{r}\tilde f_{{01}} -{\frac {l\left( l+1 \right) {e^{-\nu}}\tilde     f_{{02}} }{{r}^{2}}}-{\frac {Q}{{r}^{2}}}K{}' +4\pi v\ ,\\
  \label{eq6RN} 
0&=&\tilde f_{{01}} - \tilde f_{{02}}{}'\ ,
\end{eqnarray}
since
\begin{eqnarray}
H_0= H_2\equiv W\ , \qquad 
H_1\equiv 0\ , \qquad
\tilde f_{{12}}\equiv 0\ .
\end{eqnarray}

We are dealing with a system of $6$ coupled ordinary differential equations for $4$ unknown functions: $K$, $W$, $\tilde f_{{01}} $ and $\tilde f_{{02}}$.
Compatibility of the system requires that these equations not be independent. We see that the Eq.~(\ref{eq3RN}) can be obtained from Eqs. (\ref{eq2RN}) and (\ref{eq4RN}): in fact it is enough to consider Eqs.~(\ref{eq2RN}) and  (\ref{eq4RN}), solving them for $K{}'$ and $W{}'$, together with the corresponding equations obtained by differentiation with respect to $r$, solving those for $K{}''$ and $W{}''$ and finally substituting all these quantities into Eq.~(\ref{eq3RN}), which is then identically satisfied.

Next another equation must be eliminated. It is useful to introduce the the new combinations 
\begin{eqnarray}
\label{newfunctions}
X=K-W\ , \qquad Y=K+W\ ,
\end{eqnarray}
so that the system we have to solve for the unknown functions $X$, $Y$, $\tilde f_{01}$ and $\tilde f_{02}$ is the following
\begin{eqnarray}
  \label{eq1RNxy}  
0&=&{e^{2\nu}} \left[ X{}''+Y{}''-\frac1r(X{}'-Y{}')+\left(\frac{\nu{}'}2+\frac3r\right)(X{}'+Y{}')-    2\left(\frac1{r^2}+\frac{\nu{}'}r\right)(X-Y) \right]-\frac{2\lambda e^{\nu}}{r^2}Y   
    -{\frac {{Q}^{2}{e^{\nu}}}{{r}^{4}}}(X-Y) \nonumber\\
    &&-4{\frac{Q{e^{\nu}}\tilde     f_{{01}}}{{r}^{2}}}+A_{{00}}\ , \\
  \label{eq2RNxy} 
0&=&-\frac2r X{}'-\frac{\nu{}'}2(X{}'+Y{}') +\frac{2\lambda e^{-\nu}}{r^2}X-        {\frac{{Q}^{2}{e^{-\nu}}(X-Y) }{{r}^{4}}}+4{\frac {Q{e^{-\nu}}\tilde f_{{01}} }{{r}^{2}}}\ ,\\
  \label{eq4RNxy} 
0&=&X{}'  -\frac{\nu{}'}2(X-Y) +4{\frac {Q{e^{-\nu}}\tilde f_{{02}} }{{r}^{2}}}\ ,\\
  \label{eq5RNxy} 
0&=&\tilde f_{{01}}{}'+\frac2{r}\tilde f_{{01}} -{\frac {l\left( l+1 \right) {e^{-\nu}}\tilde     f_{{02}} }{{r}^{2}}}-{\frac {Q}{2{r}^{2}}}(X{}'+Y{}') +4\pi v\ ,\\
  \label{eq6RNxy} 
0&=&\tilde f_{{01}} - \tilde f_{{02}}{}'\ .
\end{eqnarray}
First we solve algebraically Eqs.~(\ref{eq2RNxy}) and (\ref{eq4RNxy}) for $\tilde f_{01}$ and $\tilde f_{02}$ respectively. Then we substitute the quantities so obtained into Eqs.~(\ref{eq1RNxy}), (\ref{eq5RNxy}) and (\ref{eq6RNxy}), involving only the gravitational perturbation functions $X$ and $Y$
\begin{eqnarray}
  \label{effe01}
\tilde f_{{01}}&=&\frac{2r^2-3{\mathcal M}r+Q^2}{4rQ}X{}'+\frac{{\mathcal M}r-Q^2}{4rQ}Y{}'-\frac{2\lambda r^2+Q^2}{4Qr^2}X+\frac{Q}{4r^2}Y\ , \\
  \label{effe02}
\tilde f_{{02}}&=&-\frac{r^2-2{\mathcal M}r+Q^2}{4Q}X{}'+\frac{{\mathcal M}r-Q^2}{4rQ}(Y-X)\ , \\
  \label{eqYcompat}
0&=&Y{}''+ \frac2{r}Y{}'-\frac{2(\lambda+1)}{r^2-2{\mathcal M}r+Q^2}Y+\frac{r^4A_{00}}{(r^2-2{\mathcal M}r+Q^2)^2} - \frac{2(r+Q)(r-Q)}{r(r^2-2{\mathcal M}r+Q^2)}X{}'+\frac{2(2\lambda +1)}{r^2-2{\mathcal M}r+Q^2}X\ , \\
  \label{eqX}
0&=&(r^2-2{\mathcal M}r+Q^2)X{}''+4(r-{\mathcal M})X{}'-2\lambda X\ , \\
  \label{eqY}
0&=&Y{}''+ \frac2{r}Y{}'-\frac{2(\lambda+1)}{r^2-2{\mathcal M}r+Q^2}Y+\frac{16\pi rQ}{({\mathcal M}r-Q^2)}v - \frac{2(r+Q)(r-Q)}{r(r^2-2{\mathcal M}r+Q^2)}X{}'+\frac{2(2\lambda +1)}{r^2-2{\mathcal M}r+Q^2}X\ .
\end{eqnarray}

\subsection{The compatibility and Bonnor's condition}

Direct comparison between Eqs.~(\ref{eqYcompat}) and (\ref{eqY}) shows that they are  compatible only if the following stability condition holds
\begin{equation}
\label{compcondnextr}
m=qQ\frac{b f(b)^{1/2}}{{\mathcal M}b-Q^2}\ .
\end{equation} 

The same condition can be obtained by imposing conservation of the stress-energy tensor: 
\begin{equation}
\label{compatib}
0=A_{00}-32\pi \frac{Q e^{\nu}}{r^2\nu{}'}v\ .
\end{equation}
So Eq.~(\ref{eq1RN}) becomes
\begin{equation}
\label{eq1newRN}  
0={e^{2\nu}} \left[ 2K{}''-\frac2rW{}'+\left(\nu{}'+\frac6r\right) { K{}'}-    4\left(\frac1{r^2}+\frac{\nu{}'}r\right)W \right]-\frac{2\lambda e^{\nu}}{r^2}(W+K)   
    -2{\frac {{Q}^{2}{e^{\nu}}W }{{r}^{4}}}-4{\frac{Q{e^{\nu}}\tilde     f_{{01}}}{{r}^{2}}}-32\pi \frac{Q e^{\nu}}{r^2\nu{}'}v\ .
\end{equation}
This equation is identically satisfied, seen by substituting into it the quantities $K{}'$, $W{}'$, and $K{}''$ and $W{}''$ obtained from Eqs.~(\ref{eq2RN}), (\ref{eq4RN}) and their derivatives, and $\tilde f_{{01}}{}'$ and $\tilde f_{{02}}{}'$ from Eqs.~(\ref{eq5RN}) and (\ref{eq6RN}) respectively.

The stability condition (\ref{compcondnextr}) coincides exactly with the equilibrium condition (\ref{bonnoreqcond}) for such a system, which has been discussed by Bonnor \cite{bonnor} in the case of a test field approximation.
Therefore the static configuration of a particle at rest near the black hole remains an equilibrium configuration as a result of the perturbation only for certain positions of the particle which are completely determined by the charge to mass ratio of the black hole and of the particle itself: in particular, if we require that $Q/{\mathcal M}<1$, the particle must be overcritically charged, i.e. $q/m>1$. However, the choice of equilibrium configurations is a special one, since it forces us to consider particle and black hole charges of the same sign only. Hence, if we want to consider the more general case of charges of opposite sign, the nonvanishing contribution to the stress-energy tensor due to some external force should be taken into account in solving the full Einstein-Maxwell system (\ref{EM}).
In the extreme black hole case ($Q={\mathcal M}$), instead the equilibrium condition (\ref{compcondnextr}) becomes simply
\begin{equation}
\label{compcondextr}
m=q\ ,
\end{equation} 
implying that all the configurations are equilibrium configurations as a result of the perturbation, but the particle also is forced to have the critically charged ratio $q/m=1$.

\subsubsection{Discussion of the equilibrium condition}

Let us examine the equilibrium condition (\ref{compcondnextr}) in more detail; it can be rewritten in the form
\begin{equation}
\label{Fdib}
\frac{q}{m}\frac{Q}{\mathcal M}=\frac{b/{\mathcal M}-(Q/{\mathcal M})^2}{[(b/{\mathcal M})^2-2b/{\mathcal M}+(Q/{\mathcal M})^2]^{1/2}}\equiv {\mathcal F}(b/{\mathcal M}; Q/{\mathcal M})\ ,
\end{equation}
with ${\mathcal F}(b/{\mathcal M}; Q/{\mathcal M})\geq1$ for all values of $|Q/{\mathcal M}|\leq1$ and $b\geq r_+$.
Thus the system turns out to be coupled in such a way that, for a fixed value of $Q/{\mathcal M}$, the charge to mass ratio $q/m$ of the particle must decrease as the separation parameter $b/{\mathcal M}$ 
increases, while it must be larger and larger as the particle approaches the black hole horizon.
On the other hand, for a fixed distance parameter $b/{\mathcal M}$, the value of $q/m$ must increase for decreasing values of $Q/{\mathcal M}$, in order to oppose the gravitational attraction and so maintain the equilibrium,
while it must decrease until it reaches the value $1$ as the black hole approaches the extreme condition, since ${\mathcal F}(b/{\mathcal M}; Q/{\mathcal M}\to1)\to1$ irrespective of the value of $b/{\mathcal M}$.
Obviously this is the limiting situation of two critically charged bodies which are in equilibrium at any separation distance. 

\begin{figure} 
\typeout{*** EPS figure indhor}
\begin{center}
$\begin{array}{c@{\hspace{1in}}c}
\includegraphics[scale=0.33]{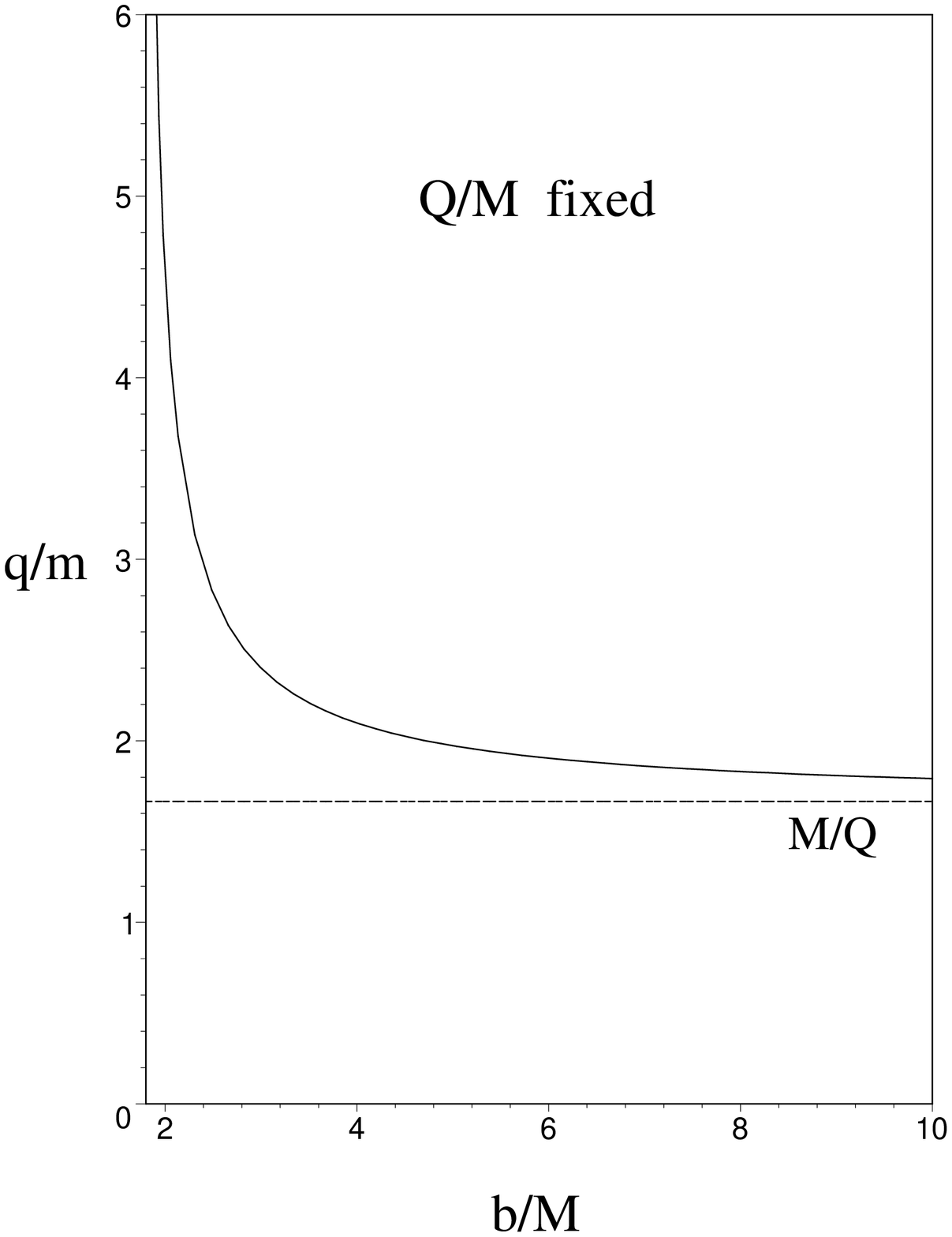}&
\includegraphics[scale=0.33]{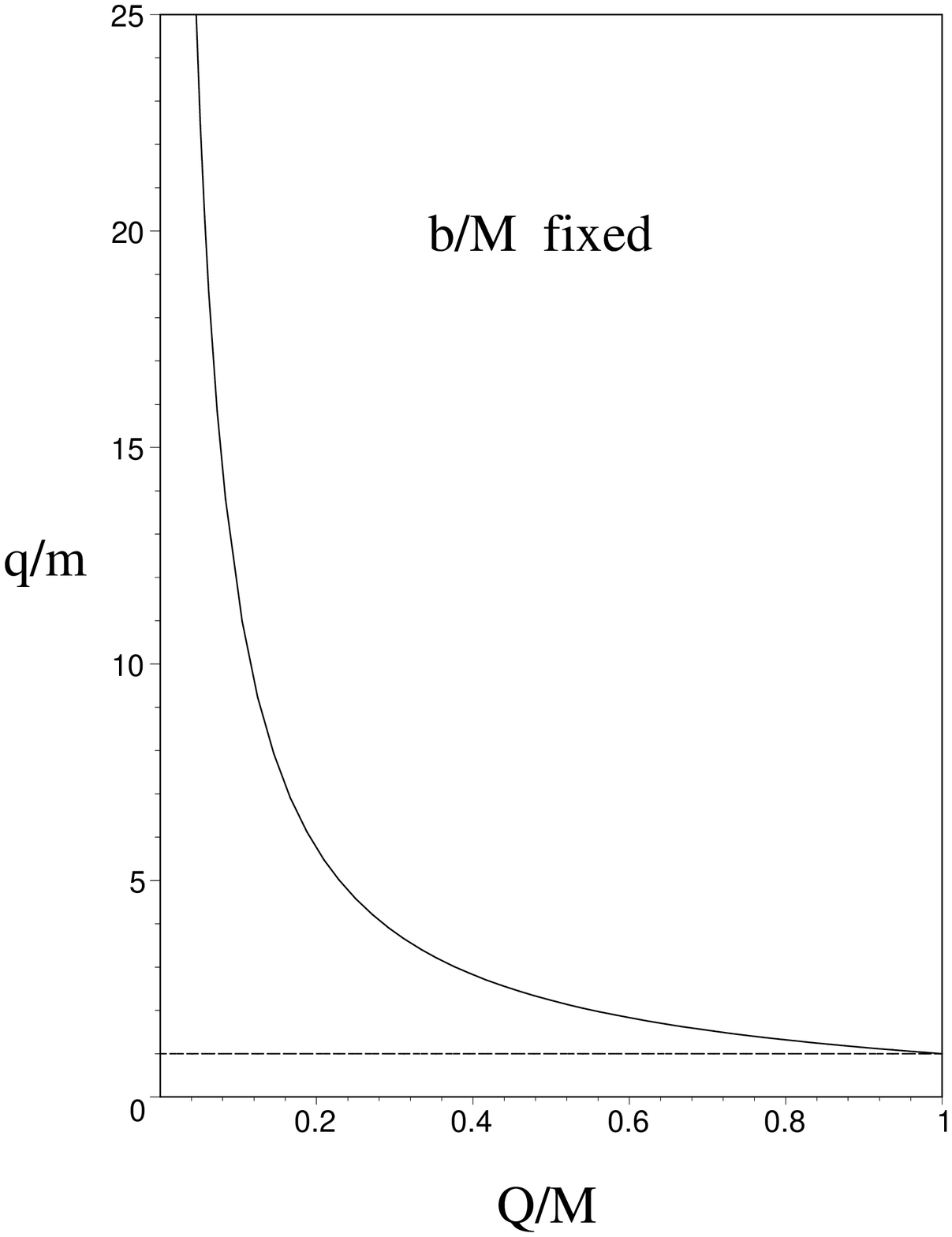}\\[0.4cm]
\mbox{(a)} & \mbox{(b)}\\
\end{array}$
\end{center}
\caption{Fig. (a) shows the behaviour of the charge to mass ratio $q/m$ of the particle as a function of the separation parameter $b/{\mathcal M}$, according to the equilibrium condition (\ref{Fdib}), for a fixed value of $Q/{\mathcal M}=0.6$.
For $b\to\infty$ we have that $q/m\to{\mathcal M}/Q\approx1.67$. 
Instead in Fig. (b) the parameter $q/m$ is plotted as a function of the charge to mass ratio $Q/{\mathcal M}$ of the black hole, at a fixed distance $b=8{\mathcal M}$.
For $Q/{\mathcal M}\to 1$, we have that $q/m\to1$ as well.}
\label{fig:indhor}
\end{figure}

Summarizing (see also Fig. \ref{fig:indhor}):

\begin{itemize}

\item[a)]
$Q/{\mathcal M}$ fixed:

\begin{eqnarray}
&&\mbox{for} \qquad b\to r_+\ , \qquad  q/m\to\infty\ ; \nonumber\\
&&\mbox{for} \qquad b\to \infty\ , \qquad  q/m\to{\mathcal M}/Q\ ;
\end{eqnarray}

\item[b)]
$b/{\mathcal M}$ fixed:

\begin{eqnarray}
&&\mbox{for} \qquad Q/{\mathcal M}\to 0\ , \qquad  q/m\to\infty\ ; \nonumber\\
&&\mbox{for} \qquad Q/{\mathcal M}\to 1\ , \qquad  q/m\to1\ .
\end{eqnarray}

\end{itemize}

% figure indhor

\subsection{Solutions of the perturbation equations: the extreme case ($Q={\mathcal M}$)}

Consider first the cases $l=0, 1$.

\subsubsection{The $l=0$ case}

The relevant equations come from the first order perturbation quantities (\ref{RNlambda00})--(\ref{RNdeltaF2nu}) appearing in the field equations, which do not contain angular derivatives and which reduce to the follwing for $l=0$
\begin{eqnarray}
  \label{RNlambda00leq0extr}  
0&=&(r-{\mathcal M})^2K{}''+\frac{3r^2-5{\mathcal M}r+2{\mathcal M}^2}r K{}'-\frac{(r-{\mathcal M})^2}r H_2{}'+K-H_2+\frac{{\mathcal M}^2}{r^2}H_0 +\frac{r^4}{2(r-{\mathcal M})^2}{A_{{00}}} \nonumber\\
&&- 2{\mathcal M}\tilde f_{01}\ , \\
  \label{RNlambda11leq0extr} 
0&=&\frac{(r-{\mathcal M})^2}rH_0{}'-(r-{\mathcal M})K{}'+H_2-K-\frac{{\mathcal M}^2}{r^2}H_0 + 2{\mathcal M}\tilde f_{01}\ , \\
  \label{RNlambda22leq0extr} 
0&=&(r-{\mathcal M})^2[K{}''-H_0{}'']+ (r-{\mathcal M})[2K{}'-H_2{}']-\frac{r^2+{\mathcal M}r-2{\mathcal M}^2}rH_0{}'-\frac{2{\mathcal M}^2}{r^2}H_2+4{\mathcal M}\tilde f_{01}\ ,\\
  \label{RNdeltaF0nuleq0extr} 
0&=&\tilde f_{{01}}{}'+\frac2{r}\tilde f_{{01}}-\frac{{\mathcal M}}{r^2}K{}'+4\pi v\ .
\end{eqnarray}
Let us look for a solution of the form $H_0= H_2\equiv W$ and $K=W$. Then the preceding equations reduce to the following 
\begin{eqnarray}
  \label{eq1RNleq0sol2extr}  
0&=&(r-{\mathcal M})^2W{}''+\frac{2r^2-3{\mathcal M}r+{\mathcal M}^2}rW{}'+\frac{{\mathcal M}^2}{r^2}W+\frac{r^4}{2(r-{\mathcal M})^2}{A_{{00}}} - 2{\mathcal M}\tilde f_{01}\ , \\
  \label{eq2RNleq0sol2extr} 
0&=&\frac{{\mathcal M}(r-{\mathcal M})}rW{}'+\frac{{\mathcal M}^2}{r^2}W-2{\mathcal M}\tilde f_{01}\ , \\
  \label{eq3RNleq0sol2extr} 
0&=&\tilde f_{{01}}{}'+\frac2{r}\tilde f_{{01}}-\frac{{\mathcal M}}{r^2}W{}'+4\pi v\ .
\end{eqnarray}
By solving Eq.~(\ref{eq2RNleq0sol2extr}) for $\tilde f_{{01}}$ and substituting it into the other equations we obtain
\begin{eqnarray}
  \label{eq1RNleq0sol2extrnew}  
0&=&W{}''+\frac2rW{}'+\frac{r^4}{2(r-{\mathcal M})^4}{A_{{00}}}\ , \\
  \label{eq3RNleq0sol2extrnew} 
0&=&W{}''+\frac2rW{}'+\frac{8\pi r}{r-{\mathcal M}} v\ .
\end{eqnarray}
These are actually the same equation, as follows from Eqs.~(\ref{sorgexp}) and (\ref{compcondextr}), and the solution is simply 
\begin{equation}
\label{RNleq0solWextr}
W=4\sqrt{\pi}\frac{m}{b-{\mathcal M}}\left[\frac{r-{\mathcal M}}r\vartheta(b-r)+\frac{b-{\mathcal M}}r\vartheta(r-b)\right]\ ,
\end{equation}
after a suitable choice of the integration constants; then from Eq.~(\ref{eq2RNleq0sol2extr}) 
we obtain
\begin{equation}
\label{RNleq0solf01extr}
\tilde f_{01}=2\sqrt{\pi}\frac{q}{r^3}\left[2{\mathcal M}\frac{r-{\mathcal M}}{b-{\mathcal M}}\vartheta(b-r)-(r-2{\mathcal M})\vartheta(r-b)\right]\ .
\end{equation}
It is worth noting that the function $\tilde f_{02}$ turns out to be undetermined since it does not appear in the system (\ref{RNlambda00leq0extr})--(\ref{RNdeltaF0nuleq0extr}). However, the tensor harmonic expansion (\ref{RWempert}) of the electromagnetic field shows that the angular factor corresponding to the component $\tilde f_{02}$ vanishes identically for $l=0$.

\subsubsection{The $l=1$ case}

The derivation of the relevant equations from Eqs.~(\ref{RNlambda00})--(\ref{RNdeltaF2nu}) requires particular acare in this case. In fact, the two angular factors in the expression (\ref{RNlambda22}) for ${\tilde G}_{22}$ are not independent when $l=1$, since ${\partial^2 Y_{10}}/{\partial \theta^2}=-Y_{10}$, so the two separate terms in the corresponding field equations collapse to one. 
The relevant equations are thus given by
\begin{eqnarray}
  \label{RNlambda00leq1extr}  
0&=&(r-{\mathcal M})^2K{}''+\frac{3r^2-5{\mathcal M}r+2{\mathcal M}^2}r K{}'-\frac{(r-{\mathcal M})^2}r H_2{}'-2H_2+\frac{{\mathcal M}^2}{r^2}H_0\nonumber\\
&& +\frac{r^4}{2(r-{\mathcal M})^2}{A_{{00}}}- 2{\mathcal M}\tilde f_{01}\ , \\
  \label{RNlambda11leq1extr} 
0&=&\frac{(r-{\mathcal M})^2}rH_0{}'-(r-{\mathcal M})K{}'+H_2-\left[1+\frac{{\mathcal M}^2}{r^2}\right]H_0 + 2{\mathcal M}\tilde f_{01}\ , \\
  \label{RNlambda22leq1extr} 
0&=&(r-{\mathcal M})^2[K{}''-H_0{}'']+ (r-{\mathcal M})[2K{}'-H_2{}']-\frac{r^2+{\mathcal M}r-2{\mathcal M}^2}rH_0{}'+H_0\nonumber\\
&&-\left[1+2\frac{{\mathcal M}^2}{r^2}\right]H_2+4{\mathcal M}\tilde f_{01}\ ,\\
  \label{RNlambda12leq1extr} 
0&=&(r-{\mathcal M})^2[K{}'-H_0{}']-(r-{\mathcal M})H_2+\frac{r^2-3{\mathcal M}r+2{\mathcal M}^2}{r}H_0 + 4{\mathcal M}\tilde f_{02}\ ,\\
 \label{RNdeltaF0nuleq1extr} 
0&=&\tilde f_{{01}}{}'+\frac2{r}\tilde f_{{01}}-\frac2{(r-{\mathcal M})^2}\tilde f_{{02}}-\frac{{\mathcal M}}{r^2}K{}'+4\pi v\ , \\
 \label{RNdeltaF3nuleq1extr} 
0&=&\tilde f_{{01}} - \tilde f_{{02}}{}'\ .
\end{eqnarray}
We look for a solution of the form $H_0= H_2\equiv W$ and $K=W$. Thus the preceding equations reduce to
\begin{eqnarray}
  \label{eq1RNleq1sol2extr}  
0&=&(r-{\mathcal M})^2W{}''+\frac{2r^2-3{\mathcal M}r+{\mathcal M}^2}rW{}'-\left[2-\frac{{\mathcal M}^2}{r^2}\right]W+\frac{r^4}{2(r-{\mathcal M})^2}{A_{{00}}} - 2{\mathcal M}\tilde f_{01}\ , \\
  \label{eq2RNleq1sol2extr} 
0&=&\frac{{\mathcal M}(r-{\mathcal M})}rW{}'+\frac{{\mathcal M}^2}{r^2}W-2{\mathcal M}\tilde f_{01}\ , \\
 \label{eq3RNleq1sol2extr} 
0&=&-\frac{r-{\mathcal M}}rW+2\tilde f_{02}\ , \\
  \label{eq4RNleq1sol2extr} 
0&=&\tilde f_{{01}}{}'+\frac2{r}\tilde f_{{01}}-\frac2{(r-{\mathcal M})^2}\tilde f_{{02}}-\frac{{\mathcal M}}{r^2}W{}'+4\pi v\ , \\
  \label{eq5RNleq1sol2extr} 
0&=&\tilde f_{{01}} - \tilde f_{{02}}{}'\ .
\end{eqnarray}
By solving Eq.~(\ref{eq2RNleq1sol2extr}) for $\tilde f_{{01}}$ and Eq.~(\ref{eq3RNleq1sol2extr}) for $\tilde f_{{02}}$, and substituting the results into the other equations we obtain
\begin{eqnarray}
  \label{eq1RNleq1sol2extrnew}  
0&=&W{}''+\frac2rW{}'-\frac2{(r-{\mathcal M})^2}W+\frac{r^4}{2(r-{\mathcal M})^4}{A_{{00}}} \ ,\\
  \label{eq3RNleq1sol2extrnew} 
0&=&W{}''+\frac2rW{}'-\frac2{(r-{\mathcal M})^2}W+\frac{8\pi r}{r-{\mathcal M}} v\ .
\end{eqnarray}
These are actually the same equation, as follows from Eqs. (\ref{sorgexp}) and (\ref{compcondextr}), and the solution is simply
\begin{equation}
\label{RNleq1solWextr}
W=\frac{4\sqrt{\pi}}{\sqrt{3}}\frac{m}{r}\left[\left(\frac{r-{\mathcal M}}{b-{\mathcal M}}\right)^2\vartheta(b-r)+\frac{b-{\mathcal M}}{r-{\mathcal M}}\vartheta(r-b)\right]\ ,
\end{equation}
after a suitable choice of the integration constants; then from Eqs. (\ref{eq2RNleq1sol2extr}) and (\ref{eq3RNleq1sol2extr}) we obtain
\begin{eqnarray}
  \label{RNleq1solf01extr}
\tilde f_{01}&=&\frac{2\sqrt{\pi}}{\sqrt{3}}\frac{q}{r^3}\left[\left(\frac{r-{\mathcal M}}{b-{\mathcal M}}\right)^2(r+2{\mathcal M})\vartheta(b-r)-2(b-{\mathcal M})\vartheta(r-b)\right]\ ,\\
  \label{RNleq1solf02extr}
\tilde f_{02}&=&\frac{2\sqrt{\pi}}{\sqrt{3}}q\frac{r-{\mathcal M}}{r^2}\left[\left(\frac{r-{\mathcal M}}{b-{\mathcal M}}\right)^2\vartheta(b-r)+\frac{b-{\mathcal M}}{r-{\mathcal M}}\vartheta(r-b)\right]\ . 
\end{eqnarray}

\subsubsection{The $l\geq2$ case}

Finally, consider the general case $l\geq2$. The system of equations to be solved for the unknown functions $X$, $Y$, $\tilde f_{01}$ and $\tilde f_{02}$ coming from Eqs. (\ref{eq2RNxy})--(\ref{eq6RNxy}) is
\begin{eqnarray}
  \label{effe01extr}
\tilde f_{{01}}&=&\frac{2r^2-3{\mathcal M}r+{\mathcal M}^2}{4{\mathcal M}r}X{}'+\frac{r-{\mathcal M}}{4r}Y{}'-\frac{2\lambda r^2+{\mathcal M}^2}{4{\mathcal M}r^2}X+\frac{{\mathcal M}}{4r^2}Y\ , \\
  \label{effe02extr}
\tilde f_{{02}}&=&-\frac{(r-{\mathcal M})^2}{4{\mathcal M}}X{}'+\frac{r-{\mathcal M}}{4r}(Y-X)\ , \\
  \label{eqXextr}
0&=&(r-{\mathcal M})^2X{}''+4(r-{\mathcal M})X{}'-2\lambda X\ , \\
  \label{eqYextr}
0&=&Y{}''+ \frac2{r}Y{}'-\frac{2(\lambda+1)}{(r-{\mathcal M})^2}Y+\frac{16\pi r}{(r-{\mathcal M})}v - \frac{2(r+{\mathcal M})}{r(r-{\mathcal M})}X{}'+\frac{2(2\lambda +1)}{(r-{\mathcal M})^2}X\ ,
\end{eqnarray}
where the relation (\ref{compcondextr}) must be taken into account once the solution is found.

Eq. (\ref{eqXextr}) involves the function $X$ only and can be solved exactly
\begin{eqnarray}
X=c_1 (r-{\mathcal M})^{l-1} + c_2 (r-{\mathcal M})^{-(l+2)}\ ,
\end{eqnarray}
which can be re-expressed as
\begin{eqnarray}
X=(r-{\mathcal M})^{-1}[c_1 \Upsilon_l(r)+c_2\Upsilon_{-l-1}(r)]\ 
\end{eqnarray}
by introducing the following notation 
\begin{eqnarray}
\label{Upsilon}
\Upsilon_l(r)=(r-{\mathcal M})^{l}\ .
\end{eqnarray}
At this point, it is enough to solve Eq. (\ref{eqYextr}) to obtain the complete solution of the system. 
The general solution of the homogeneous equation is simply
\begin{eqnarray}
Y_{\rm hom}=\frac{r-{\mathcal M}}{r}\left[c_3\Upsilon_l(r)+c_4\Upsilon_{-l-1}(r)\right]\ ,
\end{eqnarray}
so that the solution of Eq. (\ref{eqYextr}) becomes
\begin{eqnarray}
\label{solYextrnonomog}
Y=Y_{\rm hom}+{\mathcal B}\frac{r-{\mathcal M}}{r}\left[\Upsilon_l(b)\Upsilon_{-l-1}(r)-\Upsilon_l(r)\Upsilon_{-l-1}(b)\right]\vartheta(r-b)+c_1\Omega_l(r)+c_2\Omega_{-l-1}(r)\ ,
\end{eqnarray}
where the quantities ${\mathcal B}$ and $\Omega_l(r)$ stand for
\begin{eqnarray}
{\mathcal B}&=&\frac{8\sqrt{\pi}}{\sqrt{2l+1}}q\ , \nonumber\\ 
\Omega_l(r)&=&\frac1{2l+1}\left[{\mathcal M}(l^2-l+1)\Upsilon_{l-1}(r)+l\Upsilon_l(r)\right]\ .
\end{eqnarray}
The functions $\Omega_l(r)$ and $\Omega_{-l-1}(r)$ are not regular at the horizon nor at infinity, so that we must set the constants $c_1=0=c_2$. This means that there is only one solution which satisfies the suitable regularity condition: the solution with $X\equiv0$, and thus $K=W=Y/2$, from the relations (\ref{newfunctions}). 

The remaining undetermined constants $c_3$ and $c_4$ appearing in the solution (\ref{solYextrnonomog}) can again be found easily by imposing regularity conditions on the horizon and at infinity, giving the following final form for the solution
\begin{eqnarray}
\label{solYfinaleextr}
Y={\mathcal B}\frac{r-{\mathcal M}}{r}\left[\Upsilon_l(r)\Upsilon_{-l-1}(b)\vartheta(b-r)+\Upsilon_l(b)\Upsilon_{-l-1}(r)\vartheta(r-b)\right]\ .
\end{eqnarray}

The electromagnetic perturbation functions $\tilde f_{01}$ and $\tilde f_{02}$ are then easily determined from the relations (\ref{effe01extr}) and (\ref{effe02extr}) respectively, which for $X=0$ become
\begin{eqnarray}
  \label{effe01extrXzero}
\tilde f_{{01}}&=&\frac{r-{\mathcal M}}{4r}Y{}'+\frac{{\mathcal M}}{4r^2}Y\ , \\
  \label{effe02extrXzero}
\tilde f_{{02}}&=&\frac{r-{\mathcal M}}{4r}Y\ .
\end{eqnarray}

\subsubsection{The analytic solution summed over all values of $l$}

In order to reconstruct the solution for gravitational and electromagnetic perturbation functions for all values of $l$, it is important to note that the both solutions (\ref{RNleq0solWextr}) and (\ref{RNleq1solWextr}) for $W$ correspond to the expression (\ref{solYfinaleextr}) for the function $Y/2$ evaluated for $l=0$ and $l=1$ respectively. The same holds for the electromagnetic perturbation functions $\tilde f_{01}$ and $\tilde f_{02}$: if we set $l=0,1$ in (\ref{effe01extrXzero}) and (\ref{effe02extrXzero}), we obtain exactly (\ref{RNleq0solf01extr}), (\ref{RNleq1solf01extr}) and (\ref{RNleq1solf02extr}). Therefore, we can take the expressions (\ref{solYfinaleextr}), (\ref{effe01extrXzero}) and (\ref{effe02extrXzero}) as the solution for all values of $l$.  

The gravitational field harmonics $h_{\mu \nu }$ reduce to 
\begin{eqnarray}
 \label{hmunuextrY}
 ||h_{\mu \nu }||=\left[ 
\begin {array}{cccc} 
{e^{\nu}}\frac{Y}2Y_{l0}&0&0&0
\\\noalign{\medskip}0&{e^{-\nu}}\frac{Y}2Y_{ l0}&0&0
\\\noalign{\medskip}0&0&{r}^{2}\frac{Y}2Y_{l0}&0\\\noalign{\medskip}0&0
&0&{r}^{2} \sin ^{2}\theta \frac{Y}2Y_{l0}
\end {array} 
\right]\ ,
\end{eqnarray}
since we have $K=W=Y/2$.
Now, if we denote the sum of the radial function $Y$ over the spherical harmonics by $y$, it is easy to show that this quantity can be summed exactly using the properties of the Legendre polynomials $P_n(x)$, whose a generating function is just  
\begin{equation}
\label{genfunct}
g(t,x)\equiv[1-2xt+t^2]^{-1/2}=\sum_{n=0}^{\infty}P_n(x)t^n\ , \qquad |t|<1\ .
\end{equation}
In fact, setting $t=(b-{\mathcal M})/(r-{\mathcal M})$ and $x=\cos\theta$ in the generating function (\ref{genfunct})
leads to  
\begin{equation}
\frac1{{\mathcal D}}=\sum_{n=0}^{\infty}P_n(\cos\theta)\frac{(b-{\mathcal M})^n}{(r-{\mathcal M})^{n+1}}\ ,
\end{equation}
for $r>b$, where 
\begin{equation}
\label{Dpert}
{\mathcal D} = [(r-{\mathcal M})^2+(b-{\mathcal M})^2 - 2(r -{\mathcal M})(b-{\mathcal M})\cos\theta]^{1/2}\ ; 
\end{equation}
alternatively, choosing $t=(r-{\mathcal M})/(b-{\mathcal M})$ implies
\begin{equation}
\frac1{{\mathcal D}}=\sum_{n=0}^{\infty}P_n(\cos\theta)\frac{(r-{\mathcal M})^n}{(b-{\mathcal M})^{n+1}}\ ,
\end{equation}
for $r<b$. 
Therefore, the following representation formula holds 
\begin{eqnarray}
 \label{sommaextr} 
\frac1{{\mathcal D}}=\sum_{n=0}^{\infty}\left[\frac{(r-{\mathcal M})^n}{(b-{\mathcal M})^{n+1}}\vartheta(b-r)+\frac{(b-{\mathcal M})^n}{(r-{\mathcal M})^{n+1}}\vartheta(r-b)\right]P_n(\cos\theta)\ ,
\end{eqnarray} 
leading to the result
\begin{equation}
\label{ysum}
y=\sum_l Y Y_{l0}=4q\frac{r-{\mathcal M}}{r}\sum_{l=0}^{\infty}\left[\frac{(r-{\mathcal M})^l}{(b-{\mathcal M})^{l+1}}\vartheta(b-r)+\frac{(b-{\mathcal M})^l}{(r-{\mathcal M})^{l+1}}\vartheta(r-b)\right]P_l(\cos\theta)
=4q\frac{r-{\mathcal M}}{r}\frac1{\mathcal D}\ .
\end{equation}
At this point, after recalling the definitions (\ref{newfunctions}) for $K$ and $W$ (from which $K=W=Y/2$, for $X=0$) and the relation (\ref{compcondextr}), we have that the solution summed over the harmonics for the perturbed gravitational field turns out to be completely determined by the function
\begin{equation}
\label{hzerilli}
{\mathcal H}=\frac{y}2=2{m}\frac{r-{\mathcal M}}{r}\frac1{\mathcal D}\ ,
\end{equation}  
so that the new line element $d{\tilde s}^2$ from the first of relations (\ref{pertrelations}) and Eq. (\ref{hmunuextrY}) is then
\begin{equation}
\label{lineelemextr}
d{\tilde s}^2=-[1-{\mathcal H}]f(r)dt^2 + [1+{\mathcal H}][f(r)^{-1}dr^2+r^2(d\theta^2 +\sin ^2\theta d\phi^2)]\ .
\end{equation}  
It can be shown that this perturbed metric is spatially conformally flat; moreover, the solution remains valid as long as the condition $|{\mathcal H}|\ll1$ is satisfied. 

Next by using the relations (\ref{effe01extrXzero}) and (\ref{effe02extrXzero}), the knowledge of the gravitational perturbation function ${\mathcal H}$ (or $y$) in closed form permits us to give an exact reconstruction of the perturbed electromagnetic field $f_{\mu\nu}$ as well, namely
\begin{eqnarray}
  \label{effe01extrreconstr}
f_{{01}}&=&\sum_l \tilde f_{{01}}Y_{l0}=\frac{r-{\mathcal M}}{4r}\frac{\partial y}{\partial r}+\frac{{\mathcal M}}{4r^2}y\ , \\
  \label{effe02extrreconstr}
f_{{02}}&=&\sum_l \tilde f_{{02}}\frac{\partial Y_{l0}}{\partial \theta}=\frac{r-{\mathcal M}}{4r}\frac{\partial y}{\partial \theta}\ ,
\end{eqnarray}
so that the electric field components $E_r$ and $E_{\theta}$ are given by
\begin{eqnarray} 
\label{Zeremtensorpert} 
E_r&=&-f_{01}=\frac{q(r-{\mathcal M})}{r^3{\mathcal D}}\left\{-2{\mathcal M}+\frac{r(r-{\mathcal M})[(r-{\mathcal M})-(b-{\mathcal M})\cos\theta]}{{\mathcal D}^2}\right\}\ ,
\nonumber\\ 
E_{\theta}&=&-f_{02}=\frac{q(r-{\mathcal M})^3}{r^2{\mathcal D}^3}(b-{\mathcal M})\sin\theta\ ; 
\end{eqnarray} 
the total electromagnetic field to first order in the perturbations is then (from the second of relations (\ref{pertrelations})) just
\begin{equation}
\label{RNemfieldpertextr}
\tilde F=-\left[\frac{{\mathcal M}}{r^2}+E_r\right]dt\wedge dr - E_{\theta}dt\wedge d\theta\ .
\end{equation}
Let us verify that Gauss's theorem is satisfied by integrating the dual of the electromagnetic form (\ref{RNemfieldpertextr}) over a surface $S$ containing both the charges $Q={\mathcal M}$ and $q$
\begin{equation}
\label{gausspert}
\Phi=\int_{S}{}^*{\tilde F}\wedge dS=4\pi(Q+q)\ .
\end{equation}
We only need to calculate the component ${}^*{\tilde F}_{\theta\phi}$, which results to be (to first order in the perturbation)
\begin{equation}
{}^*{\tilde F}_{\theta\phi}=r^2\sin\theta\left[(1+{\mathcal H})\frac{{\mathcal M}}{r^2}+E_r\right]\ .
\end{equation}
Let us calculate separately the two different contributions to the integral (\ref{gausspert}) due to the background and particle electric fields. The former one is given by
\begin{eqnarray}
\label{fluxRN}
\Phi_{\rm RN}=2\pi\int_{0}^{\pi}r^2\sin\theta	(1+{\mathcal H})\frac{{\mathcal M}}{r^2} d\theta
=4\pi {\mathcal M} + 4\pi {\mathcal M}q \frac{r-{\mathcal M}}{r}\int_{0}^{\pi}\frac{\sin\theta}{{\mathcal D}}d\theta\ ,
\end{eqnarray}
recalling the definition (\ref{hzerilli}) for the gravitational perturbation function ${\mathcal H}$. Next it is easy to show from (\ref{Dpert}) that 
\begin{eqnarray}
\label{Dintegral}
\int_{0}^{\pi}\frac{\sin\theta}{{\mathcal D}}d\theta&=&\frac1{(r-{\mathcal M})(b-{\mathcal M})}\int_{0}^{\pi}\frac{\partial {\mathcal D}}{\partial \theta}d\theta
=\frac{{\mathcal D}(\pi)-{\mathcal D}(0)}{(r-{\mathcal M})(b-{\mathcal M})}  \nonumber\\
&=&\frac2{(r-{\mathcal M})(b-{\mathcal M})}[(r-{\mathcal M})\vartheta(b-r)+(b-{\mathcal M})\vartheta(r-b)]\ .
\end{eqnarray}
Thus the flux (\ref{fluxRN}) becomes
\begin{eqnarray}
\label{fluxRNnew}
\Phi_{\rm RN}=4\pi {\mathcal M} + \frac{8\pi {\mathcal M}q}{r(b-{\mathcal M})}[(r-{\mathcal M})\vartheta(b-r)+(b-{\mathcal M})\vartheta(r-b)]\ .
\end{eqnarray}
The charged particle contribution to the flux is given by
\begin{eqnarray}
\label{fluxpart}
\Phi_{\rm part}&=&2\pi\int_{0}^{\pi}r^2\sin\theta	E_r d\theta
=-4\pi r^2 \frac{\partial}{\partial r}\left[\frac{r-{\mathcal M}}{4r}\int_{0}^{\pi}{\mathcal H}\sin\theta d\theta\right]\nonumber\\
&=&-4\pi q r^2 \frac{\partial}{\partial r}\left[\frac{(r-{\mathcal M})^2}{r^2}\int_{0}^{\pi}\frac{\sin\theta}{{\mathcal D}}d\theta\right]\ ,
\end{eqnarray}
so that using the expression  (\ref{Dintegral}) it reduces to
\begin{eqnarray}
\label{fluxpartnew}
\Phi_{\rm part}=\left[4\pi q - \frac{8\pi {\mathcal M}q}{r}\right]\vartheta(r-b)-\frac{8\pi {\mathcal M}q}{r} \frac{r-{\mathcal M}}{b-{\mathcal M}}\vartheta(b-r)\ .
\end{eqnarray}
Therefore the total flux $\Phi$ is
\begin{equation}
\Phi=\Phi_{\rm RN}+\Phi_{\rm part}=4\pi {\mathcal M}+4\pi q\vartheta(r-b)\ ,
\end{equation} 
so Gauss's theorem (\ref{gausspert}) is satisfied.

As a matter of fact there exists an alternative way to treat the problem, at least for the case of particle and black hole charges of the same sign, which permits one to deal {\it ab initio} with closed form expressions both for the perturbed gravitational and electromagnetic field, as we will see in the next section. That leads to enormous simplification in the treatment of the perturbations of extreme black holes, and it is a powerful tool for the analysis of higher order perturbations, since one can stop in the perturbation theory at any order one desires.

\subsection{Perturbations of the extreme black hole: an alternative approach}

Although the spacetime of a black hole and a charged particle is not an electro-vacuum one, we start from an exact solution of the Einstein-Maxwell field equations without sources, and by linearizing it, we will obtain an exact solution of the coupled perturbation equations.

The exact solution from which we start is the Majumdar-Papapetrou solution describing a system of many extreme Reissner-Nordstr\"om black holes. We note that by using the Israel-Wilson-Perj\'es approach one could also include rotation in the present formulation.
In particular, we will be interested in the two body problem, which is discussed in detail in the monograph of Chandrasekhar \cite{chandrasekhar}.

The physical scenario we have in mind is the one in which one black hole has a mass (charge) absolutely smaller than the one of the companion object. In this limit and assuming a separation distance of the two objects not comparable with their gravitational radii, we can interpret the system as an extreme charged black hole interacting nonlinearly with a very small charged object. The charge-to-mass ratio $q/m=1$ of such a particle is not a problem for the effect we are going to discuss. In fact in the equations of Zerilli the charge and mass of the particle appear in the source terms of the coupled ODE system only as a common multiplicative factor. The particular integral built with the solutions of the homogeneous system will be proportional to such quantities, and this makes the mathematics of the problem (but not the physics, of course) naively insensitive to a particular choice of such parameters, provided always that the perturbative regime is valid.
Clearly this assumption is very delicate, because the black hole with a particle is not a vacuum spacetime and we mimic it with an electro-vacuum one. In some sense the idea is to use an analogy such that the particle and a very small extreme black hole can be thought as the same kind of object and have the same fields at far distances from themselves. The Majumdar-Papapetrou solution is included in the class of electro-vacuum and conformally static solutions of the Einstein-Maxwell system.

The three-dimensional conformally flat geometry in this case is allowed only for extreme black holes, and this fact demonstrates the special nature of Reissner-Nordstr\"om black holes with $Q=\mathcal M$. 
In this situation the complicate system of non linear PDEs collapses into a simple complex Laplace equation, which acts as master equation for building up infinite solutions; however, apart from the Majumdar-Papapetrou spacetime, the other solutions give rise to naked singularities. It is important to note that, because of the linearity of the Laplace equation, a superposition of many relativistic bodies is possible.

We are now ready to discuss quantitatively the Majumdar-Papapetrou spacetime. 
In fact, by introducing the quantity 
\begin{equation} 
U=1+\frac{{\mathcal M}}{R}+\frac{m}{\sqrt{R^2+b^2-2Rb\cos\theta}}\ , 
\end{equation} 
it is easy to show that a solution of the electro-vacuum Einstein-Maxwell system is given by 
\begin{equation} 
\label{MPmetric}
ds^2=-\frac{dt^2}{U^2}+U^2[dR^2+R^2(d\theta^2+\sin^2\theta d\phi^2)]\ ,  
\end{equation} 
where $U$ satisfies the 3-dimensional Laplace equation $\nabla^2 U=0$, and the electrostatic potential is given by $A_t^{\rm MP}=U^{-1}$.
The quantity $b$ is the separation distance between the two bodies, and the quantities ${\mathcal M}$, $m$ can be identified with the masses, if the sources are far enough apart. 
The charges of the system are equal to these masses and have the same sign. This is a limit of the analogy we are using in this section, because we can study an extreme black hole and a charged particle provided that both have the same sign of the charge. Bodies with different signed charge would require the study of the full Zerilli system, as we have done in the previous section.

The manifold under consideration has a coordinate interpretation which is absolutely nontrivial, in particular in relation with the location of the horizons and the singularities. However, because we will be interested in the study of the perturbations of such an object, we can use a coordinate transformation $R=r-{\mathcal M}$, which in the case $m=0$ gives immediately the usual Reissner-Nordstr\"om solution, and consequently we can construct the gravitational perturbation with respect to $m$ referred to such coordinates by expanding the metric to first order in $m$:
\begin{eqnarray} 
\label{MPgrpert} 
h_{tt}^{\rm MP}&=&\frac{2m}{\Sigma}\frac{(r-{\mathcal M})^3}{r^3}\ , \nonumber\\ 
h_{rr}^{\rm MP}&=&\frac{2m}{\Sigma}\frac{r}{r-{\mathcal M}}\ , \nonumber\\ 
h_{\theta\theta}^{\rm MP}&=&\frac{2m}{\Sigma}r(r-{\mathcal M})\ , \nonumber\\ 
h_{\phi\phi}^{\rm MP}&=&\frac{2m}{\Sigma}r(r-{\mathcal M})\sin^2\theta\ , 
\end{eqnarray} 
where
\begin{equation} 
\Sigma=\left[(r-{\mathcal M})^2 + (b-{\mathcal M})^2 -2(r-{\mathcal M})(b-{\mathcal M})\cos\theta\right]^{1/2}\ ,
\end{equation} 
and the following representation holds (see Eq. (\ref{sommaextr}))
\begin{equation}
\frac{1}{\Sigma}= \sum_l C_l (r) Y_{l0}\ , \qquad C_l(r)=\frac{2\sqrt{\pi}}{\sqrt{2l+1}}
\left[\frac{(r-{\mathcal M})^{l}}{(b-{\mathcal M})^{l+1}}\vartheta(b-r)+ 
\frac{(b-{\mathcal M})^l}{(r-{\mathcal M})^{l+1}}\vartheta(r-b)\right]\ ;
\end{equation}
moreover, recalling the notation (\ref{Upsilon}) introduced in the previous section, one finds
\begin{eqnarray}
C_l(r)=\frac{2\sqrt{\pi}}{\sqrt{2l+1}}
\left[\Upsilon_l(r)\Upsilon_{-l-1}(b)\vartheta(b-r)+
\Upsilon_l(b)\Upsilon_{-l-1}(r)\vartheta(r-b)\right]\ .
\end{eqnarray}
We can now suitably scale the gravitational perturbation functions (\ref{MPgrpert}) as in the Zerilli approach
\begin{eqnarray*}
h_{tt}^{\rm MP}=e^{\nu}{\bar h}_{tt}\ , \qquad h_{rr}^{\rm MP}=e^{-\nu}{\bar h}_{rr}\ , \qquad h_{\theta\theta}^{\rm MP}=r^2{\bar h}_{\theta\theta}\ , \qquad h_{\phi\phi}^{\rm MP}=r^2\sin^2\theta {\bar h}_{\phi\phi}\ ,
\end{eqnarray*}
finding that in this case
\begin{equation}
{\bar h}_{tt}={\bar h}_{rr}={\bar h}_{\theta\theta}={\bar h}_{\phi\phi}\equiv {\bar h}\ ,
\end{equation}
with
\begin{eqnarray}
\label{hbarlegendre}
{\bar h}&=&\frac{2m}{\Sigma}\frac{r-{\mathcal M}}r =\frac{2m(r-{\mathcal M})}r \sum_l C_l (r) Y_{l0}\ .
\end{eqnarray}
This quantity is equal to the gravitational perturbation function (\ref{hzerilli}) obtained in the preceding section following the approach of Zerilli, since $q=m$ in the present treatment, and $\Sigma\equiv {\mathcal D}$.

The nonvanishing components of the effective electromagnetic tensor (that is, the total electromagnetic tensor, from which the contribution $F_{(1)}=-({\mathcal M}/r^2)\,dt\wedge dr$ of the \lq\lq background'' field of the black hole of mass ${\mathcal M}$ is subtracted) are easily evaluated from the metric (\ref{MPmetric}) and the vector potential $A_\mu^{\rm MP}=U^{-1}dt$
\begin{eqnarray}
F_{tr}^{\rm MP}
&=&-\frac{m(r-{\mathcal M})}{2r^2\Sigma}\frac{\Sigma[r(r-4{\mathcal M})\Sigma-2m{\mathcal M}(r-{\mathcal M})]+r^2[(r-{\mathcal M})^2-(b-{\mathcal M})^2]}{[r\Sigma+m(r-{\mathcal M})]^2}\ , \nonumber\\
F_{t\theta}^{\rm MP}
&=&-\frac{m(r-{\mathcal M})}{\Sigma}\frac{(b-{\mathcal M})}{[r\Sigma+m(r-{\mathcal M})]^2}\sin\theta\ .
\end{eqnarray}
By expanding the previous expressions to first order in $m$ we also obtain the electromagnetic perturbation 
\begin{eqnarray} 
\label{MPemtensorpert} 
f_{tr}^{\rm MP}&=&-\frac{m(r-{\mathcal M})}{r^3\Sigma}\left\{-2{\mathcal M}+\frac{r(r-{\mathcal M})[r-{\mathcal M}-(b-{\mathcal M})\cos\theta]}{\Sigma^2}\right\}\ ,
\nonumber\\ 
f_{t\theta}^{\rm MP}&=&-\frac{m(r-{\mathcal M})^3}{r^2\Sigma^3}(b-{\mathcal M})\sin\theta\ . 
\end{eqnarray} 
As in the case of the metric perturbations, the previous quantities are the same as those (see Eq. (\ref{Zeremtensorpert})) obtained in the preceding section, with the identifications
\begin{eqnarray*}
E_r\longleftrightarrow -f_{tr}^{\rm MP}\ , \qquad E_{\theta}\longleftrightarrow -f_{t\theta}^{\rm MP}\ , \qquad q\longleftrightarrow m\ , \qquad {\mathcal D}\longleftrightarrow \Sigma\ .
\end{eqnarray*}

Although the metric perturbation $h_{\mu\nu}^{\rm MP}$ continues to be singular for $r\rightarrow {\mathcal M}$ as in the background case, 
inspection of the curvature invariant $\mathcal K=R_{\alpha\beta}R^{\alpha\beta}$ shows that its perturbative expansion with respect to $m$ there gives ${\mathcal K}^{(0)}=4{\mathcal M}^{-4}$ and ${\mathcal K}^{(1)}=0$. Moreover, the electromagnetic perturbation $f_{\mu\nu}^{\rm MP}$ in the same limit is zero. This fact, in agreement with the \lq\lq No-Hair Theorem'', implies that the extreme black hole does not feel the external charged particle and preserves the shape of its horizon. This is clear from the stability theorems for the electro-vacuum perturbations of the Reissner-Nordstr\"om background, which require regularity on the horizon for all the modes and consequently for the fields built from them.

We have thus demonstrated the complete equivalence between this new approach with the standard one of Zerilli; in addition, the result is more general, since the Majumdar-Papapetrou solution is an exact solution of the Einstein-Maxwell equations. 

It is possible to reach the same conclusion directly, by verifying that the quantities (\ref{MPgrpert}) and (\ref{MPemtensorpert}) satisfy the Zerilli equations for the Reissner-Nordstr\"om perturbations, as shown in the following. The perturbed effective electrostatic potential $a_t^{\rm MP}$ is easily determined, by expanding $A_t^{\rm MP}=U^{-1}$ to first order in $m$, and by subtracting the contribution $A_{t\,(1)}={\mathcal M}/r$ due to the black hole of mass ${\mathcal M}$:
\begin{eqnarray}
\label{MPelpot}
a_t^{\rm MP}&=&1-\frac{m(r-{\mathcal M})^2}{r^2\Sigma}= 1-\frac{m(r-{\mathcal M})^2}{r^2}\sum_l C_l (r) Y_{l0}\ ,
\end{eqnarray}
where the latter expression is obtained by expanding the former one in spherical harmonics. Consequently the electromagnetic perturbation function $f_{tr}^{\rm MP}$ is given by 
\begin{eqnarray}
\label{ftrlegendre}
f_{tr}^{\rm MP}=\frac{\partial }{\partial r}a_t^{\rm MP}\ . 
\end{eqnarray}
At this point, denoting by $\tilde f_{tr}$ and $\tilde h$ the $\theta$-independent parts of the expansions (\ref{ftrlegendre}) and (\ref{hbarlegendre}) respectively (so that the meaning of the notation $\tilde f_{t\theta}$ is also clear), it is easy to show that these quantities satisfy the Zerilli equations (\ref{eq2RN}), (\ref{eq4RN}), (\ref{eq5RN}) and (\ref{eq6RN}) for the extreme case together with the stability condition (\ref{compcondextr}), with the identifications 
\begin{eqnarray*}
q=m\ , \qquad K=W={\tilde h}\ , \qquad \tilde f_{{01}}=\tilde f_{{tr}}\ , \qquad \tilde f_{{02}}=\tilde f_{{t\theta}}\ .
\end{eqnarray*}

\subsection{Solutions of the perturbation equations: the general non-extreme case ($Q<{\mathcal M}$)}

Let us consider first the cases $l=0, 1$.

\subsubsection{The $l=0$ case}

The relevant equations come from quantities (\ref{RNlambda00})--(\ref{RNdeltaF2nu}) which do not contain angular derivatives, and which in this case become
\begin{eqnarray}
  \label{RNlambda00leq0}  
0&=&(r^2-2{\mathcal M}r+Q^2)K{}''+\frac{3r^2-5{\mathcal M}r+2Q^2}r K{}'-\frac{r^2-2{\mathcal M}r+Q^2}r H_2{}'+K-H_2+\frac{Q^2}{r^2}H_0\nonumber\\
&&+\frac{r^4}{2(r^2-2{\mathcal M}r+Q^2)}A_{{00}} - 2Q\tilde f_{01}\ , \\
  \label{RNlambda11leq0} 
0&=&\frac{r^2-2{\mathcal M}r+Q^2}rH_0{}'-(r-{\mathcal M})K{}'+H_2-K-\frac{Q^2}{r^2}H_0 + 2Q\tilde f_{01}\ , \\
  \label{RNlambda22leq0} 
0&=&(r^2-2{\mathcal M}r+Q^2)[K{}''-H_0{}'']+ (r-{\mathcal M})[2K{}'-H_2{}']-\frac{r^2+{\mathcal M}r-2Q^2}rH_0{}'\nonumber\\
&&-\frac{2Q^2}{r^2}H_2+4Q\tilde f_{01}\ ,\\
  \label{RNdeltaF0nuleq0} 
0&=&\tilde f_{{01}}{}'+\frac2{r}\tilde f_{{01}}-\frac{Q}{r^2}K{}'+4\pi v\ .
\end{eqnarray}
Let us look for a solution of the form $H_0= H_2\equiv W$ and $K=W$. Then the preceding equations reduce to the following ones
\begin{eqnarray}
  \label{eq1RNleq0sol2}  
0&=&(r^2-2{\mathcal M}r+Q^2)W{}''+\frac{2r^2-3{\mathcal M}r+Q^2}rW{}'+\frac{Q^2}{r^2}W+\frac{r^4}{2(r^2-2{\mathcal M}r+Q^2)}A_{{00}}\nonumber\\
&& - 2Q\tilde f_{01}\ , \\
  \label{eq2RNleq0sol2} 
0&=&\frac{{\mathcal M}r-Q^2}rW{}'+\frac{Q^2}{r^2}W-2Q\tilde f_{01}\ , \\
  \label{eq3RNleq0sol2} 
0&=&\tilde f_{{01}}{}'+\frac2{r}\tilde f_{{01}}-\frac{Q}{r^2}W{}'+4\pi v\ .
\end{eqnarray}
By solving Eq.~(\ref{eq2RNleq0sol2}) for $\tilde f_{{01}}$, and substituting it into the other equations we obtain
\begin{eqnarray}
  \label{eq1RNleq0sol2new}  
0&=&W{}''+\frac2rW{}'+\frac{r^4}{2(r^2-2{\mathcal M}r+Q^2)^2} A_{00}\ , \\
  \label{eq3RNleq0sol2new} 
0&=&W{}''+\frac2rW{}'+\frac{8\pi Qr}{{\mathcal M}r-Q^2} v\ .
\end{eqnarray}
These are actually the same equation, as follows from relations (\ref{sorgexp}) and (\ref{compcondnextr}), and the solution is simply
\begin{equation}
\label{RNleq0solW}
W=4\sqrt{\pi}\frac{m}{b}f(b)^{-1/2}\frac1r\left[(r-{\mathcal M})\vartheta(b-r)+(b-{\mathcal M})\vartheta(r-b)\right]\ ,
\end{equation}
after a suitable choice of the integration constants; moreover, from Eq.~(\ref{eq2RNleq0sol2}) 
we obtain
\begin{eqnarray}
\label{RNleq0solf01}
\tilde f_{01}=2\sqrt{\pi}\frac{q}{{\mathcal M}b-Q^2}\frac1{r^3}\big\{\left[{\mathcal M}({\mathcal M}r-2Q^2)+rQ^2\right]\vartheta(b-r)-({\mathcal M}r-2Q^2)(b-{\mathcal M})\vartheta(r-b)\big\}\ .
\end{eqnarray}

\subsubsection{The $l=1$ case}

The relevant equations coming from quantities (\ref{RNlambda00})--(\ref{RNdeltaF2nu}) are given by
\begin{eqnarray}
  \label{RNlambda00leq1}  
0&=&(r^2-2{\mathcal M}r+Q^2)K{}''+\frac{3r^2-5{\mathcal M}r+2Q^2}rK{}'-\frac{r^2-2{\mathcal M}r+Q^2}rH_2{}'-2H_2+\frac{Q^2}{r^2}H_0\nonumber\\
&&+\frac{r^4}{2(r^2-2{\mathcal M}r+Q^2)} A_{{00}} - 2Q\tilde f_{01}\ , \\
  \label{RNlambda11leq1} 
0&=&\frac{r^2-2{\mathcal M}r+Q^2}rH_0{}'-(r-{\mathcal M})K{}'+H_2-\left[1+\frac{Q^2}{r^2}\right]H_0 + 2Q\tilde f_{01}\ ,\\
  \label{RNlambda22leq1} 
0&=&(r^2-2{\mathcal M}r+Q^2)[K{}''-H_0{}'']+ (r-{\mathcal M})[2K{}'-H_2{}']-\frac{r^2+{\mathcal M}r-2Q^2}rH_0{}'\nonumber\\
&&+H_0-\left[1+\frac{2Q^2}{r^2}\right]H_2+4Q\tilde f_{01}\ ,\\
 \label{RNlambda12leq1} 
0&=&(r^2-2{\mathcal M}r+Q^2)[K{}'-H_0{}']-(r-{\mathcal M})H_2+\frac{r^2-3{\mathcal M}r+2Q^2}{r}H_0 + 4Q\tilde f_{02}\ ,\\
 \label{RNdeltaF0nuleq1} 
0&=&\tilde f_{{01}}{}'+\frac2{r}\tilde f_{{01}}-\frac2{r^2-2{\mathcal M}r+Q^2}\tilde f_{{02}}-\frac{Q}{r^2}K{}'+4\pi v\ , \\
 \label{RNdeltaF3nuleq1} 
0&=&\tilde f_{{01}} - \tilde f_{{02}}{}'\ .
\end{eqnarray}
Let us look for a solution of the form $H_0= H_2\equiv W$ and $K=W$. Thus the preceding equations reduce to the following ones
\begin{eqnarray}
  \label{eq1RNleq1sol2}  
0&=&(r^2-2{\mathcal M}r+Q^2)W{}''+\frac{2r^2-3{\mathcal M}r+Q^2}rW{}'-\left[2-\frac{Q^2}{r^2}\right]W+\frac{r^4}{2(r^2-2{\mathcal M}r+Q^2)}A_{{00}}\nonumber\\
&& - 2Q\tilde f_{01}\ , \\
  \label{eq2RNleq1sol2} 
0&=&\frac{{\mathcal M}r-Q^2}rW{}'+\frac{Q^2}{r^2}W-2Q\tilde f_{01}\ , \\
 \label{eq3RNleq1sol2} 
0&=&-\frac{{\mathcal M}r-Q^2}rW+2Q\tilde f_{02}\ , \\
  \label{eq4RNleq1sol2} 
0&=&\tilde f_{{01}}{}'+\frac2{r}\tilde f_{{01}}-\frac2{r^2-2{\mathcal M}r+Q^2}\tilde f_{{02}}-\frac{Q}{r^2}W{}'+4\pi v\ , \\
  \label{eq5RNleq1sol2} 
0&=&\tilde f_{{01}} - \tilde f_{{02}}{}'\ .
\end{eqnarray}
By solving Eq.~(\ref{eq2RNleq1sol2}) for $\tilde f_{{01}}$ and Eq.~(\ref{eq3RNleq1sol2}) for $\tilde f_{{02}}$, and then substituting the results into the other equations we obtain
\begin{eqnarray}
  \label{eq1RNleq1sol2new}  
0&=&W{}''+\frac2rW{}'-\frac2{r^2-2{\mathcal M}r+Q^2}W+\frac{r^4}{2(r^2-2{\mathcal M}r+Q^2)^2}A_{{00}}\ , \\
  \label{eq3RNleq1sol2new} 
0&=&W{}''+\frac2rW{}'-\frac2{r^2-2{\mathcal M}r+Q^2}W+\frac{8\pi rQ}{{\mathcal M}r-Q^2} v\ .
\end{eqnarray}
These are actually the same equation, as follows from relations (\ref{sorgexp}) and (\ref{compcondnextr}), and the solution is simply
\begin{eqnarray}
\label{RNleq1solW}
W&=&\sqrt{3\pi}\frac{m}{\Gamma^3}f(b)^{-1/2}\frac{b-r_-}{b}\frac{r-r_-}{r}\nonumber\\
&&\times\bigg\{\left[2\Gamma(b-{\mathcal M})-(b-r_+)(b-r_-)\ln{\left(\frac{b-r_-}{b-r_+}\right)}\right]\frac{r-r_+}{b-r_-}\vartheta(b-r)\nonumber\\
&&+\left[2\Gamma(r-{\mathcal M})-(r-r_+)(r-r_-)\ln{\left(\frac{r-r_-}{r-r_+}\right)}\right]\frac{b-r_+}{r-r_-}\vartheta(r-b)\bigg\}\ ,
\end{eqnarray}
after a suitable choice of the integration constants; then from Eqs. (\ref{eq2RNleq1sol2}) and (\ref{eq3RNleq1sol2}) we obtain
\begin{eqnarray}
  \label{RNleq1solf01}
\tilde f_{01}&=&\frac{\sqrt{3\pi}}2\frac{q}{\Gamma^3}\frac{b-r_-}{{\mathcal M}b-Q^2}\bigg\{\frac{{\mathcal M}r(r^2-3Q^2)-2r_+^2r_-^2}{r^3(b-r_-)}\bigg[2\Gamma(b-{\mathcal M})\nonumber\\
&& +(b-r_+)(b-r_-)\ln{\left(\frac{b-r_-}{b-r_+}\right)}\bigg]\vartheta(b-r)+\bigg[2\Gamma{\mathcal M}(r^2-2{\mathcal M}r-2Q^2)\nonumber\\
&& +[{\mathcal M}r(r^2-3Q^2)-2r_+^2r_-^2]\ln{\left(\frac{r-r_-}{r-r_+}\right)}\bigg](b-r_+)\vartheta(r-b)\bigg\}\ , \nonumber\\
  \label{RNleq1solf02}
\tilde f_{02}&=&\frac{\sqrt{3\pi}}2\frac{q}{\Gamma^3}\frac{{\mathcal M}r-Q^2}{{\mathcal M}b-Q^2}\frac{(r-r_-)(b-r_-)}{r^2}\bigg\{\bigg[2\Gamma(b-{\mathcal M})\nonumber\\
&&-(b-r_+)(b-r_-)\ln{\left(\frac{b-r_-}{b-r_+}\right)}\bigg]\frac{r-r_+}{b-r_-}\vartheta(b-r)\nonumber\\
&&+\left[2\Gamma(r-{\mathcal M})-(r-r_+)(r-r_-)\ln{\left(\frac{r-r_-}{r-r_+}\right)}\right]\frac{b-r_+}{r-r_-}\vartheta(r-b)\bigg\}\ . 
\end{eqnarray}

\subsubsection{The $l\geq2$ case}

Let us look for the solution of the system consisting of Eqs. (\ref{effe01}), (\ref{effe02}), (\ref{eqX}) and (\ref{eqY}) for the general non-extreme case, together with the relation (\ref{compcondnextr}).

Eq. (\ref{eqX}) involves only the function $X$, and can be solved exactly.
Putting $X=f(r)^{-1/2}w(r)/r$ and making the transformation $z=(r-{\mathcal M})/\Gamma$ Eq. (\ref{eqX}) becomes
\begin{equation}
0=(1-z^2)w{}''-2zw{}'+\left[l(l+1)-\frac1{1-z^2}\right]w\ ,
\end{equation} 
where primes now denote differentiation with respect to the new variable $z$.
The general solutions are the associated Legendre functions of the first and second kind $P_l^1(z)$ and $Q_l^1(z)$, implying that 
\begin{equation}
X=\frac{f(r)^{-1}}{r}[c_1f_l(r)+c_2g_l(r)]
\equiv c_1X_1+c_2X_2\ ,
\end{equation}
where the functions $f_l(r)$ and $g_l(r)$ have been defined in (\ref{multipoliRN}).
The arbitrary constants $c_1$ and $c_2$ must be set both equal to zero in order that the function $X$ to satisfy regularity conditions on the horizon and at infinity, in agreement with the analysis of the solution we have done in the extreme case.
At this point, it is enough to solve Eq.~(\ref{eqY}) with the condition $X=0$ and thus $K=W=Y/2$ to obtain the complete solution of the system.
 
Let us consider Eq.~(\ref{eqY}) in which we set $X=0$, namely 
\begin{eqnarray}
\label{eqYXeq0}
0=(r^2-2{\mathcal M}r+Q^2)Y{}''+ \frac2{r}(r^2-2{\mathcal M}r+Q^2)Y{}'-l(l+1)Y+\frac{16\pi b^3f(b)Q}{{\mathcal M}b-Q^2}v \ .
\end{eqnarray} 
Putting $Y=f(r)^{1/2}w(r)$ and making the transformation $z=(r-{\mathcal M})/\Gamma$ the previous equation becomes
\begin{equation}
 \label{eqYXeq0leg} 
0=(1-z^2)w{}''-2zw{}'+\left[l(l+1)-\frac1{1-z^2}\right]w-8\sqrt{\pi}\frac{q}{\Gamma}\sqrt{2l+1}Q\frac{\sqrt{\beta^2-1}}{{\mathcal M}\beta+\Gamma}\delta(z-\beta)\ ,
\end{equation} 
where primes now denote differentiation with respect to the new variable $z$ and $\beta=(b-{\mathcal M})/\Gamma$.
This equation is formally identical to Eq. (\ref{eqleg1l}) apart from a different coefficient in front of the delta function. The general solutions of the corresponding homogeneous equation are thus the associated Legendre functions of the first and second kind $P_l^1(z)$ and $Q_l^1(z)$. 
Taking the functions (\ref{multipoliRN}) as the two linearly independent solutions of the homogeneous equation, the solution of Eq. (\ref{eqYXeq0leg}) turns out to be given by
\begin{eqnarray}
\label{solYfinale}
Y&=&-8\sqrt{\pi}\frac{\sqrt{2l+1}}{l(l+1)}q\frac{Q}{{\mathcal M}b-Q^2}\frac{1}{\Gamma r}(b-r_+)(b-r_-)(r-r_+)(r-r_-)\left[\frac{dQ_l(z(r))}{dr}\bigg\vert_{r=b}\frac{dP_l(z(r))}{dr}\vartheta(b-r)\right.\nonumber\\
&&\left.+\frac{dP_l(z(r))}{dr}\bigg\vert_{r=b}\frac{dQ_l(z(r))}{dr}\vartheta(r-b)\right]\ .
\end{eqnarray}
Note that for $l=1$ this solution reduces to Eq. (\ref{RNleq1solW}), since $W=Y/2$.

The electromagnetic perturbation functions $\tilde f_{01}$ and $\tilde f_{02}$ are then easily evaluated from the relations (\ref{effe01}) and (\ref{effe02}) respectively, which for $X=0$ become
\begin{eqnarray}
  \label{effe01Xzero}
\tilde f_{{01}}&=&\frac{{\mathcal M}r-Q^2}{4rQ}Y{}'+\frac{Q}{4r^2}Y\ , \\
  \label{effe02Xzero}
\tilde f_{{02}}&=&\frac{{\mathcal M}r-Q^2}{4rQ}Y\ .
\end{eqnarray}

\subsubsection{The analytic solution summed over all values of $l$}

As in the extreme case we are able to completely reconstruct the solution for all gravitational and electromagnetic perturbation functions in closed form also in this case. 

\begin{figure} 
\typeout{*** EPS figure 2}
\begin{center}
\includegraphics[scale=0.5]{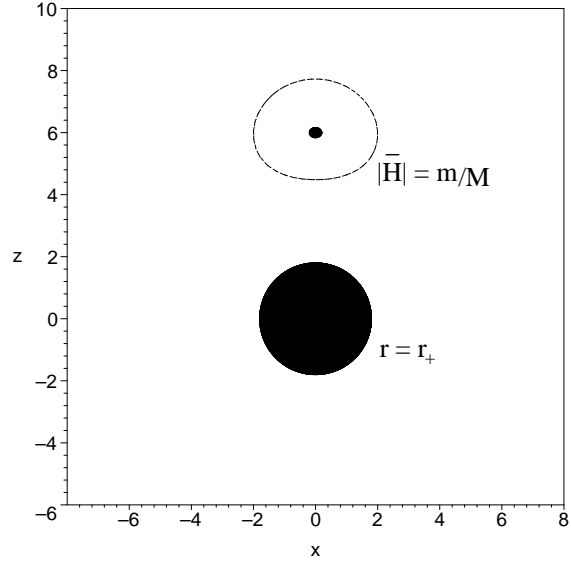}
\end{center}
\caption{The surface $|{\bar {\mathcal H}}|= m/{\mathcal M}$ is plotted as a curve in the polar coordinate plane with the choice of parameters $Q/{\mathcal M}=0.6$ and $b/{\mathcal M}=6$ (dashed curve). 
The whole region inside this limiting surface is not accessible since the condition $|{\bar {\mathcal H}}|\ll1$ is not satisfied there (in fact, in this region we have that $|{\bar {\mathcal H}}|>m/{\mathcal M}$). 
The black circle represents the black hole horizon. 
The \lq\lq induced'' horizon around the particle location is also shown (shaded region) for the further choice $m/{\mathcal M}=0.1$.
}
\label{fig:chap5fig2}
\end{figure}

Consider first the gravitational perturbation function $Y$, whose solution is given by Eq. (\ref{solYfinale}) for $l\geq1$ and Eq. (\ref{RNleq0solW}), with $W=Y/2$, for $l=0$. The sum over all multipoles turns out to be
\begin{eqnarray}
\label{bary}
\bar y&=&\sum_{l=0}^\infty Y Y_{l0}=\sum_{l=1}^\infty Y Y_{l0}+\frac{1}{2\sqrt{\pi}}Y\vert_{l=0}\nonumber\\
&=&-4q\frac{Q}{{\mathcal M}b-Q^2}\frac{1}{\Gamma r}(b-r_+)(b-r_-)(r-r_+)(r-r_-)\nonumber\\
&&\times\sum_{l=1}^\infty\frac{2l+1}{l(l+1)}\left[\frac{dQ_l(z(r))}{dr}\bigg\vert_{r=b}\frac{dP_l(z(r))}{dr}\vartheta(b-r)+\frac{dP_l(z(r))}{dr}\bigg\vert_{r=b}\frac{dQ_l(z(r))}{dr}\vartheta(r-b)\right]P_l(\cos\theta)\nonumber\\
&&+4\frac{m}{b}f(b)^{-1/2}\frac1r\left[(r-{\mathcal M})\vartheta(b-r)+(b-{\mathcal M})\vartheta(r-b)\right]\nonumber\\
&=&4q\frac{Q}{{\mathcal M}b-Q^2}\frac1{r}\frac{(r-{\mathcal M})(b-{\mathcal M}) -\Gamma^2\cos\theta}{{\bar {\mathcal D}}}\ ,
\end{eqnarray}
where the quantity ${\bar {\mathcal D}}=D_{\rm{RN}}$ is equal to
\begin{equation}
\label{barD}
{\bar {\mathcal D}} = [(r-{\mathcal M})^2+(b-{\mathcal M})^2 - 2(r -{\mathcal M})(b-{\mathcal M})\cos\theta - \Gamma^2\sin^2\theta]^{1/2}\ , 
\end{equation}
and the representation formula (\ref{sommacopson}) has been used.
Since $K=W=Y/2$ and taking into account the relation (\ref{compcondnextr}), we have that the solution summed over the harmonics for the perturbed gravitational field turns out to be completely determined by the function
\begin{equation}
\label{barH}
{\bar {\mathcal H}}=\frac{{\bar y}}2 =2\frac{m}{br}f(b)^{-1/2}\frac{(r-{\mathcal M})(b-{\mathcal M}) -\Gamma^2\cos\theta}{{\bar {\mathcal D}}}\ ,
\end{equation}  
so that the new line element $d{\tilde s}^2$ from the first of relations (\ref{pertrelations}) and Eq. (\ref{hmunuextrY}) is then 
\begin{equation}
\label{lineelemnonextr}
d{\tilde s}^2=-[1-{\bar {\mathcal H}}]f(r)dt^2 + [1+{\bar {\mathcal H}}][f(r)^{-1}dr^2+r^2(d\theta^2 +\sin ^2\theta d\phi^2)]\ .
\end{equation}  
The asymptotic mass measured at large distances by the Schwarzschild-like behaviour of the metric of the whole system consisting of black hole and particle is given by
\begin{equation}
M_{\rm eff}={\mathcal M}+m+E_{\rm int}\ ,
\end{equation} 
where the interaction energy turns out to be
\begin{equation}
E_{\rm int}=-m\left[1-\left(1-\frac{{\mathcal M}}{b}\right)f(b)^{-1/2}\right]\ .
\end{equation} 
It can be shown that this perturbed metric is spatially conformally flat; moreover, the solution remains valid as long as the condition $|{\bar {\mathcal H}}|\ll1$ is satisfied. Let us choose the mass $m$ itself as a smallness indicator.
Figure~\ref{fig:chap5fig2} thus shows the regions where our perturbative treatment fails, assuming that $|{\bar {\mathcal H}}|\approx m/{\mathcal M}$. 

Introduce isotropic coordinates $\{t,\rho,\theta,\phi\}$ such that
\begin{equation}
r(\rho)={\mathcal M}+\rho+\frac{\Gamma^2}{4\rho}\ .
\end{equation}  
The perturbed metric (\ref{lineelemnonextr}) then becomes
\begin{eqnarray}
\label{metrisotr}
d{\tilde s}^2=-[1-{\bar {\mathcal H}}(\rho)]A(\rho)dt^2 + [1+{\bar {\mathcal H}}(\rho)]B(\rho)^2[d\rho^2+\rho^2(d\theta^2 +\sin ^2\theta d\phi^2)]\ ,
\end{eqnarray}
where 
\begin{equation}
A(\rho)=\frac{1}{r(\rho)^2}\left[\rho^2-\frac{\Gamma^2}{4}\right]^2\ , \qquad B(\rho)=\frac{r(\rho)}{\rho}\ ,
\end{equation}
and 
\begin{equation}
{\bar {\mathcal H}}(\rho)=-2\frac{m}{r(\rho)}\frac{\beta}{4\beta^2-\Gamma^2}\left[\left(\frac{D_2}{D_1}\right)^{1/2}+4\Gamma^2\left(\frac{D_1}{D_2}\right)^{1/2}\right]\ ,
\end{equation} 
with
\begin{eqnarray}
D_1 =\rho^2+\beta^2 - 2\rho\beta\cos\theta\ , \qquad
D_2 =16\rho^2\beta^2 - 8\Gamma^2\rho\beta\cos\theta+\Gamma^4\ , \qquad
2\beta=-{\mathcal M}+b+bf(b)^{1/2}\ . 
\end{eqnarray}
By inspection of (\ref{metrisotr}) we see immediately that the spatial 3-metric 
\begin{equation}
{}^{(3)}d{\tilde s}^2=\Psi[d\rho^2+\rho^2(d\theta^2 +\sin ^2\theta d\phi^2)]\ ,
\end{equation}
is conformally flat, with conformal factor 
\begin{equation}
\Psi=[1+{\bar {\mathcal H}}(\rho)]B(\rho)^2\ .
\end{equation}

The condition ${\tilde g}_{tt}=0$ defines the new coordinate horizons for the system \lq\lq black hole + particle'':
\begin{equation}
\label{horRNnonextr}
0=f(r)[1-{\bar {\mathcal H}}]\ ,
\end{equation}
where the quantity ${\bar {\mathcal H}}$ is given by Eq.~(\ref{barH}).
In addition to the usual black hole inner and outer horizons $r_-$ and $r_+$ there would exist another one, implicitly defined by the equation ${\bar {\mathcal H}}=1$, as a function of the variables $r$ and $\theta$.
Hence, the presence of the black hole seems to induce the formation of a small horizon around the particle, which however would lie inside the nonaccessible region (see Fig. \ref{fig:chap5fig2}). 
Therefore, such a conclusion cannot be drawn since our (perturbative) treatment is no longer valid there.

% figure 2

The reconstruction of the perturbed electromagnetic field $f_{\mu\nu}$ by means of relations (\ref{effe01Xzero}) and (\ref{effe02Xzero}) and by using the closed form expression (\ref{barH}) for the gravitational perturbation function ${\bar {\mathcal H}}$ (or Eq. (\ref{bary}) for ${\bar{y}}$) is the following
\begin{eqnarray}
  \label{effe01nonextrreconstr}
f_{{01}}&=&\sum_l \tilde f_{{01}}Y_{l0}=\frac{{\mathcal M}r-Q^2}{4rQ}\frac{\partial {\bar y}}{\partial r}+\frac{Q}{4r^2}{\bar y}\ , \\
  \label{effe02nonextrreconstr}
f_{{02}}&=&\sum_l \tilde f_{{02}}\frac{\partial Y_{l0}}{\partial \theta}=\frac{{\mathcal M}r-Q^2}{4rQ}\frac{\partial {\bar y}}{\partial \theta}\ ,
\end{eqnarray}
so that the electric field components $E_r$ and $E_{\theta}$ are given by
\begin{eqnarray} 
\label{Zeremtensorpertnonextr} 
E_r=-f_{01}&=&\frac{q}{r^3}\frac{{\mathcal M}r-Q^2}{{\mathcal M}b-Q^2}\frac1{{\bar {\mathcal D}}}\bigg\{-\bigg[{\mathcal M}(b-{\mathcal M})+\Gamma^2\cos\theta+[(r-{\mathcal M})(b-{\mathcal M})-\Gamma^2\cos\theta]\frac{Q^2}{{\mathcal M}r-Q^2}\bigg]\nonumber\\
&&+\frac{r[(r-{\mathcal M})(b-{\mathcal M})-\Gamma^2\cos\theta]
[(r-{\mathcal M}) -(b-{\mathcal M})\cos\theta]}{{\bar {\mathcal D}}^2}\bigg\}\ ,
\nonumber\\ 
E_{\theta}=-f_{02}&=&
q\frac{{\mathcal M}r-Q^2}{{\mathcal M}b-Q^2}
\frac{b^2f(b)f(r)}{{\bar {\mathcal D}}^3}\sin\theta\ ; 
\end{eqnarray} 
the total electromagnetic field to first order in the perturbations is then (from the second of relations (\ref{pertrelations})) 
\begin{equation}
\label{RNemfieldpertnonextr}
\tilde F=-\left[\frac{Q}{r^2}+E_r\right]dt\wedge dr - E_{\theta}dt\wedge d\theta\ .
\end{equation}

We are left to verify that Gauss's theorem (\ref{gausspert}) is satisfied also in this general non-extreme case. 
The only non-vanishing component ${}^*{\tilde F}_{\theta\phi}$ of the dual of the electromagnetic form (\ref{RNemfieldpertnonextr}) is given by (to first order in the perturbation)
\begin{equation}
{}^*{\tilde F}_{\theta\phi}=r^2\sin\theta\left[(1+{\bar {\mathcal H}})\frac{Q}{r^2}+E_r\right]\ .
\end{equation}
Let us calculate separately the two different contributions to the integral (\ref{gausspert}) due to the background and particle electric fields. The former one is given by
\begin{eqnarray}
\label{fluxRNnextr}
\Phi_{\rm RN}=2\pi\int_{0}^{\pi}r^2\sin\theta	(1+{\bar {\mathcal H}})\frac{Q}{r^2} d\theta
=4\pi Q + 2\pi Q\int_{0}^{\pi}{\bar {\mathcal H}}\sin\theta d\theta\ .
\end{eqnarray}
From relations (\ref{compcondnextr}), (\ref{barD}) and (\ref{barH}) it is easy to show that
\begin{eqnarray}
\label{barHintegral}
\int_{0}^{\pi}{\bar {\mathcal H}}\sin\theta d\theta&=&\frac{2qQ}{{\mathcal M}b-Q^2}\frac1r\int_{0}^{\pi}\frac{\partial {\bar {\mathcal D}}}{\partial \theta}d\theta
=\frac{2qQ}{{\mathcal M}b-Q^2}\frac1r [{\bar {\mathcal D}}(\pi)-{\bar {\mathcal D}}(0)]\ .
\end{eqnarray}
Thus the flux (\ref{fluxRNnextr}) becomes
\begin{eqnarray}
\label{fluxRNnextrnew}
\Phi_{\rm RN}=4\pi Q + \frac{4\pi qQ^2}{{\mathcal M}b-Q^2}\frac1r [{\bar {\mathcal D}}(\pi)-{\bar {\mathcal D}}(0)]\ .
\end{eqnarray}
Relation (\ref{effe02nonextrreconstr}) leads to an expression for the perturbed electrostatic potential 
\begin{equation}
\label{Vpert}
V=\frac{{\mathcal M}r-Q^2}{2rQ}{\bar {\mathcal H}}\ ,
\end{equation}
where ${\bar {\mathcal H}}={\bar y}/2$ from Eq. (\ref{barH}). 
Hence the contribution to the flux due to the charged particle is given by
\begin{eqnarray}
\label{fluxnextrpart}
\Phi_{\rm part}&=&2\pi\int_{0}^{\pi}r^2\sin\theta	E_r d\theta
=-2\pi r^2 \frac{\partial}{\partial r}\left[\frac{{\mathcal M}r-Q^2}{2rQ}\int_{0}^{\pi}{\bar {\mathcal H}}\sin\theta d\theta\right]\nonumber\\
&=&-\frac{2\pi q}{{\mathcal M}b-Q^2}r^2 \frac{\partial}{\partial r}\left\{\frac{{\mathcal M}r-Q^2}{r^2}[{\bar {\mathcal D}}(\pi)-{\bar {\mathcal D}}(0)]\right\}\ ,
\end{eqnarray}
after using the expression (\ref{barHintegral}). Therefore the total flux $\Phi$ turns out to be
\begin{eqnarray}
\Phi=\Phi_{\rm RN}+\Phi_{\rm part}&=&4\pi Q+\frac{2\pi q}{{\mathcal M}b-Q^2}\left\{{\mathcal M}[{\bar {\mathcal D}}(\pi)-{\bar {\mathcal D}}(0)]-({\mathcal M}r-Q^2)\frac{\partial}{\partial r}[{\bar {\mathcal D}}(\pi)-{\bar {\mathcal D}}(0)]\right\}\nonumber\\
&=&4\pi Q-4\pi q\left[\frac{\Gamma^2}{{\mathcal M}b-Q^2}\vartheta (b-r)-\frac{{\mathcal M}(b-{\mathcal M})}{{\mathcal M}b-Q^2}\vartheta (r-b)\right]\nonumber\\
&=&4\pi Q+4\pi q\vartheta (r-b)-4\pi q\frac{\Gamma^2}{{\mathcal M}b-Q^2}\ ,
\end{eqnarray} 
since
\begin{equation}
{\bar {\mathcal D}}(\pi)-{\bar {\mathcal D}}(0)=2[(r-{\mathcal M})\vartheta(b-r)+(b-{\mathcal M})\vartheta(r-b)]\ ,
\end{equation}
from Eq. (\ref{barD}). So Gauss's theorem is not satisfied. However, it is enough to add to the perturbed electrostatic potential (\ref{Vpert}) the following term
\begin{equation}
\label{barV}
{\bar V}=q\frac{\Gamma^2}{{\mathcal M}b-Q^2}\frac1r\ ,
\end{equation}
which vanishes in the extreme case and whose contribution to the flux is just
\begin{equation}
\label{barPhi}
{\bar \Phi}=4\pi q\frac{\Gamma^2}{{\mathcal M}b-Q^2}\ .
\end{equation}
Hence the total flux $\Phi$ is given by
\begin{eqnarray}
\Phi=\Phi_{\rm RN}+\Phi_{\rm part}+{\bar \Phi}=4\pi Q+4\pi q\vartheta (r-b)\ ,
\end{eqnarray} 
and Gauss's theorem (\ref{gausspert}) is satisfied.
It is worth noting that the addition of the term (\ref{barV}) also leads to a change in the perturbed metric functions: from Eqs. (\ref{eq1RNleq0sol2})--(\ref{eq3RNleq0sol2}) it is easy to show that the solution for $W$ changes simply by a constant, implying a modification of the function ${\bar {\mathcal H}}$ by the constant term $-2q\Gamma^2/[Q({\mathcal M}b-Q^2)]$. However this term can be eliminated by a suitable gauge transformation of the perturbed metric. 

Let us denote by $V_{\rm test}$ the electrostatic potential of the particle alone obtained within the test-field approximation by Leaute and Linet \cite{leaute}, and by
\begin{equation}
V_{\rm tot}=V+{\bar V}+V^{\rm{BH}}\ 
\end{equation}
the total perturbed electrostatic potential obtained by summing (\ref{Vpert}) and (\ref{barV}) plus the contribution (\ref{BHpot}) of the black hole itself. 
Direct comparison between the potentials shows that they are related as follows
\begin{equation}
\label{relperttest}
V_{\rm tot}=V_{\rm test}+\left[1+\frac12\left(1-\frac{r}b\right){\bar {\mathcal H}}
+\frac{qQ}{{\mathcal M}b-Q^2}\left(1-\frac{{\mathcal M}}b\right)\right]V^{\rm{BH}}\ . 
\end{equation}
The second and third terms in the bracketed expression of (\ref{relperttest}) represent the \lq\lq gravitationally induced'' and \lq\lq electromagnetically induced'' electrostatic potential respectively and the equilibrium condition (\ref{bonnoreqcond}) has been conveniently used.

\subsection{Comparison with the Weyl class double Reissner-Nordstr\"om solution}

The solution for two Reissner-Nordstr\"om black holes belonging to the Weyl class, once linearized with respect to the mass $m$ and charge $q$ of one of them, is given by
\begin{equation}
\label{twoRNweylsol}
d{\tilde s}^2=-f(r)[1-{\bar h}^{\rm w}_0]dt^2+f(r)^{-1}[1+{\bar h}^{\rm w}_1]dr^2+r^2[1+{\bar h}^{\rm w}_2]d\theta^2+r^2\sin\theta^2[1+{\bar h}^{\rm w}_3]d\phi^2\ ,
\end{equation}
where
\begin{eqnarray}
\label{weylgeomfunctsRN}
{\bar h}^{\rm w}_0&=&\frac{2m}{{\mathcal D}_{\rm w}}\frac{{\mathcal M}r-Q^2}{{\mathcal M}r}-2\frac{Q}{{\mathcal M}r}(q{\mathcal M}-Qm)\frac{r-{\mathcal M}+{\mathcal M}\cos^2\theta}{(r-{\mathcal M})^2-\Gamma^2\cos^2\theta}\ ,\nonumber\\
{\bar h}^{\rm w}_1&=&{\bar h}^{\rm w}_0-\frac{4m}{b^2-\Gamma^2}\frac{\Gamma^2}{{\mathcal M}}\left[1-\frac{r-{\mathcal M}-b\cos\theta}{{\mathcal D}_{\rm w}}\right]+4\frac{Q}{{\mathcal M}}\Gamma^2(q{\mathcal M}-Qm)\frac{\cos^2\theta\sin^2\theta}{[(r-{\mathcal M})^2-\Gamma^2\cos^2\theta]^2}\ ,\nonumber\\
{\bar h}^{\rm w}_2&=&{\bar h}^{\rm w}_1\ ,\nonumber\\
{\bar h}^{\rm w}_3&=&{\bar h}^{\rm w}_0\ ,
\end{eqnarray}
and
\begin{equation}
V_{\rm w}=\frac{Q}{r}+\frac{Qr}{{\mathcal M}r-Q^2}\left[\frac{f(r)}{2}{\bar h}^{\rm w}_0+\frac{q{\mathcal M}-Qm}{Qr}\right]\ , 
\end{equation}
with
\begin{equation}
{\mathcal D}_{\rm w}= [(r-{\mathcal M})^2+b^2 - 2(r -{\mathcal M})b\cos\theta - \Gamma^2\sin^2\theta]^{1/2}\ . 
\end{equation}
Direct comparison of this solution with our solution shows that they do not coincide.
In addition, the metric (\ref{twoRNweylsol}) is characterized by the presence of a conical singularity between the bodies, which can be removed only if both of them are critically charged (see Appendix B); our solution is instead totally free of singularities.

\section{Conclusions}

The problem of the interaction of a massive charged particle at rest with a Reissner-Nordstr\"om black hole has been studied taking into account both electromagnetic and gravitational perturbations on the background fields due to the presence of the particle.
Following Zerilli's approach to the perturbations of a charged static black hole, we derived the corresponding solutions of the linearized Einstein-Maxwell equations for both the electromagnetic and gravitational perturbation functions. We were
able to exactly reconstruct these functions summing over all multipoles, giving closed form expressions for the components of the perturbed metric as well as the electromagnetic field.  
A detailed analysis of the properties of this solution will be presented elsewhere.

\begin{acknowledgments}
The authors are indebted to V. Belinski, C. Cherubini, G. Cruciani,  R.T. Jantzen and B. Mashhoon for useful discussions.
\end{acknowledgments}

\appendix

\section{The Weyl class two-body solution}

Axisymmetric static vacuum solutions of the Einstein field equations can be described by the Weyl formalism \cite{weyl}. The line element in coordinates $(t,\rho,z,\phi)$ is given by
\begin{equation}
\label{weylmetric}
ds^2=-e^{2\psi}dt^2+e^{2(\gamma-\psi)}[d\rho^2+dz^2]+\rho^2e^{-2\psi}d\phi^2\ ,
\end{equation}
where the function $\psi$ and $\gamma$ depend on coordinates $\rho$ and $z$ only. 
The vacuum Einstein field equations in Weyl coordinates reduce to 
\begin{eqnarray}
\label{einsteqs}
0&=&\psi_{,\rho\rho}+\frac1\rho\psi_{,\rho}+\psi_{,zz}\ ,\nonumber\\
0&=&\gamma_{,\rho}-\rho[\psi_{,\rho}^2-\psi_{,z}^2]\ ,\nonumber\\
0&=&\gamma_{,z}-2\rho\psi_{,\rho}\psi_{,z}\ .
\end{eqnarray}
The first equation is the 3-dimensional Laplace equation
in cylindrical coordinates; so the function $\psi$ plays
the role of a newtonian potential. The linearity of that equation
allows to find explicit solutions representing superpositions of
two or more axially symmetric bodies. In general, these solutions
correspond to configurations which are not gravitationally stable because of
the occurrence of gravitationally inert singular structures
(``struts'' and ``membranes'') that keep the bodies apart making the configuration as stable.
In the case of collinear distributions 
of matter displaced along the symmetry axis, this fact is revealed by 
the presence of a conical singularity on the
axis, the occurrence of which is related to the non-vanishing
of the function $\gamma(\rho,z)$ on the portion of the axis
between the sources or outside them.
In the former case we can interpret the singular segment of the axis as a strut holding the bodies apart, while in the latter one as a pair of cords on which the bodies are suspended. 

For the static axisymmetric vacuum solutions the regularity condition on the axis of symmetry (``elementary flatness'') is given by
\begin{equation}
\label{regcond}
\lim_{\rho\rightarrow0}\gamma=0\ .
\end{equation}

\subsubsection{Superposition of a Schwarzschild black hole and a Chazy-Curzon particle}

The solution corresponding to a point particle of mass $m$ at rest on the symmetry axis at $z=b$ above the horizon of a Schwarzschild black hole with mass ${\mathcal M}$ at the origin is given by the metric (\ref{weylmetric}) with functions
\begin{eqnarray}
\label{psigammaSCb}
\psi=\psi_{\rm S}+\psi_{\rm C_b}\ , \qquad \gamma=\gamma_{\rm S}+\gamma_{\rm C_b}+\gamma_{\rm SC_b}\ ,
\end{eqnarray}
where 
\begin{eqnarray}
\label{ccsolbw}
\psi_{\rm C_b}=-\frac{m}R_{\rm C_b}\ , \qquad \gamma_{\rm C_b}=-\frac12\frac{m^2\rho^2}{R_{\rm C_b}^4}\ , \qquad R_{\rm C_b}=\sqrt{\rho^2+(z-b)^2}\ , 
\end{eqnarray}
and
\begin{eqnarray}
\label{Ssolw}
\psi_{\rm S}&=&\frac12\ln{\left[\frac{R_1+R_2-2{\mathcal M}}{R_1+R_2+2{\mathcal M}}\right]}\ ,\qquad 
\gamma_{\rm S}=\frac12\ln{\left[\frac{(R_1+R_2)^2-4{\mathcal M}^2}{4R_1R_2}\right]}\ ,\nonumber\\
R_1&=&\sqrt{\rho^2+(z-{\mathcal M})^2}\ ,\qquad\qquad\, 
R_2=\sqrt{\rho^2+(z+{\mathcal M})^2}\ ,
\end{eqnarray}
while $\gamma_{\rm SC_b}$ can be obtained by solving Einstein's equations (\ref{einsteqs})
\begin{equation}
\gamma_{\rm SC_b}=-\frac{m}{b^2-{\mathcal M}^2}\frac{(b+{\mathcal M})R_1-(b-{\mathcal M})R_2}{R_{\rm C_b}}+C\ .
\end{equation}
The value of arbitrary constant $C$ can be determined by imposing the elementary flatness condition (\ref{regcond}).
We can make $\gamma_{\rm SC_b}$ zero between the mass and the Schwarzschild source by choosing $C=-2m{\mathcal M}/(b^2-{\mathcal M}^2)$, or outside them by taking $C$ to have the opposite sign, but $C$ can not be chosen so that $\gamma_{\rm SC_b}$ vanishes on the whole $z$ axis, giving rise to the well known conical singularity, corresponding to a strut in compresson which holds the black hole and particle apart. 

The relation with standard Boyer-Lindquist coordinates in the Schwarzschild case is simply
\begin{equation}
\label{wtobl}
\rho=\sqrt{r^2-2{\mathcal M}r}\sin\theta\ , \qquad z=(r-{\mathcal M})\cos\theta\ .
\end{equation}
In Boyer-Lindquist coordinates, under the transformation (\ref{wtobl}) the metric (\ref{weylmetric}) re-writes as
\begin{eqnarray}
\label{Schwmetric}
ds^2&=&-e^{2\psi}dt^2+e^{2(\gamma-\psi)}[(r-{\mathcal M})^2-{\mathcal M}^2\cos^2\theta]\left[\frac{dr^2}{r(r-2{\mathcal M})}+d\theta^2\right]+e^{-2\psi}(r^2-2{\mathcal M}r)\sin^2\theta d\phi^2\ ,
\end{eqnarray}
with the functions (\ref{psigammaSCb}) given by
\begin{eqnarray}
\label{exSCsolftsbl}
\psi_{\rm S}&=&\frac12\ln{\left(1-\frac{2{\mathcal M}}r\right)}\ ,\qquad
\gamma_{\rm S}=\frac12\ln{\left[\frac{r(r-2{\mathcal M})}{(r-{\mathcal M})^2-{\mathcal M}^2\cos^2\theta}\right]}\ ,\nonumber\\
\psi_{\rm C_b}&=&-\frac{m}{R}\ ,\qquad\qquad\qquad\quad
\gamma_{\rm C_b}=-\frac12\frac{m^2r(r-2{\mathcal M})\sin^2\theta}{R^4}\ ,\nonumber\\
\gamma_{\rm SC_b}&=&-\frac{2m{\mathcal M}}{b^2-{\mathcal M}^2}\left[\frac{r-{\mathcal M}-b\cos\theta}{R}-1\right]\ ,\nonumber\\
R&=&[(r-{\mathcal M})^2+b^2-2(r-{\mathcal M})b\cos\theta-{\mathcal M}^2\sin^2\theta]^{1/2}\ ,
\end{eqnarray}
by virtue of the choice $C=2m{\mathcal M}/(b^2-{\mathcal M}^2)$ for the arbitrary costant $C$.
It is easy to show that the linearization of this exact solution with respect to $m$ is just the metric (\ref{weylsol}).

\subsubsection{Superposition of two Schwarzschild black holes}

The solution corresponding to a linear superposition of two Schwarzschild black hole with masses ${\mathcal M}$ and $m$ and positions $z=0$ and $z=b$ on the $z$ axis respectively is given by metric (\ref{weylmetric}) with functions
\begin{eqnarray}
\label{psigammaSSb}
\psi=\psi_{\rm S}+\psi_{\rm S_b}\ , \qquad \gamma=\gamma_{\rm S}+\gamma_{\rm S_b}+\gamma_{\rm SS_b}\ ,
\end{eqnarray}
where $\psi_{\rm S}$ and $\gamma_{\rm S}$ are given by (\ref{Ssolw}), and
\begin{eqnarray}
\psi_{\rm S_b}&=&\frac12\ln{\left[\frac{R_2^{+}+R_2^{-}-2m}{R_2^{+}+R_2^{-}+2m}\right]}\ ,\qquad
\gamma_{\rm S_b}=\frac12\ln{\left[\frac{(R_2^{+}+R_2^{-})^2-4m^2}{4R_2^{+}R_2^{-}}\right]}\ ,\nonumber\\
\gamma_{\rm SS_b}&=&\frac12\ln{\left[\frac{E_{(1^{+},2^{-})}E_{(1^{-},2^{+})}}{E_{(1^{+},2^{+})}E_{(1^{-},2^{-})}}\right]}+C\ , \qquad
E_{(1^{\pm},2^{\pm})}=\rho^2+R_1^{\pm}R_2^{\pm}+Z_1^{\pm}Z_2^{\pm}\ ,\nonumber\\
R_1^{\pm}&=&\sqrt{\rho^2+(Z_1^{\pm})^2}\ ,\qquad
R_2^{\pm}=\sqrt{\rho^2+(Z_2^{\pm})^2}\ ,\nonumber\\
Z_1^{\pm}&=&z\pm{\mathcal M}\ ,\qquad
Z_2^{\pm}=z-(b\mp m)\ ,
\end{eqnarray}
the function $\gamma_{\rm SS_b}$ being obtained by solving Einstein's equations (\ref{einsteqs}).
The value of arbitrary constant $C$ can be determined by imposing the regularity condition (\ref{regcond}).
A unique choice of the arbitrary constant $C$ allowing to make zero the function $\gamma_{\rm SS_b}$ on the whole $z$ axis does not exist: in fact it vanishes on the segment ${\mathcal M}<z<b-m$ between the sources for $C=-\ln{([b^2-({\mathcal M}+m)^2]/[b^2-({\mathcal M}-m)^2])}$, and outside them (that is, for $z>b+m$ and $z<-{\mathcal M}$) if $C$ is chosen to be equal to zero. 

In Boyer-Lindquist coordinates, under the transformation (\ref{wtobl}) the metric (\ref{weylmetric}) is given by (\ref{Schwmetric}) with functions 
\begin{eqnarray}
\label{exSSsolftsbl}
\psi_{\rm S}&=&\frac12\ln{\left(1-\frac{2{\mathcal M}}r\right)}\ ,\qquad
\gamma_{\rm S}=\frac12\ln{\left[\frac{r(r-2{\mathcal M})}{R_1^{+}R_1^{-}}\right]}\ ,\nonumber\\
\psi_{\rm S_b}&=&\frac12\ln{\left[\frac{R_2^{+}+R_2^{-}-2m}{R_2^{+}+R_2^{-}+2m}\right]}\ ,\qquad
\gamma_{\rm S_b}=\frac12\ln{\left[\frac{(R_2^{+}+R_2^{-})^2-4m^2}{4R_2^{+}R_2^{-}}\right]}\ ,\nonumber\\
\gamma_{\rm SS_b}&=&\frac12\ln{\left[\frac{E_{(1^{+},2^{-})}E_{(1^{-},2^{+})}}{E_{(1^{+},2^{+})}E_{(1^{-},2^{-})}}\right]}\ ,\qquad
E_{(1^{\pm},2^{\pm})}=r(r-2{\mathcal M})\sin^2\theta+R_1^{\pm}R_2^{\pm}+Z_1^{\pm}Z_2^{\pm}\ ,\nonumber\\
R_1^{\pm}&=&r-{\mathcal M}\pm{\mathcal M}\cos\theta\ ,\qquad
R_2^{\pm}=[(r-{\mathcal M})^2+(b\mp m)^2-2(r-{\mathcal M})(b\mp m)\cos\theta-{\mathcal M}^2\sin^2\theta]^{1/2}\ ,\nonumber\\
Z_1^{\pm}&=&(r-{\mathcal M})\cos\theta\pm{\mathcal M}\ ,\qquad
Z_2^{\pm}=(r-{\mathcal M})\cos\theta-(b\mp m)\ ,
\end{eqnarray}
by virtue of the choice $C=0$ of the arbitrary costant $C$.
It is easy to show that also in this case the linearization of the solution with respect to $m$ is just the metric (\ref{weylsol}).

\section{The Weyl class charged two-body solution}

Axisymmetric static electrovacuum solutions of the Einstein field equations can be described by the Weyl formalism \cite{weyl}. The line element in coordinates $(t,\rho,z,\phi)$ is given by Eq.~(\ref{weylmetric}).
The electrovacuum Einstein-Maxwell field equations in Weyl coordinates reduce to 
\begin{eqnarray}
\label{eveinsteqs}
0&=&\nabla^2\psi-e^{-2\psi}[V_{,\rho}^2+V_{,z}^2]\ ,\nonumber\\
0&=&\nabla^2V-2[\psi_{,\rho}V_{,\rho}+\psi_{,z}V_{,z}]\ ,\nonumber\\
0&=&\gamma_{,\rho}-\rho\{\psi_{,\rho}^2-\psi_{,z}^2-e^{-2\psi}[V_{,\rho}^2-V_{,z}^2]\}\ ,\nonumber\\
0&=&\gamma_{,z}-2\rho[\psi_{,\rho}\psi_{,z}-e^{-2\psi}V_{,\rho}V_{,z}]\ ,
\end{eqnarray}
where 
\begin{equation}
\nabla^2X=X_{,\rho\rho}+\frac1\rho X_{,\rho}+X_{,zz}\ ,
\end{equation}
and the 4-potential $A_{\mu}$ is determined by the electrostatic potential $V$ only
\begin{equation}
A_{\mu}=-V(\rho, z)dt\ .
\end{equation}
Thus, once the first two equations are solved for $\psi$ and $V$ the last two equations serve to determine $\gamma$.

The solutions belonging to the Weyl class are characterized by the metric function $\psi$ which is a function of the electrostatic potential, 
i.e. $\psi=\psi(V)$, so that the gravitational and electrostatic equipotential surfaces overlap.
Weyl \cite{weyl} showed that for asymptotically flat boundary conditions the unique functional relationship between $\psi$ and $V$ is given by
\begin{equation}
\label{overlap}
e^{2\psi}=1-2\frac{M_{\rm tot}}{Q_{\rm tot}}V+V^2\ ,
\end{equation}
where $M_{\rm tot}$ and $Q_{\rm tot}$ are the total mass and charge of the system, respectively.
The solution can thus be written in terms of only one function, say $f(\rho, z)$, as follows (see e.g. \cite{coopdelacruz})
\begin{equation}
\label{psieVsols}
\psi=\frac12\ln{\left[\frac{f(a^2-1)^2}{(a^2f-1)^2}\right]}\ , \qquad
V=a\frac{f-1}{a^2f-1}\ , 
\end{equation}
where the parameter $a$ is defined by
\begin{equation}
\label{adef}
\frac{1+a^2}{a}=2\frac{M_{\rm tot}}{Q_{\rm tot}}\ .
\end{equation}

The solution representing two Reissner-Nordstr\"om black holes separated by a distance $b$ is given by \cite{perry}
\begin{eqnarray}
f&=&\left(\frac{R_1^{+}+R_1^{-}-2L_1}{R_1^{+}+R_1^{-}+2L_1}\right)\left(\frac{R_2^{+}+R_2^{-}-2L_2}{R_2^{+}+R_2^{-}+2L_2}\right)\ , \nonumber\\
R_1^{\pm}&=&\sqrt{\rho^2+(Z_1^{\pm})^2}\ , \qquad 
R_2^{\pm}=\sqrt{\rho^2+(Z_2^{\pm})^2}\ , \nonumber\\
Z_1^{\pm}&=&z\pm L_1\ , \qquad 
Z_2^{\pm}=z-(b\mp L_2)\ .
\end{eqnarray}
The constants $L_{1,2}$ are the lenghts of the Weyl rods generating the solution, while the parameter $a$ is related to the total mass $M_{\rm tot}=M_1+M_2$ and charge $Q_{\rm tot}=Q_1+Q_2$ of the system through Eq.~(\ref{adef}); in addition, one has 
\begin{eqnarray}
&&M_1=\frac{1+a^2}{1-a^2}L_1\ , \qquad Q_1=\frac{2a}{1-a^2}L_1\ , \nonumber\\
&&M_2=\frac{1+a^2}{1-a^2}L_2\ , \qquad Q_2=\frac{2a}{1-a^2}L_2\ ,
\end{eqnarray}
so that the constants $L_1$ and $L_2$ turn out to be given by 
\begin{equation}
\label{L12sols}
L_1=\sqrt{M_1^2-Q_1^2}\ , \qquad L_2=\sqrt{M_2^2-Q_2^2}\ .
\end{equation}
The metric function $\psi$ and the electrostatic potential $V$ are found through Eq.~(\ref{psieVsols}), 
while the metric function $\gamma$ can be obtained by solving the field equations (\ref{eveinsteqs})
\begin{eqnarray}
\gamma&=&\frac12\ln\left[\left(\frac{(R_1^{+}+R_1^{-})^2-4L_1^2}{4R_1^{+}R_1^{-}}\right)\left(\frac{(R_2^{+}+R_2^{-})^2-4L_2^2}{4R_2^{+}R_2^{-}}\right)\left(\frac{E_{(1^{+},2^{-})}E_{(1^{-},2^{+})}}{E_{(1^{+},2^{+})}E_{(1^{-},2^{-})}}\right)\right]+C\ , \nonumber\\
E_{(1^{\pm},2^{\pm})}&=&\rho^2+R_1^{\pm}R_2^{\pm}+Z_1^{\pm}Z_2^{\pm}\ .
\end{eqnarray}
The value of arbitrary constant $C$ can be determined by imposing the elementary flatness condition (\ref{regcond})
We have that a unique choice of $C$ allowing to make zero the function $\gamma$ on the whole $z$ axis does not exist: in fact, it vanishes on the segment $L_1<z<b-L_2$ between the sources for $C=-\ln{([b^2-(L_1+L_2)^2]/[b^2-(L_1-L_2)^2])}$, and outside them (that is, for $z>b+L_2$ and $z<-L_1$) if $C$ is chosen to be equal to zero. 
The value of $\gamma$ on the symmetry axis between the sources is thus given by
\begin{equation}
\label{axislim2RN}
\lim_{\rho\rightarrow0}\gamma=\ln{\left[\frac{b^2-(L_1+L_2)^2}{b^2-(L_1-L_2)^2}\right]}\ ,
\end{equation}
and can be made vanishing only by imposing that $L_1L_2=0$, which implies that both sources are critically charged, from Eq.~(\ref{L12sols}). 

The relation with standard Boyer-Lindquist coordinates in the Reissner-Nordstr\"om case is simply
\begin{equation}
\label{wtoblRN}
\rho=rf(r)^{1/2}\sin\theta\ , \qquad z=(r-{\mathcal M})\cos\theta\ .
\end{equation}

\end{document}